Discontinuities of energy derivatives in spin-density functional theory.

Ali M. Malek, Robert Balawender*

Institute of Physical Chemistry, Polish Academy of Sciences,
Kasprzaka 44/52, PL-01-224 Warsaw, Poland

Standard approximations for the exchange-correlation functional are known to deviate from linear dependence of the energy on the electron and spin numbers (in $(\mathcal{N},\mathcal{S})$-space). Violation of this flat-plane condition underlies the failure of all known approximate functionals to describe band gaps in strongly correlated systems. Then crucial for further development of functionals is recognition of the behavior of the energy as a function of $(\mathcal{N},\mathcal{S})$, and its derivatives (derivatives discontinuity pattern at 0K limit). In this Letter, the energy derivatives are analyzed thoroughly. It is shown that apart from the well-known type of discontinuity pattern, three other patterns are possible in the vicinity of the singlet state point. The zero temperature limits of energy derivatives at this point are derived, found different for various directions. Existence of all discontinuity patterns is illustrated on the example of diatomic molecules set.

*to whom correspondence should be addressed: rbalawender@ichf.edu.pl

The formally exact ground-state Kohn-Sham (KS) density-functional theory (DFT)[1] is a self-consistent orbital theory in which the only unknown functional of the total electron density $\rho(\mathbf{r})$ is the exchange-correlation energy $E_{xc}[\rho]$; its functional derivative $\delta E_{xc}[\rho]/\delta\rho(\mathbf{r})$ is the exchange-correlation potential $v_{xc}(\mathbf{r};\rho)$. The efficiency and typically useful accuracy of this theory have made it the most popular computational approach in electronic structure calculations. The great success of DFT is clouded by spectacular failures in practical calculations[2]. Missing of some important condition for the exact energy in currently used approximations to $E_{xc}[\rho]$ is vital for these failures. The conventional analysis shows that the discontinuity of the exchange-correlation potential plays a critical role in the correct prediction of band gaps, or the chemical hardness.[3, 4] This discontinuity originates from the generalization of the ground-state DFT to the thermodynamic equilibrium state[5] and the domain extension to a noninteger electron number.[6] The equilibrium-state energy function for a fractional electron number system with an external potential $v(\mathbf{r})$ is a combination of two ground-state energies of integer electron systems, $E[\mathcal{N}=\mathcal{Z}\pm\omega,v]=\omega E[\mathcal{Z}\pm 1,v]+(1-\omega)E[\mathcal{Z},v]$ with $0\leq\omega\leq 1$.[6] It reveals an exact condition for a fractional system, namely, the linearity in $\mathcal{N}$ condition. In the spin extension of the DFT, the linearity condition is combined with the constancy condition: the energy of the ensemble constructed from the components of the spin doublet is independent of the spin number.[7] Standard approximations for the exchange-correlation functional have been found to give big errors for the linearity condition of fractional charges, leading to delocalization error, and for the constancy condition of fractional spins, leading to static correlation error.

In general, extension for fractional state can be done either on ensemble[8] or on pure state theories[9]. In this Letter, discontinuities of the energy derivatives will be investigated in the framework of thermodynamic (ensemble) extension of the spin-density functional theory (SDFT) at zero temperature limit. In the spin-canonical ensemble, the expectation value of the spin-free Hamiltonian $\hat{H}$ is given by

$$E[\beta,\mathcal{N},\mathcal{S};v]=\mathrm{Tr}\,\hat{\Gamma}_{eq}^{[\beta,\mathcal{N},\mathcal{S};v]}\hat{H}[v], \qquad (1)$$

where $\hat{\Gamma}_{eq}[\beta,\mathcal{N},\mathcal{S};v]$ the equilibrium density matrix (eq-DM) dependent on the reciprocal temperature $\beta$, the electron and spin numbers $\mathcal{N}$, $\mathcal{S}$ and the external potential $v(\mathbf{r})$.

The result of maximization of the Lagrange function in the entropy representation with respect to trial DM is the Massieu function (a state function)

$$Y[\beta,\mathcal{N},\mathcal{S};v] \equiv \underset{\hat{\Gamma}}{\text{Max}}\left\{-\text{Tr}\,\hat{\Gamma}\left(\ln\hat{\Gamma}+\beta\hat{H}[v]\right)\Big|\text{Tr}\,\hat{\Gamma}=1, \text{Tr}\,\hat{\Gamma}\hat{\mathcal{N}}=\mathcal{N}, \text{Tr}\,\hat{\Gamma}\hat{\mathcal{S}}=\mathcal{S}\right\}$$

$$= -\text{Tr}\,\hat{\Gamma}_{\text{eq}}^{[\beta,\mathcal{N},\mathcal{S};v]}\left(\ln\hat{\Gamma}_{\text{eq}}^{[\beta,\mathcal{N},\mathcal{S};v]}+\beta\hat{H}[v]\right) \quad (2)$$

$$= S[\beta,\mathcal{N},\mathcal{S};v]-\beta E[\beta,\mathcal{N},\mathcal{S};v]$$

The eq-DM is the maximizer. Here $\hat{\mathcal{N}}$ and $\hat{\mathcal{S}}$ are the electron and spin number operators. Equivalently, the result of minimization of the Lagrange function in the energy representation is the spin Helmholtz function (a state function) with eq-DM as the minimizer,

$$A[\beta,\mathcal{N},\mathcal{S};v] \equiv \underset{\hat{\Gamma}}{\text{Min}}\left\{\text{Tr}\,\hat{\Gamma}\left(\hat{H}[v]+\beta^{-1}\ln\hat{\Gamma}\right)\Big|\text{Tr}\,\hat{\Gamma}=1, \text{Tr}\,\hat{\Gamma}\hat{\mathcal{N}}=\mathcal{N}, \text{Tr}\,\hat{\Gamma}\hat{\mathcal{S}}=\mathcal{S}\right\}$$

$$= \text{Tr}\,\hat{\Gamma}_{\text{eq}}^{[\beta,\mathcal{N},\mathcal{S};v]}\left(\hat{H}[v]+\beta^{-1}\ln\hat{\Gamma}_{\text{eq}}^{[\beta,\mathcal{N},\mathcal{S};v]}\right) \quad (3)$$

$$= E[\beta,\mathcal{N},\mathcal{S};v]-\beta^{-1}S[\beta,\mathcal{N},\mathcal{S};v].$$

The function $S[\beta,\mathcal{N},\mathcal{S};v]=-\text{Tr}\,\hat{\Gamma}_{\text{eq}}^{[\beta,\mathcal{N},\mathcal{S};v]}\ln\hat{\Gamma}_{\text{eq}}^{[\beta,\mathcal{N},\mathcal{S};v]}$ is the entropy of the whole system. Note that $S$ (italic, with arguments) denotes entropy (dependent variable), while $\mathcal{S}$ (calligraphic) denotes the spin number (independent variable).

The eq-DM can be rewritten in terms of the macrostate eq-DM contributions as

$$\hat{\Gamma}_{\text{eq}}^{[\beta,\mathcal{N},\mathcal{S};v]} = \sum_{I} \omega_{(I)}^{[\beta,\mathcal{N},\mathcal{S};v]} \hat{\Gamma}_{\text{eq},(I)}^{[\beta,v]}, \quad (4)$$

where $\hat{\Gamma}_{\text{eq},(I)}^{[\beta,v]}$ and $\omega_{(I)}^{[\beta,\mathcal{N},\mathcal{S};v]}$ are the (I) macrostate eq-DM and weight. Here the macrostate is defined as the set of the microstates, $\{|\Psi_K\rangle\}$, (eigenfunctions of operators $\hat{H}$, $\hat{\mathcal{N}}$ and $\hat{\mathcal{S}}$) such that:

$$(I) \equiv \left\{|\Psi_K\rangle\Big|\ (\hat{\mathcal{N}},\hat{\mathcal{S}})|\Psi_K\rangle = (\mathcal{N}_{(I)},\mathcal{S}_{(I)})|\Psi_K\rangle\right\}, \quad (5)$$

so the $(I)$ macrostate is equivalent to the $\mathcal{S}_{(I)}$-spin subspace of the $\mathcal{N}_{(I)}$-particle Hilbert space (the integers $\mathcal{N}_{(I)}$ and $\mathcal{S}_{(I)}$ are the eigenvalues of two operators, satisfying $\mathcal{N}_{(I)} \geq 0$, $\mathcal{S}_{(I)} \in \{-\mathcal{N}_{(I)}, -\mathcal{N}_{(I)}+2, ..., \mathcal{N}_{(I)}\}$). The macrostate eq-DM, $\hat{\Gamma}_{\text{eq},(I)}^{[\beta,v]}$, is the unique maximizer in the subspace spanned by the microstates belonging to the $(I)$ macrostate in the entropy representation; the $(I)$ macrostate Massieu function is

$$Y_{(I)}^{[\beta;v]} \equiv \underset{\hat{\Gamma}}{\text{Max}}\left\{-\text{Tr}\,\hat{\Gamma}\left(\ln\hat{\Gamma}+\beta\hat{H}[v]\right)\Big|\hat{\Gamma} \in \left\{\sum_{K\in(I)} p_K|\Psi_K\rangle\langle\Psi_K|, \sum_{K\in(I)} p_K = 1, p_K \geq 0\right\}\right\},$$

$$= -\text{Tr}\,\hat{\Gamma}_{\text{eq},(I)}^{[\beta,v]}\left(\ln\hat{\Gamma}_{\text{eq},(I)}^{[\beta,v]}+\beta\hat{H}[v]\right) = S_{(I)}^{[\beta;v]} - \beta E_{(I)}^{[\beta;v]}. \quad (6)$$

Simultaneously $\hat{\Gamma}_{\text{eq},(I)}^{[\beta,v]}$ is the unique minimizer in the energy representation, the macrostate spin Helmholtz function is

$$A_{(I)}^{[\beta;v]} \equiv E_{(I)}^{[\beta;v]} - \beta^{-1} S_{(I)}^{[\beta;v]} = -\beta^{-1} Y_{(I)}^{[\beta;v]}, \tag{7}$$

where the $(I)$ macrostate average energy and entropy are $E_{(I)}^{[\beta;v]} = \text{Tr}\, \hat{\Gamma}_{\text{eq},(I)}^{[\beta;v]} H[v]$ and $S_{(I)}^{[\beta;v]} = -\text{Tr}\, \hat{\Gamma}_{\text{eq},(I)}^{\beta;v} \ln \hat{\Gamma}_{\text{eq},(I)}^{\beta;v}$. So, using Eq.(44), the equilibrium energy (as an expectation value), Eq.(11), can be rewritten as

$$E[\beta, \mathcal{N}, \mathcal{S}; v] = \sum_I \omega_{(I)}^{[\beta,\mathcal{N},\mathcal{S};v]} E_{(I)}^{[\beta,v]}, \tag{8}$$

where the macrostate equilibrium weight is [see Ref [10]]

$$\omega_{(I)}^{[\beta,\mathcal{N},\mathcal{S};v]} = \exp\left(-\beta\left(\left(E_{(I)}^{[\beta;v]} - E^{[\beta,\mathcal{N},\mathcal{S};v]}\right) - \mu_{\text{N}}^{[\beta,\mathcal{N},\mathcal{S};v]}\left(\mathcal{N}_{(I)} - \mathcal{N}\right) - \mu_{\text{S}}^{[\beta,\mathcal{N},\mathcal{S};v]}\left(\mathcal{S}_{(I)} - \mathcal{S}\right)\right)\right) \\ \times \exp\left(S_{(I)}^{[\beta;v]} - S^{[\beta,\mathcal{N},\mathcal{S};v]}\right). \tag{9}$$

Here the chemical and the spin potential, $\mu_{\text{N}}$ and $\mu_{\text{S}}$, are the first derivatives of the spin Helmholtz function.

The generalization of the straight-line connection to include the spin number $\mathcal{S}$ beside $\mathcal{N}$ is a triangular connection formula.[10-12] It follows from Eq.(88) at 0K limit, when $E_{(I)}^{[\beta,v]}$ becomes just the ground state energy of the macrostate, while, for given $(\mathcal{N}, \mathcal{S})$, $\omega_{(I)}$ are nonzero only at three macrostates – the vertices of a triangle containing this point. The equilibrium energy surface in the $(\mathcal{N}, \mathcal{S})$-space is a set of adjacent flat triangles, each one spanned by three points $\left(\mathcal{N}_{(I)}, \mathcal{S}_{(I)}, E_{(I)}^{[v]}\right)$. The resulting surface is a continuous, convex hull. A common side of two triangles is a place of discontinuity of the energy derivative with respect to $\mathcal{N}$ and/or $\mathcal{S}$ variables. The set of these segments (sides) will be named the discontinuity pattern (DP). It can be determined, in principle, for each molecule from the knowledge of the set of its macrostate ground-state energies.[10] Important examples will be described below. It should be noted that the previously mentioned exact conditions for DFT functionals – linearity and constancy – are to be generalized by the flat-energy-surface condition within each $(\mathcal{N}, \mathcal{S})$ triangle.

To investigate the energy surface, Eq.(88), at 0K limit ($\beta \to \infty$), the combination of β-dependent weights, Eq.(99), of four chosen macrostates $\{k\} \equiv (I_k)$, $k = 1, 2, 3, 4$, is introduced

$$p_{\{1,2,3,4\}}^{[\beta,\mathcal{N},\mathcal{S};v]} \equiv \frac{\omega_{\{1\}}^{[\beta,\mathcal{N},\mathcal{S};v]}\omega_{\{3\}}^{[\beta,\mathcal{N},\mathcal{S};v]}}{\omega_{\{2\}}^{[\beta,\mathcal{N},\mathcal{S};v]}\omega_{\{4\}}^{[\beta,\mathcal{N},\mathcal{S};v]}} \qquad (10)$$

$$= \exp\left(-\beta\left(\Delta E_{\{1,2,3,4\}}^{[\beta,v]} - \mu_{\mathrm{N}}^{[\beta,\mathcal{N},\mathcal{S};v]}\Delta\mathcal{N}_{\{1,2,3,4\}} - \mu_{\mathrm{S}}^{[\beta,\mathcal{N},\mathcal{S};v]}\Delta\mathcal{S}_{\{1,2,3,4\}}\right) + \Delta S_{\{1,2,3,4\}}^{[\beta,v]}\right)$$

Here the notation $\Delta O_{\{1,2,3,4\}} \equiv O_{\{1\}} - O_{\{2\}} + O_{\{3\}} - O_{\{4\}}$ is introduced, where $O_{\{k\}}$ is the macrostate counterpart of $O$ defined for whole system, e.g., $A_{(I)}^{[\beta,\mathcal{N},\mathcal{S};v]}$, $S_{(I)}^{[\beta;v]}$, $E_{(I)}^{[\beta;v]}$, $\mathcal{N}_{(I)}$ for $I = I_k$. For further considerations, four macrostates representing consecutive vertices of some *parallelogram* are chosen, because this choice results in $\Delta\mathcal{N}_{\{1,2,3,4\}} = \Delta\mathcal{S}_{\{1,2,3,4\}} = 0$. The dependence on $\mu_{\mathrm{N}}[\beta,\mathcal{N},\mathcal{S};v]$ and $\mu_{\mathrm{S}}[\beta,\mathcal{N},\mathcal{S};v]$ in Eq.(1010) disappears, and the asymptotic form of $p_{\{1,2,3,4\}}^{[\beta,\mathcal{N},\mathcal{S};v]}$ at large $\beta$ is especially simple and independent of $(\mathcal{N},\mathcal{S})$:

$$p_{\{1,2,3,4\}}^{[\beta,\mathcal{N},\mathcal{S};v]} = \exp\left(-\beta A_{\{1,2,3,4\}}^{[\beta,\mathcal{N},\mathcal{S};v]}\right) \simeq \exp\left(-\beta\Delta E_{\{1,2,3,4\}}^{[v]} + \Delta S_{\{1,2,3,4\}}^{[v]}\right). \qquad (11)$$

$E_{(I)}^{[v]}$ and $S_{(I)}^{[v]}$ are the 0K limits of the $(I)$ macrostate average energy and entropy, the ground-state energy and the entropy of the system in the $(\mathcal{N}_{(I)}, \mathcal{S}_{(I)})$ Hilbert subspace (of course, independent of $(\mathcal{N},\mathcal{S})$). From possible limits of $p_{\{1,2,3,4\}}^{[\beta,\mathcal{N},\mathcal{S};v]}$ for $\beta \to \infty$, some conclusions about weights at 0K limit can be drown.

If $\Delta E_{\{1,2,3,4\}}^{[v]} > 0$ then $p_{\{1,2,3,4\}}^{[\beta,\mathcal{N},\mathcal{S};v]} \doteq p_{\{1,2,3,4\}}^{[\mathcal{N},\mathcal{S};v]} = 0$; either both the denominator and the numerator of $p_{\{1,2,3,4\}}^{[\beta,\mathcal{N},\mathcal{S};v]}$ approach zero but the denominator more slowly than the numerator, or $\omega_{\{1\}}^{[\mathcal{N},\mathcal{S};v]}\omega_{\{3\}}^{[\mathcal{N},\mathcal{S};v]} = 0$ and $\omega_{\{2\}}^{[\mathcal{N},\mathcal{S};v]}\omega_{\{4\}}^{[\mathcal{N},\mathcal{S};v]} > 0$ (a notation for the result of the 0K limit is introduced: $\lim_{\beta \to \infty} f[\beta,x] \equiv f[\infty,x] \equiv f[x]$, abbreviated further to $f[\beta,x] \doteq f[x]$). A common feature of these cases can be written as $\omega_{\{1\}}^{[\mathcal{N},\mathcal{S};v]}\omega_{\{3\}}^{[\mathcal{N},\mathcal{S};v]} = 0$, to be described as the non-coexistence property of macrostates $\{1\}$ and $\{3\}$. To see its meaning, let us consider a point $(\mathcal{N},\mathcal{S})$ laying inside the triangle $\{\{1,\},\{2\},\{4\}\}$. Then as discussed earlier, only three weights $\omega_{\{1\}}$, $\omega_{\{2\}}$, $\omega_{\{4\}}$ are nonzero, therefore $\omega_{\{3\}}$ must be zero. Similarly, when $(\mathcal{N},\mathcal{S})$ is inside the triangle $\{\{2\},\{3\},\{4\}\}$, we find $\omega_{\{3\}} \neq 0$, $\omega_{\{1\}} = 0$. So, the non-coexistence of $\{1\}$ and $\{3\}$ means that the discontinuity segment (the common side of triangles) is $\{\{2\},\{4\}\}$ segment; for $(\mathcal{N},\mathcal{S})$ laying on this segment, both $\omega_{\{1\}} = 0$ and $\omega_{\{3\}} = 0$.

When $\Delta E^{[v]}_{\{1,2,3,4\}} < 0$, then $p^{[\beta,\mathcal{N},\mathcal{S};v]}_{\{1,2,3,4\}} \doteq +\infty$, the macrostates $\{2\}$ and $\{4\}$ are non-coexisting, as it follows from the previous conclusion because $p_{\{2,1,4,3\}} = 1/p_{\{1,2,3,4\}}$. The non-accidental $\Delta E^{[v]}_{\{1,2,3,4\}} = 0$ ($p^{[\beta,\mathcal{N},\mathcal{S};v]}_{\{1,2,3,4\}} \doteq \exp\left(\Delta S^{[v]}_{\{1,2,3,4\}}\right)$) can happen when the macrostates $\{1\}$ and $\{2\}$ belong to one multiplet, while the macrostates $\{3\}$ and $\{4\}$ belong also to one multiplet (the same or a different one). Because in our analysis one of the included macrostates will be always the singlet state, this case is out of consideration, only triangle connection formula is discussed.

The analysis of the singlet state vicinity (see Figure 1) can start with a special parallelogram – a square. For vertices $\{\{1,\},\{2\},\{3\},\{4\}\} = \{[-1,+1],[0,0],[+1,+1],[0,+2]\}$, two possible discontinuity segments should be checked: joining the $[0,0]$ and $[0,+2]$ macrostates or the $[+1,+1]$ and $[-1,+1]$ macrostates (here coordinates in square brackets mean $\{I\} \to [\Delta\mathcal{N}_{(I)}, \Delta\mathcal{S}_{(I)}] = (\mathcal{N}_{(I)}, \mathcal{S}_{(I)}) - (\mathcal{Z}, 0)$, i.e., the deviation from the singlet position). In the first case, the $[+1,+1]$ and $[-1,+1]$ macrostates are the vertices of different triangles, this means that $\omega^{[\mathcal{N},\mathcal{S};v]}_{[+1,+1]} \omega^{[\mathcal{N},\mathcal{S};v]}_{[-1,+1]} = 0$. In the second case, the $[0,0]$ and $[0,+2]$ macrostates are non-coexisting. For this parallelogram, we have

$$\Delta E^{[v]}_{\{1,2,3,4\}} = \left(E^{[v]}_{[+1,+1]} + E^{[v]}_{[-1,+1]}\right) - \left(E^{[v]}_{[0,0]} + E^{[v]}_{[0,+2]}\right) = \eta^{[v]} - \Delta E^{[v]}_{ST}, \tag{12}$$

where $\eta^{[v]} = E^{[v]}_{[+1,+1]} + E^{[v]}_{[-1,+1]} - 2E^{[v]}_{[0,0]}$ is the chemical hardness — the difference between the ionization energy, $I = E^{[v]}_{[-1,+1]} - E^{[v]}_{[0,0]}$, and the electron affinity, $A = E^{[v]}_{[0,0]} - E^{[v]}_{[+1,+1]}$, and $\Delta E^{[v]}_{ST} = E^{[v]}_{[0,+2]} - E^{[v]}_{[0,0]}$ is the singlet-triplet excitation energy. Now, $\Delta E^{[v]}_{\{1,2,3,4\}} > 0$ means that the hardness is larger than the singlet-triplet excitation energy. Validity of such inequality for coulombic systems is confirmed by the experimental and calculated data,[13] and theoretical study.[10, 14, 15] Due to the symmetry with respect to $\mathcal{S}$ variable at absence of the magnetic field, the same inequality is true for the opposite sign of spins. This means that anionic and cationic doublets do not coexist at 0K limit, the singlet and triplet macrostate are joined by the energy derivative discontinuity segment. Similar analysis can be repeated e.g. for the $\{[+1,+1],[+2,0],[+1,-1],[0,0]\}$ square, resulting in the discontinuity segment joining $[+1,+1]$ and $[+1,-1]$. A common feature is the presence of discontinuity segments belonging to the lines $\mathcal{N} = \text{const} = \text{integer}$ (blue lines on Figure 1). These discontinuities are generalization of a spinless case: the derivative discontinuity for the system at an integer electron number at 0K limit.[6] For the parallelogram with vertices $\{\{1,\},\{2\},\{3\},\{4\}\}=\{[0,0],[+1,+1],[+1,-1],[0,+2]\}$ we find $\Delta E^{[v]}_{\{1,2,3,4\}}=\Delta E^{[v]}_{ST}>0$, so the discontinuity segment joins the $[0,0]$ and $[+1,+1]$ macrostates. Replacing the triplet component $[0,+2]$ by its opposite spin counterpart yields the discontinuity segment $[0,0]$, $[+1,-1]$. The same is true for the electron removing. In general, the singlet macrostate and the macrostate which is one of the ionic doublet components are always joined by the discontinuity segment (green lines on Figure 1). Finally, a parallelogram having vertices $\{\{1,\},\{2\},\{3\},\{4\}\}=\{[0,0],[+1,+1],[+1,+3],[0,+2]\}$ is considered. For it, $\Delta E^{[v]}_{\{1,2,3,2\}}$ can be rewritten as

$$\Delta E^{[v]}_{\{1,2,3,4\}}=\left(E^{[v]}_{[+1,+3]}-E^{[v]}_{[+1,+1]}\right)-\left(E^{[v]}_{[0,+2]}-E^{[v]}_{[0,0]}\right)=\Delta E^{[v],+}_{DQ}-\Delta E^{[v]}_{ST}, \qquad (13)$$

where $\Delta E^{[v],+}_{DQ}=E^{[v]}_{[+1,+3]}-E^{[v]}_{[+1,+1]}$ is the doublet-quadruplet excitation energy for anionic system. If $\Delta E^{[v]}_{\{1,2,3,4\}}>0$, what means that $\Delta E^{[v],+}_{DQ}>\Delta E^{[v]}_{ST}$, the doublet and triplet state are joined by discontinuity segment. If $\Delta E^{[v]}_{\{1,2,3,4\}}<0$ ( $\Delta E^{[v]}_{\{2,1,4,3\}}>0$ ), the singlet and quadruplet state are joined by the discontinuity segment. Other parallelograms can be analyzed similarly.

All possible DPs for a singlet state (a set of the discontinuity segments having common vertex at this reference point) are presented at Figure 2 with their labels. For the electron addition, there are two cases: with two (T) or four (F) inclined lines. The same is for the electron removing. In the spinless DFT, it was found that when the particle number crosses any integer, the energy derivative experiences a discontinuity.[3, 4, 6] In the SDFT, the situation is more complicated. The energy gradient at 0K limit

$$\left(\left(\frac{\partial E[\beta,\mathcal{N},\mathcal{S};v]}{\partial \mathcal{N}}\right)_{\beta,\mathcal{S};v},\left(\frac{\partial E[\beta,\mathcal{N},\mathcal{S};v]}{\partial \mathcal{S}}\right)_{\beta,\mathcal{N};v}\right)\doteq\left(\mu^{[\mathcal{N},\mathcal{S};v]}_{N},\mu^{[\mathcal{N},\mathcal{S};v]}_{S}\right)\equiv\mu[\mathcal{N},\mathcal{S};v] \quad (14)$$

is to be analyzed. It can be calculated from Eq.(99), using the normalization condition and the spin canonical conditions, $\sum_{I}\omega_{(I)}=1$, $\sum_{I}\omega_{(I)}\mathcal{N}_{(I)}=\mathcal{N}$ and $\sum_{I}\omega_{(I)}\mathcal{S}_{(I)}=\mathcal{S}$ .[16]

The singlet macrostate, $(\mathcal{N},\mathcal{S})=(\mathcal{Z},\Sigma=0)$ is a common vertex of various simplexes (triangles, discontinuity segments and the point itself, see Figure 2). For investigation of the energy gradient in its vicinity, the directional limit is introduced. Let $(\mathcal{N}_{(I)},\mathcal{S}_{(I)})=(\mathcal{Z},\Sigma)$ and $(\mathcal{N}, \mathcal{S}) = (\mathcal{Z} + \delta_N, \Sigma + \delta_S)$ be a vertex of interest and an interior point of the associated simplex $\{...\}$ (a triangle or a segment), respectively. If for all points $(\mathcal{Z} + \delta_N, \Sigma + \delta_S)$ the limit

$$\mu_{\{...\}}^{[v],(\mathcal{Z},\Sigma)} \equiv \lim_{\varepsilon \to 0^+} \mu[\mathcal{Z} + \varepsilon\delta_N, \Sigma + \varepsilon\delta_S; v], \quad (15)$$

exists and is independent of $(\delta_N, \delta_S)$, then $\mu_{\{...\}}^{[v],(\mathcal{Z},\Sigma)}$ will be called the energy gradient limit at the $(\mathcal{Z},\Sigma)$ point from the $\{...\}$ simplex direction (interior).

The gradient for the triangle $\{\{1,\},\{2\},\{3\}\} = \{[0,0], [\Delta\mathcal{N}_{\{2\}}, \Delta\mathcal{S}_{\{2\}}], [\Delta\mathcal{N}_{\{3\}}, \Delta\mathcal{S}_{\{3\}}]\}$ is[16]

$$\mu_{\{1,2,3\}}^{[v]} = \left( \frac{\left(E_{\{3\}}^{[v]} - E_{\{1\}}^{[v]}\right)\Delta\mathcal{S}_{\{2\}} - \left(E_{\{2\}}^{[v]} - E_{\{1\}}^{[v]}\right)\Delta\mathcal{S}_{\{3\}}}{\Delta\mathcal{N}_{\{3\}}\Delta\mathcal{S}_{\{2\}} - \Delta\mathcal{N}_{\{2\}}\Delta\mathcal{S}_{\{3\}}}, \frac{\left(E_{\{3\}}^{[v]} - E_{\{1\}}^{[v]}\right)\Delta\mathcal{N}_{\{3\}} - \left(E_{\{2\}}^{[v]} - E_{\{1\}}^{[v]}\right)\Delta\mathcal{N}_{\{2\}}}{\Delta\mathcal{N}_{\{3\}}\Delta\mathcal{S}_{\{2\}} - \Delta\mathcal{N}_{\{2\}}\Delta\mathcal{S}_{\{3\}}} \right). \quad (16)$$

It is a constant vector for the whole interior of the triangle, so is satisfies Eq.(1515).

For the point lying on the open discontinuity segment, say the diagonal $\{\{1,\},\{3\}\}$ in the parallelogram $\{\{1,\},\{2\},\{3\},\{4\}\}$, the 0K limit of the energy gradient is simply the average of the adjacent triangles results[16]

$$\mu_{\{1,3\}}^{[v]} = \frac{1}{2}\left(\mu_{\{1,2,3\}}^{[v]} + \mu_{\{1,3,4\}}^{[v]}\right). \quad (17)$$

In the case $\{...\} = \{[0,0]\}$ (0-simplex), the energy gradient at 0K limit is[16]

$$\mu[\mathcal{N} = \mathcal{Z}, \mathcal{S} = 0; v] = \left(-(I+A)/2, 0\right). \quad (18)$$

Summing up the discussion on energy derivatives discontinuities in the vicinity of the singlet point, three different values $\mu = \mu_N = -I, -(I+A)/2, -A$ are known for $\mathcal{N} < \mathcal{Z}$, $\mathcal{N} = \mathcal{Z}$, $\mathcal{N} > \mathcal{Z}$ in the spinless DFT, while in the SDFT there are many energy gradient limits at singlet state: each one depends on the simplex from which the reference point is approached and equals the simplex energy gradient (the energy gradients for all necessary triangles are collected in Table 1). For example, 17 different $\mu_{\{...\}}^{[v],(\mathcal{Z},\Sigma)}$ occur for the FT-type pattern: 8 due to triangles, 8 due segments and 1 at the point.

As follows from the above analysis, the $E[\mathcal{N}, \mathcal{S}; v]$ function is only semi-differentiable in the $(\mathcal{N}, \mathcal{S})$ domain (weaker than Gâteaux differentiable[17]). Our set of energy gradient limits is not equivalent to generalization of the concept of one-side derivatives (thus a collection of derivatives such as $(\partial E/\partial X)^{+/-}$, $X \in \{\mathcal{N}, \mathcal{S}\}$ is insufficient for full description). It should be noted that instead of using $(\mathcal{N}, \mathcal{S})$ variables, it is also possible to use $(\mathcal{N}_\alpha, \mathcal{N}_\beta)$ variables.

However not all conclusions can be transferred easily from one representation to another. From Figure 2 it is clear that such intuitive property as a derivative discontinuity at integer $\mathcal{N}_\alpha$ or $\mathcal{N}_\beta$ [12] (see for TT-type DP) is not always occurring (e.g. $\Delta \mathcal{N}_\alpha = 1$ is not discontinuity line for FT-type and FF-type DP ).

To verify that all established DPs are not only mathematical object but also chemical realities, the set of 47 diatomic molecules was chosen for testing.[16] The obtained DPs are collected in Table 2. The TT-type pattern (commonly accepted)[7, 15] is associated with the alkali-metal molecules and their hydrides and halides. The TF-type pattern is associated with the boron group hydrides and halides. The hydrogen halides show the TT-type pattern, while for the diatomic halogen molecules and the interhalogen compounds, the FT-type pattern was found. In the case of the molecule consisting of a carbon group atom and an oxygen group atom or the nitrogen group diatomic molecule (pnictide) situation is not clear. The DP for these molecules is ambiguous and depends on the functional used for calculation (however the DP for SiO and GeO is of the TF-type definitely). To characterize qualitatively the DP prediction, the delta factor based on Eq.(1313) is introduced ( $\delta^{+/-} \in (-1, +\infty)$ )

$$\delta^{+/-} = \Delta E_{DQ}^{[v],+/-} \big/ \Delta E_{ST}^{[v]} - 1. \tag{19}$$

When $\Delta E_{ST}^{[v]} > 0$ and $\Delta E_{DQ}^{[v],+/-} > 0$, the inequality $\Delta E_{\{1,2,3,4\}}^{[v]} > 0$ is equivalent to $\delta^{+/-} > 0$, the condition for the T-type pattern for addition/removing electrons, while $\delta^{+/-} < 0$ is the condition for the F-type pattern. This delta factor can be considered as a measure of departure from the accidental degeneracy, $\Delta E_{DQ}^{[v],+/-} = \Delta E_{ST}^{[v]}$, and the DP results with $\delta^{+/-} \in (-d, d)$ can be viewed as uncertain (in this letter, $d = 0.1$ was chosen). Among the pnictides, the FF-type pattern is noted for NP, NAs, P$_2$ and PAs. For nitrogen molecule, only HF and CISD results have a credible delta, while for all DFT methods absolute value of $\delta^{+/-}$ is less than 0.09. So the FF-type pattern can be assigned, based only on wavefunction methods and on similarity to another pnictides. Determination of the DPs for the diatomic molecule containing a carbon group atom and an oxygen group atom is really a challenge. However, the oxides (CO, SiO, GeO) have definitely the TF-type pattern. The sulphides show ambiguous results. The case of the carbon monosulfide is very special, the DP from DFT calculations is the TF-type, while from HF and CISD – the FF-type, both results with high delta values. The DP for CSe is the TF-type, for GeS – the FF-type one. For GeS, SiSe, GeSe, the DP is uncertain with methods used in this Letter. Concluding this illustrative part, it should be stressed that evaluation of the singlet-triple excitation energy for a given molecule and the doublet-quadruplet excitation energies for its cation and anion is necessary to establish its pattern. The quality of the results for these excitation energies is crucial for the correct recognition of the molecule discontinuity pattern.

In this Letter, the discontinuity patterns in the vicinity of a singlet state are derived. Analysis presented here shows that besides commonly recognized the TT-type pattern, three other patterns are possible (FT-, TF- and FF-type patterns). These results are consistent with the flat-plane condition for the energy dependence on the electron and spin numbers,[7] but it should be stressed that the constancy condition[7] is valid only for the triangle constructed of the singlet and two doublet points. In general it has to be replaced by the linearity condition in both variables. Correct recognition of the behavior of the energy function and its derivatives at 0K limit, obtained in this Letter, is crucial for development of better approximations that go beyond smooth orbital functionals. This is essential for the advancement of DFT towards the calculation of strongly correlated systems. Finding a functional that captures all the possible types of discontinuity seems very daunting; it appears more useful to rely on a theory that is able to target different spin states separately (see Ref. [18] for review of DFT applied to open-shell systems).
Conclusions presented here affect also the development of SDFT based chemical reactivity indices. It is clear that such concept as electrophilic or nucleophilic property (one-side derivative) is richer in the spin version of conceptual DFT (the energy gradient limit from simplex direction is an example).

This work was partially supported by the Ministry of Science and Higher Education of Poland through Grant No. N204275939, the International PhD Projects Programme of the Foundation for Polish Science, co-financed from European Regional Development Fund within Innovative Economy Operational Programme "Grants for innovation" and the Interdisciplinary Center for Mathematical and Computational Modeling computational grant. Authors are grateful to A. Holas for helpful discussions.

**Table 1.** Energy gradient at 0K limit for triangles in singlet state vicinity. Position of each triangle is indicated over some discontinuity patterns of Fig.2

| triangle | $\mu_N$ | $\mu_S$ | position |
|---|---|---|---|
| electron addition | | | |
| $\{[0,0],[+1,-1],[+1,+1]\}$ | $-A$ | $0$ | 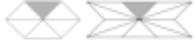 |
| $\{[0,0],[+1,\pm1],[+1,\pm3]\}$ | $-A - \Delta E_{DQ}^{[v],+}/2$ | $\pm \Delta E_{DQ}^{[v],+}/2$ | 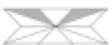 |
| $\{[0,0],[+1,\pm1],[0,\pm2]\}$ | $-A - \Delta E_{ST}^{[v],+}/2$ | $\pm \Delta E_{ST}^{[v],+}/2$ | 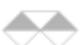 |
| $\{[0,0],[+1,\pm3],[0,\pm2]\}$ | $-A - \Delta E_{DQ}^{[v],+} - 3\Delta E_{ST}^{[v],+}/2$ | $\pm \Delta E_{ST}^{[v],+}/2$ | 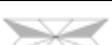 |
| electron removing | | | |
| $\{[0,0],[-1,-1],[-1,+1]\}$ | $-I$ | $0$ | 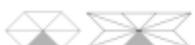 |
| $\{[0,0],[-1,\pm1],[-1,\pm3]\}$ | $-I - \Delta E_{DQ}^{[v],-}/2$ | $\pm \Delta E_{DQ}^{[v],-}/2$ | 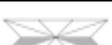 |
| $\{[0,0],[-1,\pm1],[0,\pm2]\}$ | $-I - \Delta E_{ST}^{[v],-}/2$ | $\pm \Delta E_{TS}^{[v],-}/2$ | 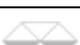 |
| $\{[0,0],[-1,\pm3],[0,\pm2]\}$ | $-I - \Delta E_{DQ}^{[v],-} - 3\Delta E_{ST}^{[v],-}/2$ | $\pm \Delta E_{ST}^{[v],-}/2$ | 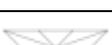 |

**Table 2.** Classification of diatomic molecules according to the type of discontinuity.

| TT-type 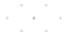 | FT-type 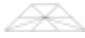 |
|---|---|
| Li$_2$, LiNa, Na$_2$, HLi, HNa, LiF, LiCl, LiBr, NaF, NaCl, NaBr, HF, HCl, HBr | HB, HAl, HGa BF, BCl, BBr AlF, AlCl, AlBr GaF, GaCl, GaBr |
| TF-type 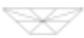 | FF-type 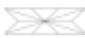 |
| F2, ClF, BrF, Cl2, BrCl, Br2 CO, SiO, GeO CS,* CSe | N$_2$,** NP, AsN, P$_2$, AsP, As$_2$ CS,** SiS,** GeSe |

\* from DFT calculations
\*\* from HF and CISD calculations

**Figure 1.** (color online) Position of macrostates (open cicrles) in vicinity of singlet one (filled red circle). Blue lines are discontinuity lines at integer $\mathcal{N}$. Green segments are common discontinuity segments for all type of the discontinuity pattern in vicinity of singlet state (e.g. $[0,0]$ and $[+1,\pm 1]$ segments or $[0,0]$ and $[-1,\pm 1]$ segments).

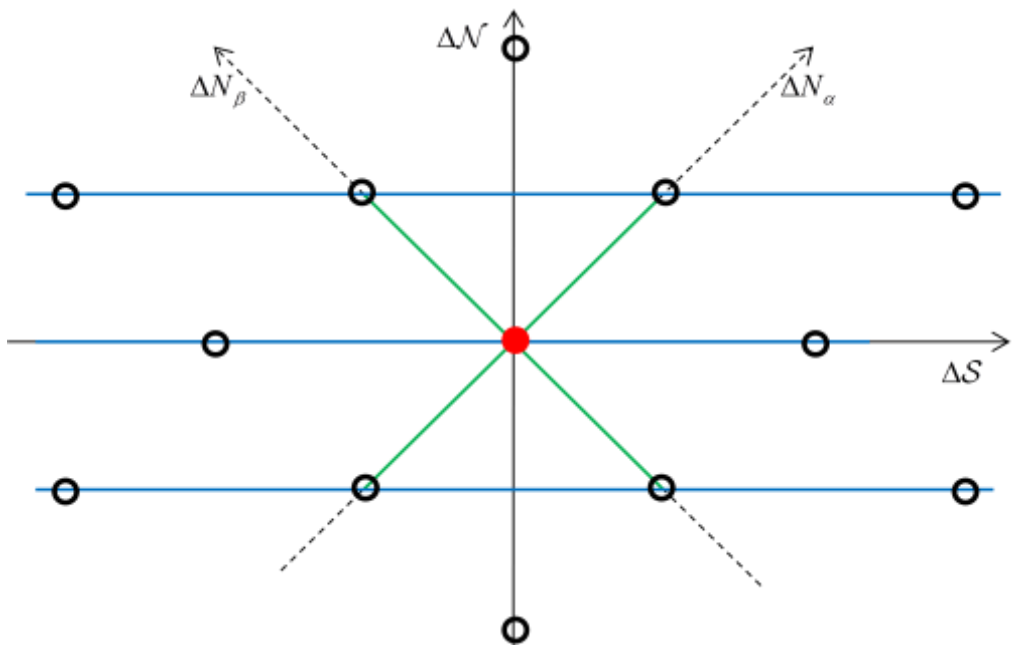

**Figure 2.** Discontinuity patterns for a singlet state. Vertices correspond to macrostate points indicated on Figure 1.

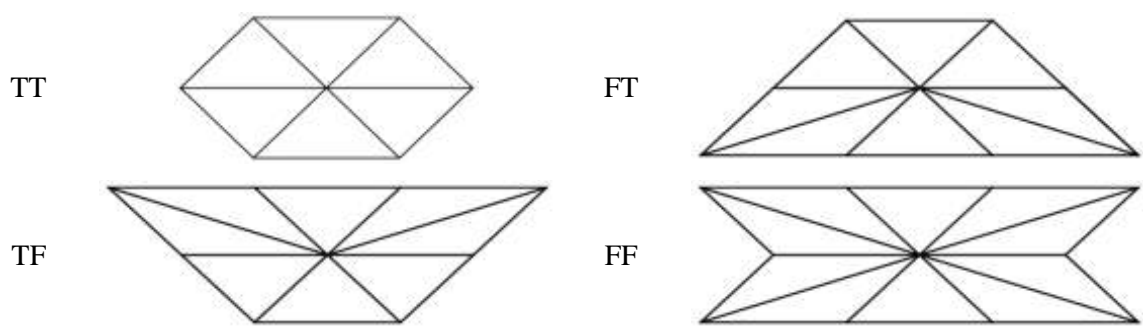

Supplemental Material for

Discontinuities of energy derivatives in spin-density functional theory.

Ali M. Malek, Robert Balawender*

Institute of Physical Chemistry, Polish Academy of Sciences,

Kasprzaka 44/52, PL-01-224 Warsaw, Poland

## Derivation of Eqs.16, 17 and 18

In the case of a point lying inside the triangle $\{\{1\},\{2\},\{3\}\}$, its macrostate zero-temperature weights have to be nonzero. This means that 0K limit of the ratio involving two of these weights has to be a finite positive number. Using Eq.(9), this ratio is

$$\frac{\omega_{(I)}^{[\beta,\mathcal{N},\mathcal{S};v]}}{\omega_{(J)}^{[\beta,\mathcal{N},\mathcal{S};v]}} = \exp\left(-\beta\left(\left(E_{(I)}^{[\beta;v]} - E_{(J)}^{[\beta;v]}\right) - \mu_N^{[\beta,\mathcal{N},\mathcal{S};v]}\left(\mathcal{N}_{(I)} - \mathcal{N}_{(J)}\right) - \mu_S^{[\beta,\mathcal{N},\mathcal{S};v]}\left(\mathcal{S}_{(I)} - \mathcal{S}_{(J)}\right)\right)\right) \quad (S1)$$
$$\times \exp\left(S_{(I)}^{[\beta;v]} - S_{(J)}^{[\beta;v]}\right).$$

Its limit is a finite positive number if and only if

$$\left(E_{(I)}^{[v]} - E_{(J)}^{[v]}\right) - \mu_N^{[\mathcal{N},\mathcal{S};v]}\left(\mathcal{N}_{(I)} - \mathcal{N}_{(J)}\right) - \mu_S^{[\mathcal{N},\mathcal{S};v]}\left(\mathcal{S}_{(I)} - \mathcal{S}_{(J)}\right) = 0. \quad (S2)$$

Using this condition for two pairs of verticess in the triangle, e.g $\{1\}$ and $\{2\}$, $\{1\}$ and $\{3\}$, we arrive at a set of equations for the unknowns $\mu_N^{[\beta\square\mathcal{N},\mathcal{S};v]}$ and $\mu_S^{[\beta,\mathcal{N},\mathcal{S};v]}$

$$\begin{cases} \mu_N^{[\beta,\mathcal{N},\mathcal{S};v]}\left(\mathcal{N}_{\{2\}} - \mathcal{N}_{\{1\}}\right) + \mu_S^{[\beta,\mathcal{N},\mathcal{S};v]}\left(\mathcal{S}_{\{2\}} - \mathcal{S}_{\{1\}}\right) = \left(E_{\{2\}}^{[\beta;v]} - E_{\{1\}}^{[\beta;v]}\right) \\ \mu_N^{[\beta,\mathcal{N},\mathcal{S};v]}\left(\mathcal{N}_{\{3\}} - \mathcal{N}_{\{1\}}\right) + \mu_S^{[\beta,\mathcal{N},\mathcal{S};v]}\left(\mathcal{S}_{\{3\}} - \mathcal{S}_{\{1\}}\right) = \left(E_{\{3\}}^{[\beta;v]} - E_{\{1\}}^{[\beta;v]}\right) \end{cases} \quad (S3)$$

The solution, at 0K limit

$$\mu_N^{[\mathcal{N},\mathcal{S};v]} = \frac{\left(E_{\{3\}}^{[v]} - E_{\{1\}}^{[v]}\right)\left(\mathcal{S}_{\{2\}} - \mathcal{S}_{\{1\}}\right) - \left(E_{\{2\}}^{[v]} - E_{\{1\}}^{[v]}\right)\left(\mathcal{S}_{\{3\}} - \mathcal{S}_{\{1\}}\right)}{\left(\mathcal{N}_{\{3\}} - \mathcal{N}_{\{1\}}\right)\left(\mathcal{S}_{\{2\}} - \mathcal{S}_{\{1\}}\right) - \left(\mathcal{N}_{\{2\}} - \mathcal{N}_{\{1\}}\right)\left(\mathcal{S}_{\{3\}} - \mathcal{S}_{\{1\}}\right)}, \quad (S4)$$

$$\mu_S^{[\mathcal{N},\mathcal{S};v]} = \frac{\left(E_{\{3\}}^{[v]} - E_{\{1\}}^{[v]}\right)\left(\mathcal{N}_{\{2\}} - \mathcal{N}_{\{1\}}\right) - \left(E_{\{2\}}^{[v]} - E_{\{1\}}^{[v]}\right)\left(\mathcal{N}_{\{3\}} - \mathcal{N}_{\{1\}}\right)}{\left(\mathcal{N}_{\{3\}} - \mathcal{N}_{\{1\}}\right)\left(\mathcal{S}_{\{2\}} - \mathcal{S}_{\{1\}}\right) - \left(\mathcal{N}_{\{2\}} - \mathcal{N}_{\{1\}}\right)\left(\mathcal{S}_{\{3\}} - \mathcal{S}_{\{1\}}\right)}, \quad (S5)$$

is equivalent to Eq.(16). The weights can be calculated directly from conditions $\sum_I \omega_{(I)} = 1$, $\sum_I \omega_{(I)} \mathcal{N}_{(I)} = \mathcal{N}$ and $\sum_I \omega_{(I)} \mathcal{S}_{(I)} = \mathcal{S}$.

Consider the case of a point lying on the open segment $\{\{1\},\{3\}\}$, the common side of the $\{\{1\},\{2\},\{3\}\}$ triangle and the $\{\{1\},\{3\},\{4\}\}$ triangle. From the spin canonical condition and the two-point formula for the line in the space, the weights are

$$\omega_{\{3\}}^{[\mathcal{N},\mathcal{S};v]} = \frac{\mathcal{S} - \mathcal{S}_{\{1\}}}{\mathcal{S}_{\{3\}} - \mathcal{S}_{\{1\}}}; \quad \omega_{\{1\}}^{[\mathcal{N},\mathcal{S};v]} = \frac{\mathcal{S}_{\{3\}} - \mathcal{S}}{\mathcal{S}_{\{3\}} - \mathcal{S}_{\{1\}}}, \quad (S6)$$

and $\omega_{\{3\}}^{[\mathcal{N},\mathcal{S};v]}/\omega_{\{1\}}^{[\mathcal{N},\mathcal{S};v]} > 0$ implies that Eq.(S2) with $I = \{3\}$, $J = \{1\}$ holds

$$\mu_N^{[\beta,\mathcal{N},\mathcal{S};v]}\left(\mathcal{N}_{\{3\}} - \mathcal{N}_{\{1\}}\right) + \mu_S^{[\beta,\mathcal{N},\mathcal{S};v]}\left(\mathcal{S}_{\{3\}} - \mathcal{S}_{\{1\}}\right) = \left(E_{\{3\}}^{[\beta;v]} - E_{\{1\}}^{[\beta;v]}\right). \quad (S7)$$

This is one equation for two unknowns $\mu_N$ and $\mu_S$. To find second equation, we include a contribution from two other vertices, the macrostates $\{3\}$ and $\{4\}$ at low temperature. For the point lying on the open segment $\{\{1\},\{3\}\}$, the relation between $\mathcal{N}$ and $\mathcal{S}$ is as follows

$$\mathcal{N} = \frac{(\mathcal{N}_{\{3\}} - \mathcal{N}_{\{1\}})}{(\mathcal{S}_{\{3\}} - \mathcal{S}_{\{1\}})}\mathcal{S} + \frac{\mathcal{N}_{\{1\}}\mathcal{S}_{\{3\}} - \mathcal{N}_{\{3\}}\mathcal{S}_{\{1\}}}{(\mathcal{S}_{\{3\}} - \mathcal{S}_{\{1\}})}. \tag{S8}$$

Inserting this into the spin canonical conditions yields

$$\begin{cases} \sum_{i=1}^{4} \omega_{\{i\}}^{[\beta,\mathcal{N},\mathcal{S};v]} \mathcal{N}_{\{i\}} = \frac{(\mathcal{N}_{\{3\}} - \mathcal{N}_{\{1\}})}{(\mathcal{S}_{\{3\}} - \mathcal{S}_{\{1\}})}\mathcal{S} + \frac{\mathcal{N}_{\{1\}}\mathcal{S}_{\{3\}} - \mathcal{N}_{\{3\}}\mathcal{S}_{\{1\}}}{(\mathcal{S}_{\{3\}} - \mathcal{S}_{\{1\}})} \\ \sum_{i=1}^{4} \omega_{\{i\}}^{[\beta,\mathcal{N},\mathcal{S};v]} \mathcal{S}_{\{i\}} = \mathcal{S} \\ \sum_{i=1}^{4} \omega_{\{i\}}^{[\beta,\mathcal{N},\mathcal{S};v]} = 1 \end{cases} \tag{S9}$$

Solving this for the unknowns $\omega_{\{2\}}^{[\beta,\mathcal{N},\mathcal{S};v]}$, $\omega_{\{3\}}^{[\beta,\mathcal{N},\mathcal{S};v]}$, $\omega_{\{4\}}^{[\beta,\mathcal{N},\mathcal{S};v]}$ in terms of $\omega_{\{1\}}$, we find

$$\omega_{\{2\}}^{[\beta,\mathcal{N},\mathcal{S};v]} = \frac{\left(\mathcal{S} - \mathcal{S}_{\{3\}} + \omega_{\{1\}}^{[\beta,\mathcal{N},\mathcal{S};v]}(\mathcal{S}_{\{3\}} - \mathcal{S}_{\{1\}})\right)}{(\mathcal{S}_{\{3\}} - \mathcal{S}_{\{1\}})} \times \frac{\left(\mathcal{N}_{\{4\}}(\mathcal{S}_{\{1\}} - \mathcal{S}_{\{3\}}) + \mathcal{N}_{\{1\}}(\mathcal{S}_{\{3\}} - \mathcal{S}_{\{4\}}) + \mathcal{N}_{\{3\}}(\mathcal{S}_{\{4\}} - \mathcal{S}_{\{1\}})\right)}{\left(\mathcal{N}_{\{4\}}(\mathcal{S}_{\{3\}} - \mathcal{S}_{\{2\}}) + \mathcal{N}_{\{3\}}(\mathcal{S}_{\{2\}} - \mathcal{S}_{\{4\}}) + \mathcal{N}_{\{2\}}(\mathcal{S}_{\{4\}} - \mathcal{S}_{\{3\}})\right)}, \tag{S10}$$

$$\omega_{\{4\}}^{[\beta,\mathcal{N},\mathcal{S};v]} = \frac{\left(\mathcal{S} - \mathcal{S}_{\{3\}} + \omega_{\{1\}}^{[\beta,\mathcal{N},\mathcal{S};v]}(\mathcal{S}_{\{3\}} - \mathcal{S}_{\{1\}})\right)}{(\mathcal{S}_{\{4\}} - \mathcal{S}_{\{3\}})} \times \frac{\left(\mathcal{N}_{\{3\}}(\mathcal{S}_{\{1\}} - \mathcal{S}_{\{2\}}) + \mathcal{N}_{\{1\}}(\mathcal{S}_{\{2\}} - \mathcal{S}_{\{3\}}) + \mathcal{N}_{\{2\}}(\mathcal{S}_{\{3\}} - \mathcal{S}_{\{1\}})\right)}{\left(\mathcal{N}_{\{4\}}(\mathcal{S}_{\{3\}} - \mathcal{S}_{\{2\}}) + \mathcal{N}_{\{3\}}(\mathcal{S}_{\{2\}} - \mathcal{S}_{\{4\}}) + \mathcal{N}_{\{2\}}(\mathcal{S}_{\{4\}} - \mathcal{S}_{\{3\}})\right)}. \tag{S11}$$

Using these weights, the ratio between them at low temperature

$$\frac{\omega_{\{2\}}^{[\beta,\mathcal{N},\mathcal{S};v]}}{\omega_{\{4\}}^{[\beta,\mathcal{N},\mathcal{S};v]}} = \frac{\left(\mathcal{N}_{\{1\}}(\mathcal{S}_{\{3\}} - \mathcal{S}_{\{4\}}) + \mathcal{N}_{\{3\}}(\mathcal{S}_{\{4\}} - \mathcal{S}_{\{1\}}) + \mathcal{N}_{\{4\}}(\mathcal{S}_{\{1\}} - \mathcal{S}_{\{3\}})\right)}{\left(\mathcal{N}_{\{1\}}(\mathcal{S}_{\{2\}} - \mathcal{S}_{\{3\}}) + \mathcal{N}_{\{2\}}(\mathcal{S}_{\{3\}} - \mathcal{S}_{\{1\}}) + \mathcal{N}_{\{3\}}(\mathcal{S}_{\{1\}} - \mathcal{S}_{\{2\}})\right)} \times \frac{(\mathcal{S}_{\{4\}} - \mathcal{S}_{\{3\}})}{(\mathcal{S}_{\{3\}} - \mathcal{S}_{\{1\}})}, \tag{S12}$$

does not depend on $\beta$. This provides an additional equation, Eq.(S2) with $I = \{4\}$, $J = \{2\}$

$$\mu_N^{[\beta,\mathcal{N},\mathcal{S};v]}(\mathcal{N}_{\{4\}} - \mathcal{N}_{\{2\}}) + \mu_S^{[\beta,\mathcal{N},\mathcal{S};v]}(\mathcal{S}_{\{4\}} - \mathcal{S}_{\{2\}}) = \left(E_{\{4\}}^{[\beta;v]} - E_{\{2\}}^{[\beta;v]}\right). \tag{S13}$$

Solving the system of Eq.(S7) and Eq.(S13) we arrive at

$$\mu_{\mathrm{N}}^{[\beta,\mathcal{N},\mathcal{S};v]} = \frac{\left(E_{\{4\}}^{[\beta;v]} - E_{\{2\}}^{[\beta;v]}\right)\left(\mathcal{S}_{\{3\}} - \mathcal{S}_{\{1\}}\right) - \left(E_{\{3\}}^{[\beta;v]} - E_{\{1\}}^{[\beta;v]}\right)\left(\mathcal{S}_{\{4\}} - \mathcal{S}_{\{2\}}\right)}{\left(\mathcal{N}_{\{4\}} - \mathcal{N}_{\{2\}}\right)\left(\mathcal{S}_{\{3\}} - \mathcal{S}_{\{1\}}\right) - \left(\mathcal{N}_{\{3\}} - \mathcal{N}_{\{1\}}\right)\left(\mathcal{S}_{\{4\}} - \mathcal{S}_{\{2\}}\right)},$$ (S14)

$$\mu_{\mathrm{S}}^{[\beta,\mathcal{N},\mathcal{S};v]} = \frac{\left(E_{\{4\}}^{[\beta;v]} - E_{\{2\}}^{[\beta;v]}\right)\left(\mathcal{N}_{\{3\}} - \mathcal{N}_{\{1\}}\right) - \left(E_{\{3\}}^{[\beta;v]} - E_{\{1\}}^{[\beta;v]}\right)\left(\mathcal{N}_{\{4\}} - \mathcal{N}_{\{2\}}\right)}{\left(\mathcal{N}_{\{4\}} - \mathcal{N}_{\{2\}}\right)\left(\mathcal{S}_{\{3\}} - \mathcal{S}_{\{1\}}\right) - \left(\mathcal{N}_{\{3\}} - \mathcal{N}_{\{1\}}\right)\left(\mathcal{S}_{\{4\}} - \mathcal{S}_{\{2\}}\right)}.$$ (S15)

At 0K limit, we have

- from Eqs.(S14) and (S15), valid only for the segment

$$\mu_{\{\{1\},\{3\}\}}^{[v]} = \left( \left\| \begin{matrix} \left(E_{\{3\}}^{[v]} - E_{\{1\}}^{[v]}\right) & \left(\mathcal{S}_{\{3\}} - \mathcal{S}_{\{1\}}\right) \\ \left(E_{\{4\}}^{[v]} - E_{\{2\}}^{[v]}\right) & \left(\mathcal{S}_{\{4\}} - \mathcal{S}_{\{2\}}\right) \end{matrix} \right\|, \left\| \begin{matrix} \left(\mathcal{N}_{\{3\}} - \mathcal{N}_{\{1\}}\right) & \left(E_{\{3\}}^{[v]} - E_{\{1\}}^{[v]}\right) \\ \left(\mathcal{N}_{\{4\}} - \mathcal{N}_{\{2\}}\right) & \left(E_{\{4\}}^{[v]} - E_{\{2\}}^{[v]}\right) \end{matrix} \right\| \right)$$
$$\times \left( \left\| \begin{matrix} \left(\mathcal{N}_{\{3\}} - \mathcal{N}_{\{1\}}\right) & \left(\mathcal{S}_{\{3\}} - \mathcal{S}_{\{1\}}\right) \\ \left(\mathcal{N}_{\{4\}} - \mathcal{N}_{\{2\}}\right) & \left(\mathcal{S}_{\{4\}} - \mathcal{S}_{\{2\}}\right) \end{matrix} \right\| \right)^{-1}$$ (S16)

- from Eqs. (S4) and (S5) for one triangle

$$\mu_{\{\{1\},\{2\},\{3\}\}}^{[v]} = \left( \left\| \begin{matrix} \left(E_{\{2\}}^{[v]} - E_{\{1\}}^{[v]}\right) & \left(\mathcal{S}_{\{2\}} - \mathcal{S}_{\{1\}}\right) \\ \left(E_{\{3\}}^{[v]} - E_{\{1\}}^{[v]}\right) & \left(\mathcal{S}_{\{3\}} - \mathcal{S}_{\{1\}}\right) \end{matrix} \right\|, \left\| \begin{matrix} \left(\mathcal{N}_{\{2\}} - \mathcal{N}_{\{1\}}\right) & \left(E_{\{2\}}^{[v]} - E_{\{1\}}^{[v]}\right) \\ \left(\mathcal{N}_{\{3\}} - \mathcal{N}_{\{1\}}\right) & \left(E_{\{3\}}^{[v]} - E_{\{1\}}^{[v]}\right) \end{matrix} \right\| \right)$$
$$\times \left( \left\| \begin{matrix} \left(\mathcal{N}_{\{2\}} - \mathcal{N}_{\{1\}}\right) & \left(\mathcal{S}_{\{2\}} - \mathcal{S}_{\{1\}}\right) \\ \left(\mathcal{N}_{\{3\}} - \mathcal{N}_{\{1\}}\right) & \left(\mathcal{S}_{\{3\}} - \mathcal{S}_{\{1\}}\right) \end{matrix} \right\| \right)^{-1}$$ (S17)

and after replacements $\{2\} \to \{3\}$, $\{3\} \to \{4\}$ in Eq.(S17), for second traingle

$$\mu_{\{\{1\},\{3\},\{4\}\}}^{[v]} = \left( \left\| \begin{matrix} \left(E_{\{3\}}^{[v]} - E_{\{1\}}^{[v]}\right) & \left(\mathcal{S}_{\{3\}} - \mathcal{S}_{\{1\}}\right) \\ \left(E_{\{4\}}^{[v]} - E_{\{1\}}^{[v]}\right) & \left(\mathcal{S}_{\{4\}} - \mathcal{S}_{\{1\}}\right) \end{matrix} \right\|, \left\| \begin{matrix} \left(\mathcal{N}_{\{3\}} - \mathcal{N}_{\{1\}}\right) & \left(E_{\{3\}}^{[v]} - E_{\{1\}}^{[v]}\right) \\ \left(\mathcal{N}_{\{4\}} - \mathcal{N}_{\{1\}}\right) & \left(E_{\{4\}}^{[v]} - E_{\{1\}}^{[v]}\right) \end{matrix} \right\| \right)$$
$$\times \left( \left\| \begin{matrix} \left(\mathcal{N}_{\{3\}} - \mathcal{N}_{\{1\}}\right) & \left(\mathcal{S}_{\{3\}} - \mathcal{S}_{\{1\}}\right) \\ \left(\mathcal{N}_{\{4\}} - \mathcal{N}_{\{1\}}\right) & \left(\mathcal{S}_{\{4\}} - \mathcal{S}_{\{1\}}\right) \end{matrix} \right\| \right)^{-1}$$ (S18)

The geometric interpretation of the Cramer's rule is that the determinant of the system of equations equals to the area of the parallelogram determined by $(a_{11}, a_{21})$ and $(a_{21}, a_{22})$ vectors. Thus the denominator in Eq. (S17) is two times area of the $\{\{1\},\{2\},\{3\}\}$ triangle, the denominator in Eq. (S18) is two times area of the $\{\{1\},\{3\},\{4\}\}$ triangle. These denominators are equal. The denominator in Eq.(S16) is the sum of previous denominators.

After adding the numerators in Eqs. (S17) and (S18), and some arithmetic manipulations, one obtains Eq.(17)

$$\mu^{[v]}_{\{1,3,4\}} + \mu^{[v]}_{\{1,2,3\}} = 2\mu^{[v]}_{\{1,2\}}. \tag{S19}$$

In the case $\{...\} = \{[0,0]\}$ (0-simplex), the system can have non-zero contributions from other (then $[0,0]$) macrostates at finite temperature. From even parity property of the energy as a function of $\mathcal{S}$ follows

$$\mu^{[\mathcal{N}=\mathcal{Z},\mathcal{S}=0;v]}_{\mathcal{S}} = 0. \tag{S19}$$

Next, the ratio between the quadruplet state and the doublet state from Eq.(S1) is

$$\frac{\omega^{[\beta,\mathcal{N}=\mathcal{Z},\mathcal{S}=0;v]}_{[+1,+3]}}{\omega^{[\beta,\mathcal{N}=\mathcal{Z},\mathcal{S}=0;v]}_{[+1,+1]}} = \exp\left(-\beta \Delta E^{[v],+}_{DQ}\right)\exp\left(S^{[\beta;v]}_{[+1,+3]} - S^{[\beta;v]}_{[+1,+1]}\right) \doteq 0, \tag{S20}$$

the doublet weights decrease with temperature more slowly than for quadruplet. Contributions $\omega_+$ and $\omega_-$ from "ions" to the electron number $\mathcal{N} = \mathcal{Z}$ has to equalize (at low temperature) and vanish at 0K. Using Eq.(S1) with Eq.(S1920), we find at finite temperature

$$1 = \frac{\omega^{[\beta,\mathcal{N},\mathcal{S};v]}_{+}}{\omega^{[\beta,\mathcal{N},\mathcal{S};v]}_{-}} \equiv \frac{\omega^{[\beta,\mathcal{N},\mathcal{S};v]}_{[+1,-1]} + \omega^{[\beta,\mathcal{N},\mathcal{S};v]}_{[+1,1]}}{\omega^{[\beta,\mathcal{N},\mathcal{S};v]}_{[-1,-1]} + \omega^{[\beta,\mathcal{N},\mathcal{S};v]}_{[-1,1]}} = \exp\left(-\beta\left(E^{[\beta,v]}_{[+1,+1]} - E^{[\beta,v]}_{[-1,+1]} - 2\mu^{[\beta,\mathcal{N}=\mathcal{Z},\mathcal{S}=0;v]}_{\mathcal{N}}\right)\right). \tag{S21}$$

This yields at 0K limit

$$\mu^{[\mathcal{N}=\mathcal{Z},\mathcal{S}=0;v]}_{\mathcal{N}} = \frac{1}{2}\left(E^{[v]}_{[+1,+1]} - E^{[v]}_{[-1,+1]}\right) = -\frac{1}{2}(I + A) \tag{S22}$$

where $I$ and $A$ are the ionization potential and the electron affinity. The set of Eqs.(S2223) and (S1920) is equivalent to Eq.(18)

## Full details of the calculations

All calculations were done using Gaussian suite of programs.[1] The energy for the neutral singlet and triplet, ionic doublets and quadruplets were calculated using the Hartre-Fock method followed by the configuration interaction with all double substitutions (CISD),[2, 3] and the DFT method. Because the non-coulombic part of exchange functionals typically dies off too rapidly and becomes very inaccurate at large distances, making them unsuitable for modeling processes such as electron excitations to high orbitals, two functionals which include long-range corrections: CAM-B3LYP,[4] LC-wPBE,[5-7] and two without corrections: B3LYP,[8, 9] M06-2X,[10] were used. These methods were combined with the correlation consistent double- and triple-ζ basis sets (along with their augmented counterparts).[11-16] Calculations for open shell systems are spin unrestricted at singlet geometry. The first observation was that for results with added diffuse functions, large discrepancy of DP types encountered, so these results were skipped in the analyses presented in our Letter. All data in Tables S2-S7 are in a.u.

**Table S1. A Discontinuity patterns associated with the alkali-metal molecules and their hydrides and halides.**

| method | HF | | | | CISD | | | | B3LYP | | | | M06-2X | | | | Cam-B3LYP | | | | LC-wPBE | | | |
|---|---|---|---|---|---|---|---|---|---|---|---|---|---|---|---|---|---|---|---|---|---|---|---|---|
| basis | cc-pVDZ | cc-pVTZ | aug-cc-pVDZ | aug-cc-pVTZ | cc-pVDZ | cc-pVTZ | aug-cc-pVDZ | aug-cc-pVTZ | cc-pVDZ | cc-pVTZ | aug-cc-pVDZ | aug-cc-pVTZ | cc-pVDZ | cc-pVTZ | aug-cc-pVDZ | aug-cc-pVTZ | cc-pVDZ | cc-pVTZ | aug-cc-pVDZ | aug-cc-pVTZ | cc-pVDZ | cc-pVTZ | aug-cc-pVDZ | aug-cc-pVTZ |
| LiLi | TT | TT | TT | TT | * | TT | * | * | TT | TT | TT | TT | TT | TT | TT | TT | TT | TT | TT | TT | TT | TT | TT | TT |
| LiNa | TT | TT | TF | TF | TT | TT | TF | TF | TT | TT | TT | TT | TT | TT | TT | TT | TT | TT | TT | TT | TT | TT | TT | TT |
| NaNa | TT | TT | TT | TT | TT | TT | TT | TT | TT | TT | TT | TT | TT | TT | TT | TT | TT | TT | TT | * | TT | TT | TT | TT |
| HLi | TT | TT | TT | TT | TT | TT | TT | TT | TT | TT | TT | TT | TT | TT | TT | TT | TT | TT | TT | TT | TT | TT | TT | TT |
| HNa | TT | TT | TT | TT | TT | TT | TT | TT | TT | TT | TT | TT | TT | TT | TT | TT | TT | TT | TT | TT | TT | TT | TT | TT |
| FLi | TT | TT | TT | TT | TT | TT | TT | TT | TT | TT | TT | TT | TT | TT | TT | TT | TT | TT | * | TT | TT | TT | TT | TT |
| ClLi | TT | TT | TT | TT | TT | TT | TT | TT | TT | TT | TT | TT | TT | TT | TT | TT | * | TT | TT | * | TT | TT | TT | TT |
| BrLi | TT | TT | TT | TT | TT | TT | TT | TT | TT | TT | TT | TT | TT | TT | TT | TT | TT | TT | * | TT | TT | TT | TT | TT |
| NaF | TT | TT | TT | TT | TT | TT | TT | TT | TT | TT | TT | TT | TT | TT | TT | TT | TT | TT | TT | TT | TT | TT | TT | TT |
| NaCl | TT | TT | TT | TT | TT | TT | TT | TT | TT | TT | TT | TT | TT | TT | TT | TT | TT | * | TT | TT | TT | TT | TT | TT |
| BrNa | TT | TT | TT | TT | TT | TT | TT | TT | TT | TT | TT | TT | TT | TT | TT | TT | TT | TT | TT | TT | TT | TT | TT | TT |

*SCF convergence problem for ionic excited state(s)

**Table S1. B Discontinuity patterns associated with the boron group hydrides and halides.**

| method | HF | | | | CISD | | | | B3LYP | | | | M06-2X | | | | Cam-B3LYP | | | | LC-wPBE | | | |
|---|---|---|---|---|---|---|---|---|---|---|---|---|---|---|---|---|---|---|---|---|---|---|---|---|
| basis | cc-pVDZ | cc-pVTZ | aug-cc-pVDZ | aug-cc-pVTZ | cc-pVDZ | cc-pVTZ | aug-cc-pVDZ | aug-cc-pVTZ | cc-pVDZ | cc-pVTZ | aug-cc-pVDZ | aug-cc-pVTZ | cc-pVDZ | cc-pVTZ | aug-cc-pVDZ | aug-cc-pVTZ | cc-pVDZ | cc-pVTZ | aug-cc-pVDZ | aug-cc-pVTZ | cc-pVDZ | cc-pVTZ | aug-cc-pVDZ | aug-cc-pVTZ |
| HB | TF | TF | TF | TT | TF | TF | TF | TT | TF | TF | TF | TF | TF | TF | TF | TF | TF | TF | TF | TF | TF | TF | TF | TF |
| HAl | TF | TF | TT | TT | TF | TF | TT | TT | TF | TF | TF | TF | TF | TF | TF | TF | TF | TF | TF | TF | TF | TF | TT | TF |
| HGa | TF | TF | TT | TT | TF | TF | TT | TT | TF | TF | TF | TF | TF | TF | TF | TF | TF | TF | TF | TF | TF | TF | TT | TT |
| BF | TF | TF | TF | TF | TF | TF | TF | TF | TF | TF | TF | TF | TF | TF | TF | TF | TF | TF | TF | TF | TF | TF | TF | TF |
| BCl | TF | TF | TT | TT | TF | TF | TT | TT | TF | TF | TF | TF | TF | TF | TF | TF | TF | TF | TF | TF | TF | TF | TF | TF |
| BBr | TF | TF | TT | TT | TF | TF | TT | TT | TF | TF | TF | TF | TF | TF | TF | TF | TF | TF | TF | TF | TF | TF | TF | TF |
| AlF | TF | TF | TF | TF | TF | TF | TF | TF | TF | TF | TF | TF | TF | TF | TF | TF | TF | TF | TF | TF | TF | TF | TF | TF |
| ClAl | TF | TF | TF | TF | TF | TF | TF | TF | TF | TF | TF | TF | TF | TF | TF | TF | TF | TF | TF | TF | TF | TF | TF | TF |
| AlBr | TF | TF | TF | TF | TF | TF | TF | TF | TF | TF | TF | TF | TF | TF | TF | TF | TF | TF | TF | TF | TF | TF | TF | TF |
| FGa | TF | TF | TF | TT | TF | TF | TF | TT | TF | TF | TF | TF | TF | TF | TF | TF | TF | TF | TF | TF | TF | TF | TF | TT |
| ClGa | TF | TF | TF | TT | TF | TF | TF | TT | TF | TF | TF | TF | TF | TF | TF | TF | TF | TF | TF | TF | TF | TF | TF | TT |

**Table S1.C Discontinuity patterns associated with the diatomic halogens, interhalogen molecules and hydrogen halides.**

| method | HF | | | | CISD | | | | B3LYP | | | | M06-2X | | | | Cam-B3LYP | | | | LC-wPBE | | | |
|---|---|---|---|---|---|---|---|---|---|---|---|---|---|---|---|---|---|---|---|---|---|---|---|---|
| basis | cc-pVDZ | cc-pVTZ | aug-cc-pVDZ | aug-cc-pVTZ | cc-pVDZ | cc-pVTZ | aug-cc-pVDZ | aug-cc-pVTZ | cc-pVDZ | cc-pVTZ | aug-cc-pVDZ | aug-cc-pVTZ | cc-pVDZ | cc-pVTZ | aug-cc-pVDZ | aug-cc-pVTZ | cc-pVDZ | cc-pVTZ | aug-cc-pVDZ | aug-cc-pVTZ | cc-pVDZ | cc-pVTZ | aug-cc-pVDZ | aug-cc-pVTZ |
| FF | FT | FT | FT | FT | FT | FT | FT | FT | FT | FT | FT | FT | FT | FT | FT | FT | FT | FT | FT | FT | FT | FT | FT | FT |
| ClF | FT | FT | FT | FT | FT | FT | * | FT | FT | FT | FT | FT | FT | FT | FT | FT | FT | FT | FT | FT | FT | FT | FT | * |
| BrF | FT | FT | FT | FT | FT | FT | FT | FT | FT | FT | FT | FT | FT | FT | FT | FT | FT | FT | FT | FT | FT | FT | FT | FT |
| ClCl | FT | FT | FT | FT | FT | FT | FT | FT | FT | FT | FT | FT | FT | FT | FT | FT | FT | FT | FT | FT | FT | FT | FT | FT |
| ClBr | FT | FT | FT | * | FT | FT | FT | * | FT | FT | FT | FT | FT | FT | FT | FT | FT | FT | FT | FT | FT | FT | FT | FT |
| BrBr | FT | FT | FT | FT | FT | FT | FT | FT | FT | FT | FT | FT | FT | FT | FT | FT | FT | FT | FT | FT | FT | FT | FT | FT |
| HF | TT | TT | TT | TT | TT | TT | TT | TT | TT | TT | TT | TT | TT | TT | TT | TT | TT | TT | TT | TT | TT | TT | TT | TT |
| HCl | TT | TT | TT | * | TT | TT | TT | TT | TT | TT | TT | TT | TT | TT | TT | TT | TT | TT | TT | TT | TT | TT | TT | TT |
| HBr | TT | TT | TT | TT | TT | TT | TT | TT | TT | TT | TT | TT | TT | TT | TT | TT | TT | TT | TT | TT | TT | TT | TT | TT |
| FF | FT | FT | FT | FT | FT | FT | FT | FT | FT | FT | FT | FT | FT | FT | FT | FT | FT | FT | FT | FT | FT | FT | FT | FT |
| ClF | FT | FT | FT | FT | FT | FT | * | FT | FT | FT | FT | FT | FT | FT | FT | FT | FT | FT | FT | FT | FT | FT | FT | * |

*SCF convergence problem for ionic excited state(s)

**Table S1.D Discontinuity patterns associated with the molecules consisting of a carbon group atom and an oxygen group atom or the nitrogen group diatomic molecules.**

| method | HF | | | | CISD | | | | B3LYP | | | | M06-2X | | | | Cam-B3LYP | | | | LC-wPBE | | | |
|---|---|---|---|---|---|---|---|---|---|---|---|---|---|---|---|---|---|---|---|---|---|---|---|---|
| basis | cc-pVDZ | cc-pVTZ | aug-cc-pVDZ | aug-cc-pVTZ | cc-pVDZ | cc-pVTZ | aug-cc-pVDZ | aug-cc-pVTZ | cc-pVDZ | cc-pVTZ | aug-cc-pVDZ | aug-cc-pVTZ | cc-pVDZ | cc-pVTZ | aug-cc-pVDZ | aug-cc-pVTZ | cc-pVDZ | cc-pVTZ | aug-cc-pVDZ | aug-cc-pVTZ | cc-pVDZ | cc-pVTZ | aug-cc-pVDZ | aug-cc-pVTZ |
| CO | TF | TF | TF | TF | FF | FF | FT | FF | TF | TF | TF | TF | TF | TF | TF | TF | TF | TF | TF | TF | TF | TF | TF | TF |
| CS | FF | FF | FT | FF | FF | FF | FT | TT | TF | TF | TF | TF | TF | TF | TF | TF | TF | TF | TF | TF | TF | TF | TF | TF |
| CSe | FF | FT | FT | FT | FF | FF | FT | FT | TF | TF | TF | TF | TF | TF | TF | TF | FF | FF | FF | FF | TF | TF | FF | TF |
| OSi | TF | TF | TF | TF | TF | TF | TF | TF | TF | TF | TF | TF | TF | TF | TF | TF | TF | TF | TF | TF | TF | TF | TF | TF |
| SSi | FF | FF | FF | FT | FF | FF | FF | FT | TF | TF | TF | TF | TF | TF | TF | TF | TF | TF | TF | TF | FF | TF | FF | TF |
| SeSi | FF | FF | FT | FT | TF | FT | TT | FT | FF | TF | FF | FT | FT | FF | FF | FF | FF | FF | FF | FF | FF | FF | FF | FF |
| GeO | TF | TF | TF | TF | TF | TF | TF | TF | TF | TF | TF | TF | TF | TF | TF | TF | TF | TF | TF | TF | TF | TF | TF | TF |
| GeS | FF | FF | FF | FF | FF | FF | FF | FF | FF | FF | FF | TF | FF | FF | TT | TF | FF | FF | FF | TF | FF | FF | FF | FF |
| GeSe | TF | TT | TT | TT | TF | FF | TT | TT | TF | TT | TT | FF | FF | TT | TT | FF | TF | FF | TF | FT | FF | TT | TT | FT |
| NN | FF | FF | FF | FF | FF | FF | FT | FF | TF | TF | FF | TF | FF | TF | TF | TF | TF | FF | TF | FF | TF | TF | FF | FF |
| NP | TF | FF | TF | FT | FF | TF | FF | FT | FF | FF | FF | FF | FF | FF | FF | FF | FF | FF | FF | FF | FF | FF | FF | FF |

## Table S2.A Data for diatomic molecules at HF/cc-pVDZ.

| | $\delta^-$ | $\delta^+$ | $\Delta E_{ST}^{[v]}$ | $\Delta E_{DQ}^{[v],-}$ | $\Delta E_{DQ}^{[v],+}$ | $I$ | $A$ |
|---|---|---|---|---|---|---|---|
| AlBr | 0.7409 | -0.4114 | 0.0460 | 0.1361 | 0.0782 | -0.3057 | 0.0151 |
| AlF | 1.5429 | -0.4250 | 0.0521 | 0.2305 | 0.0906 | -0.3229 | 0.0354 |
| AsAs | -0.3695 | -0.4329 | 0.0472 | 0.0525 | 0.0833 | -0.3207 | 0.0020 |
| AsN | -0.2164 | -0.3083 | 0.0699 | 0.0792 | 0.1011 | -0.3557 | 0.0155 |
| AsP | -0.0398 | -0.1283 | 0.0560 | 0.0616 | 0.0642 | -0.3320 | 0.0074 |
| BBr | 1.5334 | -0.5534 | 0.0215 | 0.1222 | 0.0482 | -0.3289 | 0.0273 |
| BCl | 1.8716 | -0.5317 | 0.0260 | 0.1592 | 0.0554 | -0.3360 | 0.0417 |
| BF | 1.9788 | -0.3639 | 0.0611 | 0.2863 | 0.0961 | -0.3676 | 0.0849 |
| BrBr | -0.2023 | 6.1461 | 0.6000 | 0.0670 | 0.0840 | -0.3837 | -0.0245 |
| BrF | -0.1847 | 5.9534 | 0.6356 | 0.0745 | 0.0914 | -0.4156 | 0.0094 |
| BrGa | 0.4305 | -0.3262 | 0.0650 | 0.1379 | 0.0964 | -0.3189 | 0.0146 |
| BrLi | 0.9999 | 0.3280 | 0.1877 | 0.2826 | 0.1413 | -0.3063 | -0.0155 |
| BrNa | 1.1826 | 0.3623 | 0.1681 | 0.2694 | 0.1234 | -0.2813 | -0.0206 |
| ClAl | 0.8759 | -0.4209 | 0.0469 | 0.1520 | 0.0810 | -0.3118 | 0.0201 |
| ClBr | -0.2928 | 5.8067 | 0.6148 | 0.0639 | 0.0903 | -0.3926 | -0.0144 |
| ClCl | -0.1790 | 5.2973 | 0.6791 | 0.0885 | 0.1078 | -0.4175 | 0.0011 |
| ClF | -0.1569 | 5.2124 | 0.7050 | 0.0957 | 0.1135 | -0.4468 | 0.0336 |
| ClGa | 0.4806 | -0.3438 | 0.0659 | 0.1486 | 0.1004 | -0.3288 | 0.0193 |
| ClLi | 1.1038 | 0.2525 | 0.1916 | 0.3219 | 0.1530 | -0.3220 | -0.0141 |
| CO | 0.0201 | -0.1583 | 0.1729 | 0.2095 | 0.2054 | -0.4782 | 0.1328 |
| CS | -0.3676 | -0.5598 | 0.0694 | 0.0997 | 0.1577 | -0.3671 | 0.0347 |
| CSe | -0.2187 | -0.0854 | 0.1132 | 0.0967 | 0.1238 | -0.3956 | 0.0142 |
| FF | -0.5676 | 8.6915 | 1.3132 | 0.0586 | 0.1355 | -0.5873 | 0.0742 |
| FGa | 1.2239 | -0.3683 | 0.0697 | 0.2453 | 0.1103 | -0.3451 | 0.0354 |
| FLi | 1.9472 | 0.1694 | 0.1939 | 0.4887 | 0.1658 | -0.3447 | -0.0036 |
| GeO | 0.0493 | -0.1955 | 0.0911 | 0.1189 | 0.1133 | -0.3554 | 0.0310 |
| GeS | -0.1740 | -0.2332 | 0.0787 | 0.0848 | 0.1027 | -0.3506 | 0.0080 |
| GeSe | 0.0896 | -0.1515 | 0.0777 | 0.0998 | 0.0916 | -0.3175 | 0.0008 |
| HAl | 3.1001 | -0.5999 | 0.0176 | 0.1806 | 0.0440 | -0.2659 | 0.0193 |
| HB | 14.1516 | -1.5202 | -0.0076 | 0.2200 | 0.0145 | -0.3097 | 0.0356 |
| HBr | 0.1324 | 1.3080 | 0.5176 | 0.2539 | 0.2242 | -0.3942 | 0.1136 |
| HCl | 0.1872 | 1.1378 | 0.5567 | 0.3092 | 0.2604 | -0.4252 | 0.1324 |
| HF | 0.5763 | 1.4979 | 0.8512 | 0.5372 | 0.3408 | -0.5230 | 0.1791 |
| HGa | 2.7515 | -0.4742 | 0.0280 | 0.1995 | 0.0532 | -0.2713 | 0.0227 |
| HLi | 26.2409 | 0.1932 | 0.0984 | 2.2457 | 0.0824 | -0.2563 | -0.0012 |
| HNa | 20.7939 | 0.3434 | 0.0806 | 1.3080 | 0.0600 | -0.2229 | -0.0055 |
| LiLi | 130.9659 | 0.5319 | 0.0266 | 2.2939 | 0.0174 | -0.1574 | -0.0131 |
| LiNa | 87.0899 | 0.7171 | 0.0251 | 1.2857 | 0.0146 | -0.1530 | -0.0119 |
| NaCl | 1.3168 | 0.3195 | 0.1746 | 0.3065 | 0.1323 | -0.2929 | -0.0194 |
| NaF | 2.6673 | 0.2389 | 0.1549 | 0.4585 | 0.1250 | -0.2916 | -0.0102 |
| NaNa | 107.6396 | 1.0171 | 0.0249 | 1.3427 | 0.0124 | -0.1490 | -0.0110 |
| NN | -0.2086 | -0.2011 | 0.2225 | 0.2204 | 0.2785 | -0.5589 | 0.1352 |
| NP | 0.0614 | -0.0700 | 0.0917 | 0.1047 | 0.0986 | -0.3922 | 0.0381 |
| OSi | 0.3773 | -0.2605 | 0.1005 | 0.1871 | 0.1359 | -0.3828 | 0.0392 |
| PP | -0.0199 | -0.1041 | 0.0673 | 0.0736 | 0.0751 | -0.3454 | 0.0143 |
| SeSi | -0.2569 | -0.1365 | 0.0879 | 0.0756 | 0.1018 | -0.3285 | 0.0043 |
| SSi | -0.2517 | -0.3444 | 0.0760 | 0.0867 | 0.1159 | -0.3582 | 0.0127 |

# Table S2.B Data for diatomic molecules at HF/cc-pVTZ.

| | $\delta^-$ | $\delta^+$ | $\Delta E_{\text{ST}}^{[v]}$ | $\Delta E_{\text{DQ}}^{[v],-}$ | $\Delta E_{\text{DQ}}^{[v],+}$ | $I$ | $A$ |
|---|---|---|---|---|---|---|---|
| AlBr | 0.8021 | -0.4108 | 0.0447 | 0.1367 | 0.0759 | -0.3033 | 0.0119 |
| AlF | 1.8148 | -0.4305 | 0.0495 | 0.2445 | 0.0869 | -0.3174 | 0.0321 |
| AsAs | -0.3512 | -0.4133 | 0.0470 | 0.0520 | 0.0801 | -0.3186 | -0.0092 |
| AsN | -0.1942 | -0.3013 | 0.0674 | 0.0777 | 0.0964 | -0.3553 | 0.0003 |
| AsP | -0.3085 | -0.3734 | 0.0560 | 0.0618 | 0.0893 | -0.3296 | -0.0025 |
| BBr | 1.6041 | -0.5093 | 0.0228 | 0.1211 | 0.0465 | -0.3257 | 0.0196 |
| BCl | 1.9667 | -0.4745 | 0.0284 | 0.1602 | 0.0540 | -0.3319 | 0.0346 |
| BF | 2.2137 | -0.3380 | 0.0641 | 0.3112 | 0.0968 | -0.3638 | 0.0757 |
| BrBr | -0.1550 | 3.7090 | 0.4199 | 0.0753 | 0.0892 | -0.3762 | -0.0241 |
| BrF | -0.1341 | 3.4013 | 0.4312 | 0.0848 | 0.0980 | -0.4074 | 0.0109 |
| BrGa | 0.3770 | -0.3198 | 0.0641 | 0.1297 | 0.0942 | -0.3170 | 0.0099 |
| BrLi | 0.9890 | 0.2657 | 0.1802 | 0.2832 | 0.1424 | -0.3053 | -0.0164 |
| BrNa | 1.1748 | 0.3312 | 0.1644 | 0.2686 | 0.1235 | -0.2798 | -0.0220 |
| ClAl | 1.0748 | -0.4226 | 0.0453 | 0.1627 | 0.0784 | -0.3076 | 0.0186 |
| ClBr | -0.2256 | 3.1391 | 0.3991 | 0.0747 | 0.0964 | -0.3848 | -0.0110 |
| ClCl | -0.1402 | 3.0924 | 0.4660 | 0.0979 | 0.1139 | -0.4074 | 0.0067 |
| ClF | -0.1189 | 3.2373 | 0.5132 | 0.1067 | 0.1211 | -0.4378 | 0.0392 |
| ClGa | 0.5142 | -0.3393 | 0.0645 | 0.1479 | 0.0977 | -0.3258 | 0.0161 |
| ClLi | 1.0908 | 0.2477 | 0.1939 | 0.3248 | 0.1554 | -0.3214 | -0.0133 |
| CO | 0.0331 | -0.1574 | 0.1720 | 0.2110 | 0.2042 | -0.4785 | 0.1169 |
| CS | -0.3638 | -0.5529 | 0.0695 | 0.0989 | 0.1555 | -0.3635 | 0.0253 |
| CSe | -0.0209 | 0.1140 | 0.1078 | 0.0948 | 0.0968 | -0.3932 | 0.0031 |
| FF | -0.4148 | 4.6411 | 0.8374 | 0.0869 | 0.1485 | -0.5750 | 0.0795 |
| FGa | 1.3508 | -0.3607 | 0.0684 | 0.2516 | 0.1070 | -0.3426 | 0.0292 |
| FLi | 1.8388 | 0.1454 | 0.1969 | 0.4881 | 0.1719 | -0.3492 | -0.0046 |
| GeO | 0.0396 | -0.1925 | 0.0920 | 0.1185 | 0.1140 | -0.3566 | 0.0192 |
| GeS | -0.0167 | -0.0991 | 0.0772 | 0.0843 | 0.0857 | -0.3481 | 0.0001 |
| GeSe | 0.0667 | 0.0467 | 0.0779 | 0.0794 | 0.0744 | -0.3152 | -0.0078 |
| HAl | 3.1132 | -0.5720 | 0.0187 | 0.1799 | 0.0437 | -0.2663 | 0.0159 |
| HB | 14.7280 | -1.2880 | -0.0041 | 0.2214 | 0.0141 | -0.3094 | 0.0262 |
| HBr | 0.1331 | 0.7067 | 0.3738 | 0.2481 | 0.2190 | -0.3926 | 0.0951 |
| HCl | 0.1768 | 0.5814 | 0.4041 | 0.3007 | 0.2555 | -0.4239 | 0.1128 |
| HF | 0.5985 | 0.9493 | 0.6537 | 0.5360 | 0.3353 | -0.5278 | 0.1395 |
| HGa | 2.7425 | -0.4461 | 0.0291 | 0.1969 | 0.0526 | -0.2712 | 0.0180 |
| HLi | 24.7682 | 0.1785 | 0.0983 | 2.1491 | 0.0834 | -0.2565 | -0.0029 |
| HNa | 19.7976 | 0.3090 | 0.0786 | 1.2495 | 0.0601 | -0.2220 | -0.0068 |
| LiLi | 118.9180 | 0.4197 | 0.0266 | 2.2478 | 0.0187 | -0.1582 | -0.0134 |
| LiNa | 79.8200 | 0.6360 | 0.0249 | 1.2286 | 0.0152 | -0.1533 | -0.0127 |
| NaCl | 1.3161 | 0.2640 | 0.1674 | 0.3067 | 0.1324 | -0.2911 | -0.0198 |
| NaF | 2.5135 | 0.2156 | 0.1567 | 0.4530 | 0.1289 | -0.2955 | -0.0126 |
| NaNa | 101.0512 | 0.8966 | 0.0244 | 1.3141 | 0.0129 | -0.1493 | -0.0117 |
| NN | -0.1116 | -0.0587 | 0.2499 | 0.2359 | 0.2655 | -0.5717 | 0.1384 |
| NP | -0.1680 | -0.2861 | 0.0902 | 0.1051 | 0.1263 | -0.3922 | 0.0235 |
| OSi | 0.3675 | -0.2844 | 0.0951 | 0.1818 | 0.1329 | -0.3847 | 0.0294 |
| PP | 0.0184 | -0.0767 | 0.0678 | 0.0748 | 0.0734 | -0.3426 | 0.0060 |
| SeSi | -0.1079 | -0.2099 | 0.0666 | 0.0752 | 0.0842 | -0.3437 | -0.0027 |
| SSi | -0.1548 | -0.2852 | 0.0734 | 0.0868 | 0.1026 | -0.3562 | 0.0065 |

# Table S2.C Data for diatomic molecules at HF/aug-cc-pVDZ.

| | $\delta^-$ | $\delta^+$ | $\Delta E_{ST}^{[v]}$ | $\Delta E_{DQ}^{[v],-}$ | $\Delta E_{DQ}^{[v],+}$ | $I$ | $A$ |
|---|---|---|---|---|---|---|---|
| AlBr | 0.7252 | -0.3744 | 0.0484 | 0.1335 | 0.0774 | -0.3073 | 0.0030 |
| AlF | 1.6583 | -0.3457 | 0.0573 | 0.2330 | 0.0876 | -0.3203 | 0.0153 |
| AsAs | -0.3722 | 0.4865 | 0.1207 | 0.0510 | 0.0812 | -0.3208 | -0.0148 |
| AsN | -0.1972 | -0.2966 | 0.0679 | 0.0775 | 0.0966 | -0.3568 | -0.0109 |
| AsP | -0.3325 | 0.4122 | 0.1270 | 0.0600 | 0.0899 | -0.3319 | -0.0101 |
| BBr | 1.5298 | 0.8366 | 0.0879 | 0.1211 | 0.0479 | -0.3294 | 0.0122 |
| BCl | 1.6630 | 0.5249 | 0.0836 | 0.1460 | 0.0548 | -0.3359 | 0.0249 |
| BF | 2.2136 | -0.2005 | 0.0740 | 0.2975 | 0.0926 | -0.3684 | 0.0414 |
| BrBr | -0.2032 | 1.5900 | 0.2107 | 0.0648 | 0.0813 | -0.3823 | -0.0400 |
| BrF | -0.1565 | 1.2318 | 0.2042 | 0.0772 | 0.0915 | -0.4148 | -0.0114 |
| BrGa | 0.4070 | -0.2902 | 0.0684 | 0.1357 | 0.0964 | -0.3206 | 0.0024 |
| BrLi | 0.9685 | 0.2156 | 0.1742 | 0.2821 | 0.1433 | -0.3072 | -0.0211 |
| BrNa | 1.1413 | 0.2723 | 0.1597 | 0.2688 | 0.1255 | -0.2822 | -0.0252 |
| ClAl | 0.8744 | -0.3773 | 0.0497 | 0.1495 | 0.0797 | -0.3125 | 0.0071 |
| ClBr | -0.2751 | 1.2938 | 0.2029 | 0.0641 | 0.0885 | -0.3906 | -0.0290 |
| ClCl | -0.1772 | 1.0770 | 0.2181 | 0.0864 | 0.1050 | -0.4139 | -0.0140 |
| ClF | -0.1362 | 0.8980 | 0.2116 | 0.0963 | 0.1115 | -0.4456 | 0.0096 |
| ClGa | 0.5257 | -0.3026 | 0.0697 | 0.1524 | 0.0999 | -0.3299 | 0.0064 |
| ClLi | 1.0725 | 0.1920 | 0.1857 | 0.3229 | 0.1558 | -0.3226 | -0.0190 |
| CO | 0.0343 | -0.0003 | 0.2010 | 0.2079 | 0.2010 | -0.4796 | 0.0770 |
| CS | -0.2323 | 0.0964 | 0.1415 | 0.0991 | 0.1290 | -0.3683 | 0.0146 |
| CSe | -0.2177 | 0.2987 | 0.1571 | 0.0947 | 0.1210 | -0.3955 | -0.0043 |
| FF | -0.4538 | 0.9236 | 0.2640 | 0.0750 | 0.1372 | -0.5813 | 0.0459 |
| FGa | 1.3142 | -0.2974 | 0.0763 | 0.2513 | 0.1086 | -0.3465 | 0.0142 |
| FLi | 1.8317 | 0.1160 | 0.1929 | 0.4894 | 0.1728 | -0.3497 | -0.0115 |
| GeO | 0.0209 | -0.1870 | 0.0916 | 0.1150 | 0.1126 | -0.3599 | 0.0048 |
| GeS | -0.1771 | -0.2337 | 0.0772 | 0.0829 | 0.1008 | -0.3518 | -0.0073 |
| GeSe | 0.3024 | 0.0414 | 0.0781 | 0.0977 | 0.0750 | -0.3169 | -0.0132 |
| HAl | 3.1123 | 0.6114 | 0.0704 | 0.1797 | 0.0437 | -0.2659 | 0.0062 |
| HB | 13.7191 | -0.9739 | 0.0004 | 0.2203 | 0.0150 | -0.3099 | 0.0151 |
| HBr | 0.1460 | 0.2004 | 0.2618 | 0.2499 | 0.2181 | -0.3959 | 0.0331 |
| HCl | 0.2093 | 0.1703 | 0.2929 | 0.3027 | 0.2503 | -0.4269 | 0.0336 |
| HF | 0.6471 | 0.1953 | 0.3879 | 0.5345 | 0.3245 | -0.5332 | 0.0355 |
| HGa | 2.7449 | 0.4505 | 0.0769 | 0.1985 | 0.0530 | -0.2713 | 0.0087 |
| HLi | 25.6655 | 0.1860 | 0.0980 | 2.2045 | 0.0827 | -0.2565 | -0.0094 |
| HNa | 20.6956 | 0.2969 | 0.0775 | 1.2960 | 0.0597 | -0.2218 | -0.0109 |
| LiLi | 128.0884 | 0.2041 | 0.0214 | 2.2920 | 0.0178 | -0.1577 | -0.0148 |
| LiNa | 86.0582 | -0.1560 | 0.0124 | 1.2828 | 0.0147 | -0.1531 | -0.0019 |
| NaCl | 1.2776 | 0.2498 | 0.1684 | 0.3069 | 0.1347 | -0.2935 | -0.0233 |
| NaF | 2.4305 | 0.1862 | 0.1577 | 0.4560 | 0.1329 | -0.2988 | -0.0168 |
| NaNa | 105.6499 | 1.9318 | 0.0369 | 1.3415 | 0.0126 | -0.1492 | -0.0122 |
| NN | -0.1117 | -0.0271 | 0.2500 | 0.2283 | 0.2570 | -0.5706 | 0.1115 |
| NP | 0.0823 | -0.0515 | 0.0902 | 0.1029 | 0.0951 | -0.3935 | 0.0080 |
| OSi | 0.4231 | -0.2656 | 0.0957 | 0.1855 | 0.1303 | -0.3864 | 0.0103 |
| PP | -0.2889 | 0.3403 | 0.1358 | 0.0720 | 0.1013 | -0.3451 | -0.0040 |
| SeSi | -0.2402 | 0.3961 | 0.1394 | 0.0759 | 0.0999 | -0.3278 | -0.0100 |
| SSi | -0.2554 | -0.3524 | 0.0736 | 0.0846 | 0.1137 | -0.3591 | -0.0032 |

# Table S2.D Data for diatomic molecules at HF/aug-cc-pVTZ.

|      | $\delta^-$ | $\delta^+$ | $\Delta E_{\text{ST}}^{[v]}$ | $\Delta E_{\text{DQ}}^{[v],-}$ | $\Delta E_{\text{DQ}}^{[v],+}$ | $I$ | $A$ |
|------|--------|---------|---------|---------|---------|---------|---------|
| AlBr | 0.8024 | -0.3745 | 0.0472 | 0.1361 | 0.0755 | -0.3032 | 0.0049 |
| AlF  | 1.8440 | -0.3384 | 0.0571 | 0.2453 | 0.0862 | -0.3172 | 0.0154 |
| AsAs | -0.3464 | 0.3782 | 0.1096 | 0.0520 | 0.0795 | -0.3187 | -0.0151 |
| AsN  | -0.1847 | -0.2924 | 0.0671 | 0.0774 | 0.0949 | -0.3558 | -0.0113 |
| AsP  | 0.0018 | 0.8716 | 0.1152 | 0.0617 | 0.0615 | -0.3297 | -0.0101 |
| BBr  | 1.6057 | 0.6163 | 0.0751 | 0.1210 | 0.0465 | -0.3256 | 0.0135 |
| BCl  | 1.7415 | 0.3700 | 0.0735 | 0.1471 | 0.0536 | -0.3319 | 0.0264 |
| BF   | 2.1054 | -0.1402 | 0.0822 | 0.2970 | 0.0956 | -0.3639 | 0.0391 |
| BrBr | -0.1532 | 1.0941 | 0.1855 | 0.0750 | 0.0886 | -0.3759 | -0.0289 |
| BrF  | -0.1205 | 0.7131 | 0.1649 | 0.0847 | 0.0963 | -0.4079 | -0.0016 |
| BrGa | 0.4405 | 0.1968 | 0.1124 | 0.1353 | 0.0939 | -0.3170 | 0.0036 |
| BrLi | 0.9891 | 0.2422 | 0.1769 | 0.2832 | 0.1424 | -0.3051 | -0.0202 |
| BrNa | 1.1709 | 0.2731 | 0.1575 | 0.2686 | 0.1237 | -0.2799 | -0.0247 |
| ClAl | 1.0797 | -0.3749 | 0.0486 | 0.1618 | 0.0778 | -0.3078 | 0.0090 |
| ClBr | * | * | * | * | * | * | * |
| ClCl | -0.1373 | 0.7308 | 0.1956 | 0.0975 | 0.1130 | -0.4070 | -0.0019 |
| ClF  | -0.1072 | 0.6015 | 0.1906 | 0.1062 | 0.1190 | -0.4382 | 0.0226 |
| ClGa | 0.5780 | 0.1582 | 0.1126 | 0.1535 | 0.0972 | -0.3260 | 0.0078 |
| ClLi | 1.0869 | 0.2034 | 0.1873 | 0.3248 | 0.1556 | -0.3214 | -0.0180 |
| CO   | 0.0420 | -0.0231 | 0.1977 | 0.2109 | 0.2024 | -0.4787 | 0.0698 |
| CS   | -0.2291 | -0.0095 | 0.1271 | 0.0989 | 0.1283 | -0.3640 | 0.0156 |
| CSe  | -0.2080 | 0.2597 | 0.1507 | 0.0948 | 0.1196 | -0.3934 | -0.0040 |
| FF   | -0.3973 | 0.5160 | 0.2233 | 0.0888 | 0.1473 | -0.5741 | 0.0610 |
| FGa  | 1.3564 | 0.0875 | 0.1158 | 0.2510 | 0.1065 | -0.3432 | 0.0155 |
| FLi  | 1.8184 | 0.1292 | 0.1957 | 0.4885 | 0.1733 | -0.3499 | -0.0109 |
| GeO  | 0.0432 | -0.1786 | 0.0926 | 0.1176 | 0.1127 | -0.3575 | 0.0051 |
| GeS  | -0.1646 | -0.2334 | 0.0768 | 0.0837 | 0.1002 | -0.3483 | -0.0069 |
| GeSe | 0.0721 | 0.6491 | 0.1221 | 0.0794 | 0.0741 | -0.3151 | -0.0130 |
| HAl  | 3.1177 | 0.5105 | 0.0659 | 0.1796 | 0.0436 | -0.2662 | 0.0061 |
| HB   | 14.5511 | 1.3820 | 0.0339 | 0.2215 | 0.0142 | -0.3095 | 0.0150 |
| HBr  | 0.1478 | 0.1526 | 0.2485 | 0.2475 | 0.2156 | -0.3926 | 0.0266 |
| HCl  | 0.2093 | 0.1703 | 0.2929 | 0.3027 | 0.2503 | -0.4269 | 0.0336 |
| HF   | 0.6618 | 0.1124 | 0.3579 | 0.5347 | 0.3218 | -0.5299 | 0.0304 |
| HGa  | 2.7369 | 0.2319 | 0.0648 | 0.1965 | 0.0526 | -0.2711 | 0.0090 |
| HLi  | 24.6646 | 0.1783 | 0.0983 | 2.1408 | 0.0834 | -0.2565 | -0.0090 |
| HNa  | 19.6426 | 0.2949 | 0.0779 | 1.2422 | 0.0602 | -0.2222 | -0.0111 |
| LiLi | 119.0281 | 0.1565 | 0.0216 | 2.2463 | 0.0187 | -0.1581 | -0.0147 |
| LiNa | 79.6494 | -0.1731 | 0.0126 | 1.2248 | 0.0152 | -0.1533 | -0.0021 |
| NaCl | 1.3071 | 0.2386 | 0.1646 | 0.3066 | 0.1329 | -0.2914 | -0.0230 |
| NaF  | 2.4666 | 0.1974 | 0.1564 | 0.4528 | 0.1306 | -0.2965 | -0.0166 |
| NaNa | 100.9843 | 2.1674 | 0.0408 | 1.3125 | 0.0129 | -0.1493 | -0.0125 |
| NN   | -0.1088 | -0.1148 | 0.2344 | 0.2360 | 0.2648 | -0.5722 | 0.1140 |
| NP   | -0.1588 | 0.1646 | 0.1451 | 0.1048 | 0.1246 | -0.3928 | 0.0079 |
| OSi  | 0.3814 | -0.2639 | 0.0965 | 0.1811 | 0.1311 | -0.3856 | 0.0113 |
| PP   | 0.0237 | 0.7222 | 0.1256 | 0.0746 | 0.0729 | -0.3429 | -0.0034 |
| SeSi | -0.1075 | 0.6275 | 0.1365 | 0.0749 | 0.0839 | -0.3438 | -0.0092 |
| SSi  | -0.3217 | 0.4096 | 0.1438 | 0.0692 | 0.1020 | -0.3564 | -0.0020 |

*SCF convergence problem for ionic excited state(s)

## Table S3.A Data for diatomic molecules at CISD/ cc-pVDZ.

|      | $\delta^-$ | $\delta^+$ | $\Delta E_{\text{ST}}^{[v]}$ | $\Delta E_{\text{DQ}}^{[v],-}$ | $\Delta E_{\text{DQ}}^{[v],+}$ | $I$ | $A$ |
|---|---|---|---|---|---|---|---|
| AlBr | 0.7676 | -0.4135 | 0.0457 | 0.1377 | 0.0779 | -0.3049 | 0.0158 |
| AlF  | 1.4954 | -0.4244 | 0.0522 | 0.2263 | 0.0907 | -0.3238 | 0.0347 |
| AsAs | -0.0878 | -0.1693 | 0.0396 | 0.0435 | 0.0477 | -0.3165 | -0.0026 |
| AsN  | -0.2127 | -0.2894 | 0.0624 | 0.0692 | 0.0879 | -0.3446 | 0.0045 |
| AsP  | -0.0643 | -0.1387 | 0.0477 | 0.0519 | 0.0554 | -0.3276 | 0.0024 |
| BBr  | 1.5388 | -0.5540 | 0.0215 | 0.1224 | 0.0482 | -0.3288 | 0.0274 |
| BCl  | 1.8580 | -0.5310 | 0.0260 | 0.1584 | 0.0554 | -0.3364 | 0.0414 |
| BF   | 2.0722 | -0.3665 | 0.0600 | 0.2908 | 0.0946 | -0.3696 | 0.0822 |
| BrBr | -0.1750 | 6.6002 | 0.6076 | 0.0660 | 0.0799 | -0.3855 | -0.0305 |
| BrF  | -0.1986 | 6.8166 | 0.6387 | 0.0655 | 0.0817 | -0.4172 | -0.0035 |
| BrGa | 0.4543 | -0.3301 | 0.0643 | 0.1396 | 0.0960 | -0.3183 | 0.0157 |
| BrLi | 1.0012 | 0.3288 | 0.1875 | 0.2824 | 0.1411 | -0.3062 | -0.0155 |
| BrNa | 1.1848 | 0.3635 | 0.1680 | 0.2692 | 0.1232 | -0.2811 | -0.0207 |
| ClAl | 0.9864 | -0.4221 | 0.0467 | 0.1606 | 0.0809 | -0.3111 | 0.0207 |
| ClBr | -0.3281 | 6.3209 | 0.6208 | 0.0570 | 0.0848 | -0.3937 | -0.0218 |
| ClCl | -0.1923 | 5.8216 | 0.6926 | 0.0820 | 0.1015 | -0.4203 | -0.0081 |
| ClF  | -0.1670 | 6.1270 | 0.7074 | 0.0827 | 0.0993 | -0.4495 | 0.0146 |
| ClGa | 0.5043 | -0.3463 | 0.0653 | 0.1504 | 0.1000 | -0.3281 | 0.0202 |
| ClLi | 1.1047 | 0.2529 | 0.1915 | 0.3218 | 0.1529 | -0.3219 | -0.0141 |
| CO   | -0.0295 | -0.1569 | 0.1666 | 0.1918 | 0.1976 | -0.4799 | 0.1228 |
| CS   | -0.3693 | -0.5472 | 0.0676 | 0.0942 | 0.1493 | -0.3672 | 0.0297 |
| CSe  | -0.2266 | -0.0718 | 0.1084 | 0.0903 | 0.1167 | -0.3920 | 0.0098 |
| FF   | -1.0387 | 12.0200 | 1.3474 | -0.0040 | 0.1035 | -0.6002 | 0.0253 |
| FGa  | 1.1966 | -0.3676 | 0.0699 | 0.2427 | 0.1105 | -0.3456 | 0.0350 |
| FLi  | 1.9639 | 0.1722 | 0.1928 | 0.4875 | 0.1645 | -0.3434 | -0.0039 |
| GeO  | 0.0644 | -0.1905 | 0.0859 | 0.1130 | 0.1061 | -0.3511 | 0.0273 |
| GeS  | -0.0816 | -0.1566 | 0.0822 | 0.0895 | 0.0975 | -0.3264 | 0.0055 |
| GeSe | 0.0980 | -0.1485 | 0.0741 | 0.0955 | 0.0870 | -0.3147 | -0.0014 |
| HAl  | 3.0938 | -0.5992 | 0.0177 | 0.1804 | 0.0441 | -0.2659 | 0.0193 |
| HB   | 13.2181 | -1.4364 | -0.0066 | 0.2162 | 0.0152 | -0.3102 | 0.0350 |
| HBr  | 0.1334 | 1.3438 | 0.5192 | 0.2511 | 0.2215 | -0.3939 | 0.1114 |
| HCl  | 0.1883 | 1.1730 | 0.5582 | 0.3052 | 0.2569 | -0.4248 | 0.1298 |
| HF   | 0.5789 | 1.5113 | 0.8358 | 0.5255 | 0.3328 | -0.5205 | 0.1755 |
| HGa  | 2.7559 | -0.4748 | 0.0279 | 0.1996 | 0.0531 | -0.2713 | 0.0227 |
| HLi  | 26.2424 | 0.1932 | 0.0984 | 2.2457 | 0.0824 | -0.2563 | -0.0012 |
| HNa  | 20.8928 | 0.3473 | 0.0805 | 1.3078 | 0.0597 | -0.2226 | -0.0057 |
| LiLi | * | * | * | * | * | * | * |
| LiNa | 82.2792 | 0.6128 | 0.0249 | 1.2862 | 0.0154 | -0.1536 | -0.0113 |
| NaCl | 1.3187 | 0.3203 | 0.1744 | 0.3063 | 0.1321 | -0.2927 | -0.0194 |
| NaF  | 2.7389 | 0.2509 | 0.1525 | 0.4559 | 0.1219 | -0.2887 | -0.0109 |
| NaNa | 107.9399 | 1.0228 | 0.0249 | 1.3426 | 0.0123 | -0.1489 | -0.0110 |
| NN   | -0.2092 | -0.2013 | 0.2220 | 0.2199 | 0.2780 | -0.5586 | 0.1349 |
| NP   | -0.2236 | -0.3045 | 0.0821 | 0.0916 | 0.1180 | -0.3830 | 0.0286 |
| OSi  | 0.4286 | -0.2499 | 0.0980 | 0.1867 | 0.1307 | -0.3785 | 0.0354 |
| PP   | -0.3325 | -0.3811 | 0.0585 | 0.0630 | 0.0945 | -0.3408 | 0.0089 |
| SeSi | 0.0666 | -0.1331 | 0.0839 | 0.1032 | 0.0967 | -0.3255 | 0.0018 |
| SSi  | -0.1242 | -0.2006 | 0.0750 | 0.0822 | 0.0939 | -0.3581 | 0.0099 |

*SCF convergence problem for ionic excited state(s)

# Table S3.B Data for diatomic molecules at CISD/ cc-pVTZ.

| | $\delta^-$ | $\delta^+$ | $\Delta E_{ST}^{[v]}$ | $\Delta E_{DQ}^{[v],-}$ | $\Delta E_{DQ}^{[v],+}$ | $I$ | $A$ |
|---|---|---|---|---|---|---|---|
| AlBr | 0.8438 | -0.4138 | 0.0443 | 0.1392 | 0.0755 | -0.3021 | 0.0129 |
| AlF | 1.6801 | -0.4298 | 0.0496 | 0.2331 | 0.0870 | -0.3185 | 0.0314 |
| AsAs | -0.3696 | -0.4266 | 0.0442 | 0.0486 | 0.0771 | -0.3171 | -0.0109 |
| AsN | -0.1941 | -0.2905 | 0.0625 | 0.0710 | 0.0881 | -0.3481 | -0.0064 |
| AsP | -0.3332 | -0.3905 | 0.0515 | 0.0564 | 0.0846 | -0.3272 | -0.0052 |
| BBr | 1.6890 | -0.5165 | 0.0223 | 0.1243 | 0.0462 | -0.3238 | 0.0212 |
| BCl | 1.9756 | -0.4747 | 0.0284 | 0.1606 | 0.0540 | -0.3317 | 0.0348 |
| BF | 2.1753 | -0.3408 | 0.0629 | 0.3030 | 0.0954 | -0.3656 | 0.0735 |
| BrBr | -0.1556 | 3.7251 | 0.4200 | 0.0751 | 0.0889 | -0.3763 | -0.0245 |
| BrF | -0.1444 | 3.7003 | 0.4332 | 0.0789 | 0.0922 | -0.4095 | 0.0030 |
| BrGa | 0.5104 | -0.3289 | 0.0624 | 0.1403 | 0.0929 | -0.3152 | 0.0123 |
| BrLi | 0.9850 | 0.2634 | 0.1805 | 0.2836 | 0.1429 | -0.3057 | -0.0161 |
| BrNa | 1.1710 | 0.3290 | 0.1646 | 0.2689 | 0.1239 | -0.2801 | -0.0218 |
| ClAl | 0.9944 | -0.4233 | 0.0452 | 0.1562 | 0.0783 | -0.3072 | 0.0190 |
| ClBr | -0.2342 | 3.2131 | 0.3996 | 0.0726 | 0.0949 | -0.3852 | -0.0131 |
| ClCl | -0.1451 | 3.5387 | 0.5029 | 0.0947 | 0.1108 | -0.4088 | 0.0022 |
| ClF | -0.1267 | 3.5873 | 0.5131 | 0.0977 | 0.1119 | -0.4409 | 0.0265 |
| ClGa | 0.5682 | -0.3439 | 0.0635 | 0.1518 | 0.0968 | -0.3240 | 0.0178 |
| ClLi | 1.0911 | 0.2479 | 0.1938 | 0.3248 | 0.1553 | -0.3213 | -0.0133 |
| CO | -0.0081 | -0.1569 | 0.1670 | 0.1964 | 0.1980 | -0.4798 | 0.1094 |
| CS | -0.2341 | -0.4502 | 0.0683 | 0.0952 | 0.1243 | -0.3636 | 0.0222 |
| CSe | -0.2156 | -0.0987 | 0.1060 | 0.0923 | 0.1177 | -0.3918 | 0.0015 |
| FF | -0.6911 | 5.9745 | 0.8556 | 0.0379 | 0.1227 | -0.5853 | 0.0412 |
| FGa | 1.3418 | -0.3605 | 0.0685 | 0.2508 | 0.1071 | -0.3429 | 0.0291 |
| FLi | 1.8607 | 0.1494 | 0.1954 | 0.4863 | 0.1700 | -0.3473 | -0.0051 |
| GeO | 0.0472 | -0.1839 | 0.0896 | 0.1150 | 0.1098 | -0.3541 | 0.0170 |
| GeS | -0.0140 | -0.0887 | 0.0769 | 0.0832 | 0.0844 | -0.3480 | -0.0005 |
| GeSe | -0.1135 | -0.1282 | 0.0781 | 0.0794 | 0.0896 | -0.3154 | -0.0077 |
| HAl | 3.0973 | -0.5706 | 0.0188 | 0.1795 | 0.0438 | -0.2664 | 0.0158 |
| HB | 14.1095 | -1.2451 | -0.0036 | 0.2190 | 0.0145 | -0.3097 | 0.0259 |
| HBr | 0.1330 | 0.7051 | 0.3737 | 0.2483 | 0.2192 | -0.3926 | 0.0952 |
| HCl | 0.1775 | 0.5929 | 0.4042 | 0.2988 | 0.2538 | -0.4237 | 0.1117 |
| HF | 0.5981 | 0.9538 | 0.6419 | 0.5250 | 0.3285 | -0.5254 | 0.1369 |
| HGa | 2.8197 | -0.4549 | 0.0284 | 0.1991 | 0.0521 | -0.2710 | 0.0182 |
| HLi | 24.7719 | 0.1786 | 0.0983 | 2.1491 | 0.0834 | -0.2565 | -0.0029 |
| HNa | 19.8592 | 0.3114 | 0.0785 | 1.2494 | 0.0599 | -0.2219 | -0.0068 |
| LiLi | 107.9834 | 0.2651 | 0.0261 | 2.2475 | 0.0206 | -0.1595 | -0.0121 |
| LiNa | 74.6886 | 0.5223 | 0.0247 | 1.2290 | 0.0162 | -0.1541 | -0.0120 |
| NaCl | 1.3191 | 0.2654 | 0.1672 | 0.3064 | 0.1321 | -0.2908 | -0.0199 |
| NaF | 2.5934 | 0.2306 | 0.1539 | 0.4494 | 0.1251 | -0.2919 | -0.0136 |
| NaNa | 98.5303 | 0.8491 | 0.0244 | 1.3142 | 0.0132 | -0.1495 | -0.0115 |
| NN | -0.2323 | -0.1962 | 0.2317 | 0.2213 | 0.2882 | -0.5630 | 0.1264 |
| NP | 0.1122 | -0.0276 | 0.0826 | 0.0945 | 0.0849 | -0.3853 | 0.0168 |
| OSi | 0.3923 | -0.2809 | 0.0938 | 0.1816 | 0.1304 | -0.3817 | 0.0265 |
| PP | 0.0064 | -0.0776 | 0.0615 | 0.0671 | 0.0667 | -0.3393 | 0.0021 |
| SeSi | -0.0869 | 0.0681 | 0.0880 | 0.0752 | 0.0824 | -0.3253 | -0.0038 |
| SSi | -0.2484 | -0.3463 | 0.0728 | 0.0837 | 0.1114 | -0.3562 | 0.0047 |

# Table S3.C Data for diatomic molecules at CISD/aug-cc-pVDZ.

|      | $\delta^-$ | $\delta^+$ | $\Delta E_{\text{ST}}^{[v]}$ | $\Delta E_{\text{DQ}}^{[v],-}$ | $\Delta E_{\text{DQ}}^{[v],+}$ | $I$ | $A$ |
|------|---------|---------|---------|---------|---------|---------|---------|
| AlBr | 0.6431 | -0.3748 | 0.0483 | 0.1270 | 0.0773 | -0.3071 | 0.0031 |
| AlF  | 1.6442 | -0.3485 | 0.0572 | 0.2323 | 0.0878 | -0.3222 | 0.0146 |
| AsAs | -0.4222 | 0.5855 | 0.1166 | 0.0425 | 0.0736 | -0.3168 | -0.0190 |
| AsN  | -0.0811 | 1.1671 | 0.1257 | 0.0533 | 0.0580 | -0.3609 | -0.0081 |
| AsP  | -0.0634 | 1.2683 | 0.1227 | 0.0506 | 0.0541 | -0.3276 | -0.0148 |
| BBr  | 1.5251 | 0.8379 | 0.0880 | 0.1209 | 0.0479 | -0.3295 | 0.0121 |
| BCl  | 1.6229 | 0.5414 | 0.0846 | 0.1439 | 0.0549 | -0.3370 | 0.0240 |
| BF   | 2.1385 | -0.2192 | 0.0707 | 0.2843 | 0.0906 | -0.3714 | 0.0403 |
| BrBr | -0.1485 | 1.7644 | 0.2123 | 0.0654 | 0.0768 | -0.3843 | -0.0463 |
| BrF  | -0.1749 | 1.6084 | 0.2160 | 0.0683 | 0.0828 | -0.4174 | -0.0222 |
| BrGa | 0.4196 | -0.2918 | 0.0681 | 0.1365 | 0.0961 | -0.3203 | 0.0028 |
| BrLi | 0.9743 | 0.2188 | 0.1737 | 0.2814 | 0.1425 | -0.3065 | -0.0214 |
| BrNa | 1.1484 | 0.2762 | 0.1593 | 0.2682 | 0.1248 | -0.2817 | -0.0255 |
| ClAl | 0.8605 | -0.3770 | 0.0497 | 0.1486 | 0.0799 | -0.3129 | 0.0069 |
| ClBr | -0.3154 | 1.4747 | 0.2037 | 0.0564 | 0.0823 | -0.3919 | -0.0368 |
| ClCl | -0.1929 | 1.2541 | 0.2205 | 0.0790 | 0.0978 | -0.4170 | -0.0239 |
| ClF  | * | * | * | * | * | * | * |
| ClGa | 0.5268 | -0.3027 | 0.0696 | 0.1525 | 0.0999 | -0.3299 | 0.0064 |
| ClLi | 1.0808 | 0.1957 | 0.1850 | 0.3219 | 0.1547 | -0.3216 | -0.0194 |
| CO   | -0.0188 | 0.0140 | 0.1961 | 0.1897 | 0.1934 | -0.4812 | 0.0741 |
| CS   | -0.2305 | 0.1745 | 0.1427 | 0.0935 | 0.1215 | -0.3683 | 0.0101 |
| CSe  | -0.2255 | 0.3550 | 0.1550 | 0.0886 | 0.1144 | -0.3919 | -0.0081 |
| FF   | -0.9188 | 1.7286 | 0.2795 | 0.0083 | 0.1024 | -0.5952 | -0.0040 |
| FGa  | 1.2668 | -0.2981 | 0.0765 | 0.2470 | 0.1090 | -0.3476 | 0.0137 |
| FLi  | 1.8648 | 0.1215 | 0.1904 | 0.4864 | 0.1698 | -0.3467 | -0.0121 |
| GeO  | 0.0330 | -0.1756 | 0.0872 | 0.1093 | 0.1058 | -0.3557 | 0.0017 |
| GeS  | -0.0599 | -0.1309 | 0.0830 | 0.0898 | 0.0955 | -0.3255 | -0.0097 |
| GeSe | 0.3191 | 0.0533 | 0.0748 | 0.0936 | 0.0710 | -0.3142 | -0.0153 |
| HAl  | 3.0650 | 0.6099 | 0.0707 | 0.1786 | 0.0439 | -0.2660 | 0.0061 |
| HB   | 12.8439 | -0.9299 | 0.0011 | 0.2165 | 0.0156 | -0.3104 | 0.0149 |
| HBr  | 0.1459 | 0.2082 | 0.2603 | 0.2469 | 0.2154 | -0.3957 | 0.0328 |
| HCl  | 76.7545 | 64.9419 | 0.2471 | 0.2914 | 0.0037 | -0.0069 | -0.0264 |
| HF   | 0.6408 | 0.1970 | 0.3796 | 0.5203 | 0.3171 | -0.5302 | 0.0349 |
| HGa  | 2.7332 | 0.4502 | 0.0770 | 0.1982 | 0.0531 | -0.2713 | 0.0087 |
| HLi  | 25.7500 | 0.1875 | 0.0979 | 2.2049 | 0.0824 | -0.2563 | -0.0095 |
| HNa  | 20.8719 | 0.3024 | 0.0772 | 1.2960 | 0.0593 | -0.2215 | -0.0110 |
| LiLi | * | * | * | * | * | * | * |
| LiNa | 81.6132 | -0.1738 | 0.0128 | 1.2830 | 0.0155 | -0.1537 | -0.0015 |
| NaCl | 1.2871 | 0.2540 | 0.1678 | 0.3061 | 0.1338 | -0.2927 | -0.0237 |
| NaF  | 2.5094 | 0.2005 | 0.1546 | 0.4520 | 0.1288 | -0.2950 | -0.0178 |
| NaNa | 106.6388 | 1.9562 | 0.0368 | 1.3415 | 0.0125 | -0.1491 | -0.0122 |
| NN   | -0.0592 | 0.0244 | 0.2381 | 0.2187 | 0.2324 | -0.5598 | 0.0992 |
| NP   | -0.2057 | -0.2903 | 0.0806 | 0.0902 | 0.1135 | -0.3843 | -0.0001 |
| OSi  | 0.4691 | -0.2601 | 0.0932 | 0.1851 | 0.1260 | -0.3821 | 0.0071 |
| PP   | -0.3308 | 0.4233 | 0.1313 | 0.0617 | 0.0922 | -0.3406 | -0.0091 |
| SeSi | 0.2732 | 0.7324 | 0.1373 | 0.1009 | 0.0793 | -0.3248 | -0.0124 |
| SSi  | -0.2576 | -0.3256 | 0.0727 | 0.0800 | 0.1078 | -0.3591 | -0.0059 |

*SCF convergence problem for ionic excited state(s)

## Table S3.D Data for diatomic molecules at CISD/aug-cc-pVTZ.

| | $\delta^-$ | $\delta^+$ | $\Delta E_{\text{ST}}^{[v]}$ | $\Delta E_{\text{DQ}}^{[v],-}$ | $\Delta E_{\text{DQ}}^{[v],+}$ | $I$ | $A$ |
|---|---|---|---|---|---|---|---|
| AlBr | 0.8382 | -0.3760 | 0.0469 | 0.1382 | 0.0752 | -0.3023 | 0.0055 |
| AlF | 1.7690 | -0.3409 | 0.0569 | 0.2392 | 0.0864 | -0.3187 | 0.0149 |
| AsAs | -0.3633 | 0.4103 | 0.1082 | 0.0488 | 0.0767 | -0.3172 | -0.0167 |
| AsN | -0.0345 | 0.9852 | 0.1204 | 0.0585 | 0.0606 | -0.3613 | -0.0064 |
| AsP | -0.0086 | 0.9853 | 0.1131 | 0.0565 | 0.0569 | -0.3274 | -0.0126 |
| BBr | 1.6955 | 0.5871 | 0.0732 | 0.1244 | 0.0461 | -0.3237 | 0.0149 |
| BCl | 1.7471 | 0.3677 | 0.0734 | 0.1474 | 0.0536 | -0.3317 | 0.0265 |
| BF | 2.2044 | -0.1538 | 0.0796 | 0.3015 | 0.0941 | -0.3658 | 0.0386 |
| BrBr | -0.1536 | 1.0991 | 0.1856 | 0.0748 | 0.0884 | -0.3759 | -0.0292 |
| BrF | -0.1302 | 0.8276 | 0.1661 | 0.0791 | 0.0909 | -0.4099 | -0.0083 |
| BrGa | 0.5056 | 0.1847 | 0.1098 | 0.1395 | 0.0927 | -0.3153 | 0.0057 |
| BrLi | 0.9860 | 0.2405 | 0.1771 | 0.2836 | 0.1428 | -0.3055 | -0.0200 |
| BrNa | 1.1699 | 0.2725 | 0.1576 | 0.2687 | 0.1238 | -0.2800 | -0.0247 |
| ClAl | 0.9905 | -0.3750 | 0.0486 | 0.1548 | 0.0778 | -0.3076 | 0.0091 |
| ClBr | * | * | * | * | * | * | * |
| ClCl | -0.1425 | 0.7911 | 0.1967 | 0.0942 | 0.1098 | -0.4084 | -0.0062 |
| ClF | -0.1155 | 0.7522 | 0.1924 | 0.0971 | 0.1098 | -0.4413 | 0.0114 |
| ClGa | 0.6306 | 0.1520 | 0.1111 | 0.1572 | 0.0964 | -0.3244 | 0.0091 |
| ClLi | 1.0894 | 0.2044 | 0.1871 | 0.3245 | 0.1553 | -0.3211 | -0.0181 |
| CO | -0.0005 | -0.0665 | 0.1833 | 0.1962 | 0.1963 | -0.4801 | 0.0686 |
| CS | 0.0624 | 0.0354 | 0.1279 | 0.1313 | 0.1236 | -0.3641 | 0.0128 |
| CSe | -0.2106 | 0.2782 | 0.1499 | 0.0926 | 0.1173 | -0.3921 | -0.0054 |
| FF | -0.6790 | 0.9364 | 0.2339 | 0.0388 | 0.1208 | -0.5845 | 0.0239 |
| FGa | 1.3370 | 0.0884 | 0.1161 | 0.2492 | 0.1066 | -0.3437 | 0.0153 |
| FLi | 1.8469 | 0.1226 | 0.1916 | 0.4860 | 0.1707 | -0.3474 | -0.0114 |
| GeO | 0.0505 | -0.1704 | 0.0902 | 0.1142 | 0.1087 | -0.3551 | 0.0032 |
| GeS | -0.1636 | -0.2268 | 0.0766 | 0.0829 | 0.0991 | -0.3482 | -0.0074 |
| GeSe | 0.0668 | 0.6420 | 0.1223 | 0.0795 | 0.0745 | -0.3154 | -0.0127 |
| HAl | 3.0920 | 0.5098 | 0.0660 | 0.1790 | 0.0437 | -0.2663 | 0.0060 |
| HB | 13.9165 | 1.3551 | 0.0346 | 0.2189 | 0.0147 | -0.3099 | 0.0148 |
| HBr | 0.1478 | 0.1517 | 0.2487 | 0.2478 | 0.2159 | -0.3927 | 0.0267 |
| HCl | 0.1994 | 0.1547 | 0.2856 | 0.2966 | 0.2473 | -0.4240 | 0.0285 |
| HF | 0.6563 | 0.1144 | 0.3516 | 0.5226 | 0.3155 | -0.5273 | 0.0300 |
| HGa | 2.8249 | 0.2283 | 0.0639 | 0.1989 | 0.0520 | -0.2709 | 0.0092 |
| HLi | 24.7028 | 0.1791 | 0.0982 | 2.1408 | 0.0833 | -0.2564 | -0.0090 |
| HNa | 19.7760 | 0.2994 | 0.0777 | 1.2421 | 0.0598 | -0.2219 | -0.0112 |
| LiLi | * | * | * | * | * | * | * |
| LiNa | 74.3838 | -0.1947 | 0.0131 | 1.2251 | 0.0163 | -0.1541 | -0.0015 |
| NaCl | 1.3127 | 0.2410 | 0.1643 | 0.3062 | 0.1324 | -0.2909 | -0.0232 |
| NaF | 2.5583 | 0.2140 | 0.1530 | 0.4484 | 0.1260 | -0.2923 | -0.0178 |
| NaNa | 98.3140 | 2.0932 | 0.0409 | 1.3126 | 0.0132 | -0.1495 | -0.0122 |
| NN | -0.0961 | -0.0895 | 0.2230 | 0.2214 | 0.2449 | -0.5635 | 0.1054 |
| NP | -0.1800 | 0.2215 | 0.1403 | 0.0942 | 0.1148 | -0.3859 | 0.0019 |
| OSi | 0.4069 | -0.2626 | 0.0948 | 0.1810 | 0.1286 | -0.3826 | 0.0090 |
| PP | 0.0120 | 0.8468 | 0.1223 | 0.0670 | 0.0662 | -0.3397 | -0.0071 |
| SeSi | -0.2324 | 0.3463 | 0.1319 | 0.0752 | 0.0979 | -0.3252 | -0.0101 |
| SSi | -0.2982 | 0.5231 | 0.1443 | 0.0665 | 0.0947 | -0.3564 | -0.0038 |

*SCF convergence problem for ionic excited state(s)

# Table S4.A Data for diatomic molecules at B3LYP/cc-pVDZ.

|       | $\delta^-$ | $\delta^+$ | $\Delta E_{\mathrm{ST}}^{[v]}$ | $\Delta E_{\mathrm{DQ}}^{[v],-}$ | $\Delta E_{\mathrm{DQ}}^{[v],+}$ | $I$ | $A$ |
|---|---|---|---|---|---|---|---|
| AlBr | 0.6289 | -0.1982 | 0.0872 | 0.1772 | 0.1088 | -0.3382 | 0.0035 |
| AlF  | 1.3109 | -0.2321 | 0.0969 | 0.2915 | 0.1261 | -0.3613 | 0.0237 |
| AsAs | -0.1906 | -0.2625 | 0.0777 | 0.0853 | 0.1053 | -0.3559 | -0.0111 |
| AsN  | -0.1801 | -0.1700 | 0.1069 | 0.1056 | 0.1288 | -0.4166 | 0.0145 |
| AsP  | 0.0184 | -0.0701 | 0.0859 | 0.0941 | 0.0924 | -0.3686 | -0.0070 |
| BBr  | 1.0398 | -0.1993 | 0.0647 | 0.1649 | 0.0808 | -0.3637 | 0.0128 |
| BCl  | 1.3116 | -0.2115 | 0.0689 | 0.2020 | 0.0874 | -0.3733 | 0.0243 |
| BF   | 1.8349 | -0.1831 | 0.1045 | 0.3626 | 0.1279 | -0.4083 | 0.0653 |
| BrBr | -0.1665 | 6.8861 | 0.5722 | 0.0605 | 0.0726 | -0.3904 | -0.0461 |
| BrF  | -0.1891 | 5.9423 | 0.6097 | 0.0712 | 0.0878 | -0.4367 | -0.0140 |
| BrGa | 0.3768 | -0.1411 | 0.1115 | 0.1788 | 0.1299 | -0.3554 | 0.0038 |
| BrLi | 0.8092 | 0.2354 | 0.2115 | 0.3097 | 0.1712 | -0.3476 | -0.0210 |
| BrNa | 0.9397 | 0.2945 | 0.1942 | 0.2910 | 0.1500 | -0.3240 | -0.0277 |
| ClAl | 0.7867 | -0.2124 | 0.0892 | 0.2023 | 0.1133 | -0.3464 | 0.0080 |
| ClBr | -0.1635 | 6.5426 | 0.6053 | 0.0671 | 0.0803 | -0.4065 | -0.0390 |
| ClCl | -0.1604 | 6.1092 | 0.6519 | 0.0770 | 0.0917 | -0.4257 | -0.0277 |
| ClF  | -0.1925 | 5.5442 | 0.6707 | 0.0828 | 0.1025 | -0.4682 | 0.0023 |
| ClGa | 0.4615 | -0.1613 | 0.1146 | 0.1997 | 0.1366 | -0.3673 | 0.0079 |
| ClLi | 0.8829 | 0.2029 | 0.2254 | 0.3529 | 0.1874 | -0.3679 | -0.0197 |
| CO   | 0.3024 | -0.1044 | 0.2025 | 0.2945 | 0.2261 | -0.5172 | 0.1069 |
| CS   | 0.2792 | -0.1414 | 0.1060 | 0.1580 | 0.1235 | -0.4228 | 0.0220 |
| CSe  | 0.2197 | -0.1536 | 0.0908 | 0.1308 | 0.1073 | -0.4027 | 0.0070 |
| FF   | -0.3321 | 10.2457 | 1.2292 | 0.0730 | 0.1093 | -0.5776 | 0.0369 |
| FGa  | 0.8455 | -0.2020 | 0.1208 | 0.2795 | 0.1514 | -0.3878 | 0.0250 |
| FLi  | 1.3436 | 0.1100 | 0.2566 | 0.5419 | 0.2312 | -0.4222 | -0.0097 |
| GeO  | 0.0887 | -0.0800 | 0.1308 | 0.1547 | 0.1421 | -0.4084 | 0.0189 |
| GeS  | -0.0928 | -0.1104 | 0.1167 | 0.1190 | 0.1311 | -0.3729 | -0.0041 |
| GeSe | 0.0775 | -0.1056 | 0.1064 | 0.1282 | 0.1190 | -0.3579 | -0.0097 |
| HAl  | 2.2978 | -0.2490 | 0.0521 | 0.2289 | 0.0694 | -0.3023 | 0.0044 |
| HB   | 5.1733 | -0.2668 | 0.0309 | 0.2599 | 0.0421 | -0.3564 | 0.0132 |
| HBr  | 0.1579 | 1.2711 | 0.5306 | 0.2705 | 0.2336 | -0.4315 | 0.0863 |
| HCl  | 0.2010 | 1.1085 | 0.5713 | 0.3254 | 0.2709 | -0.4663 | 0.1042 |
| HF   | 0.5479 | 1.3538 | 0.8666 | 0.5699 | 0.3682 | -0.5795 | 0.1527 |
| HGa  | 2.2097 | -0.1944 | 0.0651 | 0.2595 | 0.0808 | -0.3114 | 0.0081 |
| HLi  | 17.9151 | 0.1462 | 0.1389 | 2.2920 | 0.1212 | -0.3038 | -0.0082 |
| HNa  | 12.4673 | 0.2260 | 0.1233 | 1.3544 | 0.1006 | -0.2768 | -0.0110 |
| LiLi | 52.8763 | 0.0807 | 0.0429 | 2.1364 | 0.0397 | -0.1954 | -0.0123 |
| LiNa | 32.8009 | 0.1943 | 0.0473 | 1.3378 | 0.0396 | -0.1938 | -0.0133 |
| NaCl | 1.0344 | 0.2598 | 0.2053 | 0.3315 | 0.1630 | -0.3399 | -0.0265 |
| NaF  | 1.7229 | 0.1848 | 0.2194 | 0.5042 | 0.1852 | -0.3692 | -0.0179 |
| NaNa | 29.3594 | 0.2867 | 0.0514 | 1.2127 | 0.0399 | -0.1924 | -0.0136 |
| NN   | 0.0818 | -0.0746 | 0.2585 | 0.3022 | 0.2793 | -0.5743 | 0.1237 |
| NP   | -0.1596 | -0.1095 | 0.1310 | 0.1236 | 0.1471 | -0.4456 | 0.0265 |
| OSi  | 0.1976 | -0.1574 | 0.1272 | 0.1808 | 0.1510 | -0.4167 | 0.0248 |
| PP   | 0.0257 | -0.0617 | 0.0965 | 0.1055 | 0.1028 | -0.3833 | -0.0018 |
| SeSi | -0.1990 | -0.0943 | 0.1156 | 0.1023 | 0.1277 | -0.3685 | -0.0078 |
| SSi  | 0.0441 | -0.1507 | 0.1046 | 0.1286 | 0.1232 | -0.3836 | -0.0013 |

# Table S4.A Data for diatomic molecules at B3LYP/cc-pVDZ.

## Table S4.B Data for diatomic molecules at B3LYP/ cc-pVTZ.

|      | $\delta^-$ | $\delta^+$ | $\Delta E_{ST}^{[v]}$ | $\Delta E_{DQ}^{[v],-}$ | $\Delta E_{DQ}^{[v],+}$ | $I$ | $A$ |
|---|---|---|---|---|---|---|---|
| AlBr | 0.6554 | -0.1955 | 0.0869 | 0.1788 | 0.1080 | -0.3375 | -0.0011 |
| AlF | 1.4394 | -0.2342 | 0.0952 | 0.3034 | 0.1244 | -0.3598 | 0.0187 |
| AsAs | -0.1713 | -0.2459 | 0.0786 | 0.0864 | 0.1042 | -0.3570 | -0.0239 |
| AsN | -0.1656 | -0.1683 | 0.1063 | 0.1066 | 0.1278 | -0.4197 | -0.0038 |
| AsP | -0.1555 | -0.2315 | 0.0869 | 0.0955 | 0.1131 | -0.3694 | -0.0187 |
| BBr | 1.0795 | -0.1790 | 0.0655 | 0.1658 | 0.0797 | -0.3620 | 0.0018 |
| BCl | 1.3683 | -0.1835 | 0.0706 | 0.2047 | 0.0865 | -0.3709 | 0.0138 |
| BF | 1.9045 | -0.1769 | 0.1059 | 0.3736 | 0.1286 | -0.4074 | 0.0513 |
| BrBr | -0.1299 | 3.9877 | 0.3856 | 0.0673 | 0.0773 | -0.3891 | -0.0543 |
| BrF | -0.1416 | 3.3663 | 0.4053 | 0.0797 | 0.0928 | -0.4366 | -0.0271 |
| BrGa | 0.4209 | -0.1349 | 0.1117 | 0.1835 | 0.1291 | -0.3556 | -0.0024 |
| BrLi | 0.7745 | 0.2119 | 0.2117 | 0.3100 | 0.1747 | -0.3493 | -0.0218 |
| BrNa | 0.9036 | 0.2839 | 0.1950 | 0.2891 | 0.1519 | -0.3252 | -0.0292 |
| ClAl | 0.8352 | -0.2123 | 0.0884 | 0.2060 | 0.1123 | -0.3446 | 0.0051 |
| ClBr | -0.1293 | 3.5045 | 0.3874 | 0.0749 | 0.0860 | -0.4042 | -0.0441 |
| ClCl | -0.1296 | 3.5734 | 0.4488 | 0.0854 | 0.0981 | -0.4226 | -0.0300 |
| ClF | -0.1504 | 3.5237 | 0.4948 | 0.0929 | 0.1094 | -0.4675 | -0.0066 |
| ClGa | 0.5120 | -0.1582 | 0.1143 | 0.2053 | 0.1358 | -0.3670 | 0.0033 |
| ClLi | 0.8481 | 0.1743 | 0.2258 | 0.3553 | 0.1923 | -0.3700 | -0.0188 |
| CO | 0.3169 | -0.1035 | 0.2038 | 0.2994 | 0.2273 | -0.5206 | 0.0860 |
| CS | 0.3038 | -0.1285 | 0.1068 | 0.1598 | 0.1226 | -0.4221 | 0.0085 |
| CSe | 0.2335 | -0.1431 | 0.0912 | 0.1313 | 0.1065 | -0.4028 | -0.0075 |
| FF | -0.2883 | 5.8623 | 0.8008 | 0.0831 | 0.1167 | -0.5810 | 0.0149 |
| FGa | 0.8728 | -0.1957 | 0.1217 | 0.2834 | 0.1513 | -0.3908 | 0.0170 |
| FLi | 1.2532 | 0.0938 | 0.2632 | 0.5421 | 0.2406 | -0.4302 | -0.0105 |
| GeO | 0.1456 | -0.1413 | 0.1227 | 0.1637 | 0.1429 | -0.4062 | 0.0051 |
| GeS | -0.0948 | -0.1022 | 0.1178 | 0.1188 | 0.1312 | -0.3737 | -0.0138 |
| GeSe | 0.0219 | 0.0119 | 0.1074 | 0.1085 | 0.1061 | -0.3585 | -0.0201 |
| HAl | 2.2775 | -0.2396 | 0.0533 | 0.2297 | 0.0701 | -0.3029 | 0.0012 |
| HB | 5.2546 | -0.2259 | 0.0327 | 0.2638 | 0.0422 | -0.3565 | 0.0013 |
| HBr | 0.1564 | 0.7168 | 0.3956 | 0.2665 | 0.2305 | -0.4335 | 0.0644 |
| HCl | 0.1873 | 0.5906 | 0.4267 | 0.3185 | 0.2682 | -0.4688 | 0.0831 |
| HF | 0.5370 | 0.8396 | 0.6847 | 0.5721 | 0.3722 | -0.5935 | 0.1133 |
| HGa | 2.1957 | -0.1886 | 0.0657 | 0.2589 | 0.0810 | -0.3126 | 0.0037 |
| HLi | 16.9311 | 0.1264 | 0.1380 | 2.1967 | 0.1225 | -0.3046 | -0.0090 |
| HNa | 11.8824 | 0.1937 | 0.1202 | 1.2967 | 0.1007 | -0.2769 | -0.0115 |
| LiLi | 51.8846 | 0.0642 | 0.0421 | 2.0930 | 0.0396 | -0.1950 | -0.0127 |
| LiNa | 31.9207 | 0.1848 | 0.0461 | 1.2820 | 0.0389 | -0.1935 | -0.0139 |
| NaCl | 1.0015 | 0.2444 | 0.2057 | 0.3308 | 0.1653 | -0.3411 | -0.0272 |
| NaF | 1.5884 | 0.1554 | 0.2239 | 0.5015 | 0.1938 | -0.3783 | -0.0202 |
| NaNa | 29.2454 | 0.2725 | 0.0499 | 1.1859 | 0.0392 | -0.1922 | -0.0141 |
| NN | 0.0887 | -0.0740 | 0.2655 | 0.3122 | 0.2867 | -0.5808 | 0.1050 |
| NP | -0.1384 | -0.0996 | 0.1310 | 0.1254 | 0.1455 | -0.4490 | 0.0090 |
| OSi | 0.1383 | -0.1563 | 0.1274 | 0.1719 | 0.1510 | -0.4197 | 0.0137 |
| PP | -0.1358 | -0.2148 | 0.0976 | 0.1074 | 0.1243 | -0.3838 | -0.0119 |
| SeSi | 0.0040 | -0.1567 | 0.0967 | 0.1152 | 0.1147 | -0.3706 | -0.0163 |
| SSi | 0.0638 | -0.1478 | 0.1043 | 0.1301 | 0.1223 | -0.3833 | -0.0088 |

# Table S4.C Data for diatomic molecules at B3LYP/aug-cc-pVDZ.

|      | $\delta^-$ | $\delta^+$ | $\Delta E_{ST}^{[v]}$ | $\Delta E_{DQ}^{[v],-}$ | $\Delta E_{DQ}^{[v],+}$ | $I$ | $A$ |
|------|------|------|------|------|------|------|------|
| AlBr | 0.5574 | -0.1741 | 0.0897 | 0.1692 | 0.1086 | -0.3407 | -0.0121 |
| AlF  | 1.4260 | -0.1910 | 0.1008 | 0.3024 | 0.1246 | -0.3634 | -0.0008 |
| AsAs | -0.1862 | -0.2542 | 0.0775 | 0.0846 | 0.1040 | -0.3575 | -0.0306 |
| AsN  | -0.2228 | -0.2435 | 0.1043 | 0.1072 | 0.1379 | -0.4209 | -0.0179 |
| AsP  | 0.0205 | -0.0671 | 0.0854 | 0.0934 | 0.0915 | -0.3701 | -0.0275 |
| BBr  | 0.9167 | -0.1614 | 0.0677 | 0.1548 | 0.0808 | -0.3651 | -0.0085 |
| BCl  | 1.3167 | -0.1673 | 0.0727 | 0.2022 | 0.0873 | -0.3745 | 0.0001 |
| BF   | 1.7917 | -0.1405 | 0.1073 | 0.3486 | 0.1249 | -0.4120 | 0.0171 |
| BrBr | -0.1607 | 1.8726 | 0.2053 | 0.0600 | 0.0715 | -0.3925 | -0.0681 |
| BrF  | -0.1559 | 1.4746 | 0.2162 | 0.0738 | 0.0874 | -0.4425 | -0.0507 |
| BrGa | 0.4000 | -0.1178 | 0.1150 | 0.1825 | 0.1304 | -0.3581 | -0.0128 |
| BrLi | 0.7738 | 0.1820 | 0.2053 | 0.3080 | 0.1737 | -0.3495 | -0.0277 |
| BrNa | 0.8966 | 0.2364 | 0.1890 | 0.2899 | 0.1529 | -0.3262 | -0.0330 |
| ClAl | 0.7202 | -0.1843 | 0.0920 | 0.1940 | 0.1128 | -0.3486 | -0.0089 |
| ClBr | -0.1552 | 1.6442 | 0.2110 | 0.0674 | 0.0798 | -0.4078 | -0.0608 |
| ClCl | -0.1529 | 1.3748 | 0.2166 | 0.0773 | 0.0912 | -0.4265 | -0.0506 |
| ClF  | -0.1654 | 1.2392 | 0.2280 | 0.0850 | 0.1018 | -0.4738 | -0.0378 |
| ClGa | 0.4833 | -0.1338 | 0.1187 | 0.2033 | 0.1370 | -0.3698 | -0.0099 |
| ClLi | 0.8470 | 0.1536 | 0.2201 | 0.3524 | 0.1908 | -0.3697 | -0.0257 |
| CO   | 0.3139 | -0.0986 | 0.2015 | 0.2938 | 0.2236 | -0.5225 | 0.0446 |
| CS   | 0.2710 | -0.1279 | 0.1086 | 0.1582 | 0.1245 | -0.4261 | -0.0054 |
| CSe  | 0.2103 | -0.1408 | 0.0930 | 0.1310 | 0.1082 | -0.4059 | -0.0179 |
| FF   | -0.3014 | 1.5309 | 0.2794 | 0.0771 | 0.1104 | -0.5862 | -0.0265 |
| FGa  | 0.8301 | -0.1580 | 0.1285 | 0.2794 | 0.1527 | -0.3947 | -0.0024 |
| FLi  | 1.2661 | 0.0831 | 0.2583 | 0.5405 | 0.2385 | -0.4282 | -0.0186 |
| GeO  | 0.1989 | -0.1368 | 0.1223 | 0.1699 | 0.1417 | -0.4097 | -0.0120 |
| GeS  | -0.0849 | -0.1052 | 0.1164 | 0.1191 | 0.1301 | -0.3740 | -0.0227 |
| GeSe | 0.2061 | 0.0049 | 0.1058 | 0.1270 | 0.1053 | -0.3589 | -0.0268 |
| HAl  | 2.2834 | -0.2150 | 0.0546 | 0.2283 | 0.0695 | -0.3028 | -0.0107 |
| HB   | 5.1402 | -0.1926 | 0.0343 | 0.2612 | 0.0425 | -0.3573 | -0.0136 |
| HBr  | 0.1783 | 0.1705 | 0.2657 | 0.2675 | 0.2270 | -0.4354 | 0.0149 |
| HCl  | 0.2226 | 0.1695 | 0.3063 | 0.3202 | 0.2619 | -0.4706 | 0.0184 |
| HF   | 0.5645 | 0.1730 | 0.4268 | 0.5693 | 0.3639 | -0.6002 | 0.0219 |
| HGa  | 2.2017 | -0.1720 | 0.0669 | 0.2588 | 0.0808 | -0.3120 | -0.0089 |
| HLi  | 17.5421 | 0.1127 | 0.1349 | 2.2474 | 0.1212 | -0.3042 | -0.0157 |
| HNa  | 12.3221 | 0.1703 | 0.1177 | 1.3397 | 0.1006 | -0.2763 | -0.0158 |
| LiLi | 52.8808 | 0.0625 | 0.0421 | 2.1344 | 0.0396 | -0.1953 | -0.0148 |
| LiNa | 32.6845 | 0.1617 | 0.0460 | 1.3340 | 0.0396 | -0.1939 | -0.0153 |
| NaCl | 0.9898 | 0.2060 | 0.2009 | 0.3314 | 0.1665 | -0.3421 | -0.0313 |
| NaF  | 1.5655 | 0.1473 | 0.2247 | 0.5024 | 0.1958 | -0.3798 | -0.0253 |
| NaNa | 29.2691 | 0.2356 | 0.0495 | 1.2115 | 0.0400 | -0.1925 | -0.0153 |
| NN   | -0.0571 | -0.0840 | 0.2563 | 0.2638 | 0.2797 | -0.5814 | 0.0681 |
| NP   | -0.1482 | -0.1263 | 0.1283 | 0.1251 | 0.1468 | -0.4501 | -0.0096 |
| OSi  | 0.1049 | -0.1503 | 0.1263 | 0.1643 | 0.1487 | -0.4227 | -0.0084 |
| PP   | 0.0293 | -0.0613 | 0.0955 | 0.1047 | 0.1017 | -0.3847 | -0.0234 |
| SeSi | -0.1905 | -0.0969 | 0.1143 | 0.1025 | 0.1266 | -0.3695 | -0.0250 |
| SSi  | 0.0431 | -0.1505 | 0.1039 | 0.1276 | 0.1223 | -0.3856 | -0.0203 |

# Table S4.D Data for diatomic molecules at B3LYP/aug-cc-pVTZ.

|  | $\delta^-$ | $\delta^+$ | $\Delta E_{ST}^{[v]}$ | $\Delta E_{DQ}^{[v],-}$ | $\Delta E_{DQ}^{[v],+}$ | $I$ | $A$ |
|---|---|---|---|---|---|---|---|
| AlBr | 0.6506 | -0.1795 | 0.0884 | 0.1778 | 0.1077 | -0.3383 | -0.0082 |
| AlF | 1.5076 | -0.1912 | 0.1005 | 0.3116 | 0.1243 | -0.3609 | 0.0030 |
| AsAs | 0.0380 | -0.0684 | 0.0772 | 0.0861 | 0.0829 | -0.3575 | -0.0305 |
| AsN | -0.0497 | -0.0798 | 0.1044 | 0.1079 | 0.1135 | -0.4210 | -0.0174 |
| AsP | 0.0463 | -0.0621 | 0.0858 | 0.0958 | 0.0915 | -0.3699 | -0.0268 |
| BBr | 1.1213 | -0.1582 | 0.0656 | 0.1653 | 0.0779 | -0.3619 | -0.0064 |
| BCl | 1.4050 | -0.1618 | 0.0714 | 0.2050 | 0.0852 | -0.3710 | 0.0032 |
| BF | 1.9272 | -0.1222 | 0.1125 | 0.3750 | 0.1281 | -0.4089 | 0.0221 |
| BrBr | -0.1165 | 1.3366 | 0.1976 | 0.0747 | 0.0846 | -0.3939 | -0.0564 |
| BrF | -0.1147 | 0.9893 | 0.1906 | 0.0848 | 0.0958 | -0.4404 | -0.0404 |
| BrGa | 0.3702 | -0.1295 | 0.1113 | 0.1753 | 0.1279 | -0.3565 | -0.0094 |
| BrLi | 0.7792 | 0.1864 | 0.2091 | 0.3136 | 0.1762 | -0.3493 | -0.0249 |
| BrNa | 0.9038 | 0.2557 | 0.1934 | 0.2932 | 0.1540 | -0.3253 | -0.0308 |
| ClAl | 0.8360 | -0.1893 | 0.0906 | 0.2053 | 0.1118 | -0.3456 | -0.0044 |
| ClBr | -0.1142 | 1.1193 | 0.1970 | 0.0823 | 0.0929 | -0.4085 | -0.0483 |
| ClCl | -0.1187 | 0.9987 | 0.2103 | 0.0928 | 0.1052 | -0.4268 | -0.0367 |
| ClF | -0.1286 | 0.9097 | 0.2160 | 0.0985 | 0.1131 | -0.4713 | -0.0245 |
| ClGa | 0.4620 | -0.1482 | 0.1146 | 0.1966 | 0.1345 | -0.3684 | -0.0057 |
| ClLi | 0.8434 | 0.1559 | 0.2247 | 0.3584 | 0.1944 | -0.3706 | -0.0227 |
| CO | 0.3147 | -0.0830 | 0.2091 | 0.2997 | 0.2280 | -0.5241 | 0.0472 |
| CS | 0.2966 | -0.1177 | 0.1087 | 0.1597 | 0.1232 | -0.4239 | -0.0046 |
| CSe | 0.2317 | -0.1346 | 0.0918 | 0.1306 | 0.1060 | -0.4041 | -0.0179 |
| FF | -0.2677 | 1.0787 | 0.2581 | 0.0909 | 0.1242 | -0.5862 | -0.0128 |
| FGa | 0.8406 | -0.1709 | 0.1247 | 0.2769 | 0.1504 | -0.3936 | 0.0025 |
| FLi | 1.2248 | 0.0899 | 0.2669 | 0.5448 | 0.2449 | -0.4327 | -0.0156 |
| GeO | 0.2059 | -0.1281 | 0.1258 | 0.1739 | 0.1442 | -0.4106 | -0.0099 |
| GeS | 0.0611 | -0.0858 | 0.1074 | 0.1247 | 0.1175 | -0.3815 | -0.0216 |
| GeSe | -0.0727 | -0.0879 | 0.1067 | 0.1085 | 0.1170 | -0.3586 | -0.0263 |
| HAl | 2.3030 | -0.2111 | 0.0550 | 0.2301 | 0.0697 | -0.3022 | -0.0078 |
| HB | 5.4371 | -0.1786 | 0.0337 | 0.2644 | 0.0411 | -0.3556 | -0.0113 |
| HBr | 0.1712 | 0.1736 | 0.2699 | 0.2693 | 0.2299 | -0.4339 | 0.0155 |
| HCl | 0.2046 | 0.1704 | 0.3103 | 0.3193 | 0.2651 | -0.4696 | 0.0186 |
| HF | 0.5619 | 0.1110 | 0.4058 | 0.5705 | 0.3653 | -0.5997 | 0.0201 |
| HGa | 2.2315 | -0.1766 | 0.0656 | 0.2575 | 0.0797 | -0.3122 | -0.0055 |
| HLi | 16.6486 | 0.1247 | 0.1389 | 2.1798 | 0.1235 | -0.3044 | -0.0132 |
| HNa | 11.7121 | 0.1896 | 0.1207 | 1.2893 | 0.1014 | -0.2760 | -0.0142 |
| LiLi | 47.2032 | 0.0351 | 0.0450 | 2.0974 | 0.0435 | -0.1937 | -0.0115 |
| LiNa | 29.9998 | 0.1644 | 0.0481 | 1.2794 | 0.0413 | -0.1916 | -0.0127 |
| NaCl | 0.9902 | 0.2224 | 0.2054 | 0.3345 | 0.1681 | -0.3418 | -0.0292 |
| NaF | 1.5299 | 0.1519 | 0.2297 | 0.5045 | 0.1994 | -0.3817 | -0.0230 |
| NaNa | 29.0501 | 0.3687 | 0.0549 | 1.2053 | 0.0401 | -0.1898 | -0.0129 |
| NN | 0.0681 | -0.0754 | 0.2713 | 0.3134 | 0.2934 | -0.5894 | 0.0729 |
| NP | -0.1248 | -0.1148 | 0.1298 | 0.1283 | 0.1466 | -0.4504 | -0.0088 |
| OSi | 0.1918 | -0.1472 | 0.1306 | 0.1826 | 0.1532 | -0.4252 | -0.0053 |
| PP | -0.1227 | -0.2144 | 0.0968 | 0.1081 | 0.1232 | -0.3842 | -0.0219 |
| SeSi | -0.0024 | -0.1427 | 0.0976 | 0.1136 | 0.1138 | -0.3730 | -0.0237 |
| SSi | 0.0380 | -0.1466 | 0.1056 | 0.1284 | 0.1237 | -0.3862 | -0.0182 |

# Table S5.A Data for diatomic molecules at M06-2X/cc-pVDZ.

|      | $\delta^-$ | $\delta^+$ | $\Delta E_{ST}^{[v]}$ | $\Delta E_{DQ}^{[v],-}$ | $\Delta E_{DQ}^{[v],+}$ | $I$ | $A$ |
|------|------|------|------|------|------|------|------|
| AlBr | 0.5829 | -0.2442 | 0.0827 | 0.1733 | 0.1095 | -0.3377 | 0.0074 |
| AlF | 1.3496 | -0.2683 | 0.0942 | 0.3026 | 0.1288 | -0.3587 | 0.0296 |
| AsAs | -0.0347 | -0.0963 | 0.0676 | 0.0722 | 0.0748 | -0.3524 | -0.0173 |
| AsN | -0.1821 | -0.2444 | 0.1044 | 0.1130 | 0.1381 | -0.4148 | 0.0124 |
| AsP | -0.1997 | -0.2501 | 0.0843 | 0.0899 | 0.1124 | -0.3684 | -0.0086 |
| BBr | 0.8692 | -0.2881 | 0.0595 | 0.1563 | 0.0836 | -0.3655 | 0.0166 |
| BCl | 1.3018 | -0.2932 | 0.0639 | 0.2080 | 0.0904 | -0.3728 | 0.0299 |
| BF | 1.7852 | -0.2408 | 0.1006 | 0.3691 | 0.1325 | -0.4068 | 0.0725 |
| BrBr | -0.1587 | 5.5823 | 0.5791 | 0.0740 | 0.0880 | -0.4037 | -0.0433 |
| BrF | -0.1570 | 5.0698 | 0.6088 | 0.0845 | 0.1003 | -0.4422 | -0.0103 |
| BrGa | 0.4890 | -0.1964 | 0.0964 | 0.1785 | 0.1199 | -0.3493 | 0.0015 |
| BrLi | 0.7492 | 0.2545 | 0.2208 | 0.3079 | 0.1760 | -0.3463 | -0.0153 |
| BrNa | 0.8766 | 0.2851 | 0.1991 | 0.2908 | 0.1549 | -0.3227 | -0.0212 |
| ClAl | 0.7976 | -0.2528 | 0.0849 | 0.2043 | 0.1137 | -0.3431 | 0.0131 |
| ClBr | -0.1526 | 5.3684 | 0.6067 | 0.0807 | 0.0953 | -0.4166 | -0.0336 |
| ClCl | -0.1658 | 5.1261 | 0.6564 | 0.0894 | 0.1071 | -0.4338 | -0.0188 |
| ClF | -0.1904 | 4.5173 | 0.6756 | 0.0991 | 0.1224 | -0.4764 | 0.0148 |
| ClGa | 0.5969 | -0.2110 | 0.0987 | 0.1998 | 0.1251 | -0.3575 | 0.0069 |
| ClLi | * | * | * | * | * | * | * |
| CO | 0.1742 | -0.1282 | 0.2021 | 0.2722 | 0.2319 | -0.5166 | 0.1145 |
| CS | 0.2817 | -0.1854 | 0.1014 | 0.1595 | 0.1245 | -0.4209 | 0.0269 |
| CSe | -0.0574 | -0.2075 | 0.0858 | 0.1021 | 0.1083 | -0.4057 | 0.0078 |
| FF | -0.3184 | 8.1674 | 1.2489 | 0.0929 | 0.1362 | -0.5899 | 0.0576 |
| FGa | 0.9542 | -0.2472 | 0.1060 | 0.2751 | 0.1408 | -0.3777 | 0.0242 |
| FLi | 1.3404 | 0.1082 | 0.2570 | 0.5427 | 0.2319 | -0.4140 | -0.0020 |
| GeO | 0.1098 | -0.1865 | 0.1104 | 0.1506 | 0.1357 | -0.3973 | 0.0179 |
| GeS | -0.1668 | -0.1059 | 0.1152 | 0.1074 | 0.1289 | -0.3723 | -0.0074 |
| GeSe | 0.0848 | -0.0970 | 0.1004 | 0.1206 | 0.1112 | -0.3543 | -0.0163 |
| HAl | 2.3868 | -0.3145 | 0.0481 | 0.2378 | 0.0702 | -0.2976 | 0.0099 |
| HB | 4.9202 | -0.4327 | 0.0263 | 0.2748 | 0.0464 | -0.3508 | 0.0217 |
| HBr | 0.1316 | 1.1936 | 0.5266 | 0.2717 | 0.2401 | -0.4302 | 0.0915 |
| HCl | 0.1814 | 1.0266 | 0.5720 | 0.3335 | 0.2823 | -0.4674 | 0.1119 |
| HF | 0.5291 | 1.3018 | 0.8738 | 0.5805 | 0.3796 | -0.5799 | 0.1598 |
| HGa | 2.4822 | -0.2666 | 0.0537 | 0.2551 | 0.0733 | -0.3024 | 0.0078 |
| HLi | 17.5806 | 0.1378 | 0.1405 | 2.2939 | 0.1235 | -0.2975 | -0.0009 |
| HNa | 12.3039 | 0.2309 | 0.1254 | 1.3554 | 0.1019 | -0.2690 | -0.0057 |
| LiLi | 43.6805 | 0.0505 | 0.0516 | 2.1954 | 0.0491 | -0.1994 | -0.0065 |
| LiNa | 26.6967 | 0.1684 | 0.0564 | 1.3376 | 0.0483 | -0.1983 | -0.0095 |
| NaCl | 0.9937 | 0.2478 | 0.2109 | 0.3370 | 0.1690 | -0.3392 | -0.0192 |
| NaF | 1.7213 | 0.1719 | 0.2192 | 0.5089 | 0.1870 | -0.3614 | -0.0099 |
| NaNa | 25.7592 | 0.2839 | 0.0606 | 1.2621 | 0.0472 | -0.1971 | -0.0115 |
| NN | 0.0306 | -0.0868 | 0.2713 | 0.3062 | 0.2971 | -0.5863 | 0.1307 |
| NP | -0.2230 | -0.2021 | 0.1372 | 0.1337 | 0.1720 | -0.4471 | 0.0312 |
| OSi | 0.1182 | -0.1933 | 0.1278 | 0.1771 | 0.1584 | -0.4210 | 0.0304 |
| PP | -0.1673 | -0.2140 | 0.1029 | 0.1090 | 0.1309 | -0.3862 | 0.0016 |
| SeSi | -0.0684 | -0.1922 | 0.0947 | 0.1093 | 0.1173 | -0.3744 | -0.0065 |
| SSi | 0.0155 | -0.1937 | 0.1020 | 0.1284 | 0.1265 | -0.3849 | 0.0030 |

*SCF convergence problem for ionic excited state(s)

# Table S5.B Data for diatomic molecules at M06-2X/ cc-pVTZ.

|      | $\delta^-$ | $\delta^+$ | $\Delta E_{ST}^{[v]}$ | $\Delta E_{DQ}^{[v],-}$ | $\Delta E_{DQ}^{[v],+}$ | $I$ | $A$ |
|------|---------|---------|---------|---------|---------|---------|---------|
| AlBr | 0.6468  | -0.2343 | 0.0825  | 0.1774  | 0.1077  | -0.3355 | 0.0035  |
| AlF  | 1.4397  | -0.2532 | 0.0956  | 0.3123  | 0.1280  | -0.3574 | 0.0255  |
| AsAs | -0.0346 | -0.0992 | 0.0684  | 0.0733  | 0.0759  | -0.3515 | -0.0259 |
| AsN  | -0.1637 | -0.2441 | 0.1020  | 0.1128  | 0.1349  | -0.4164 | -0.0034 |
| AsP  | -0.1905 | -0.2399 | 0.0844  | 0.0898  | 0.1110  | -0.3678 | -0.0179 |
| BBr  | 0.8949  | -0.2706 | 0.0594  | 0.1544  | 0.0815  | -0.3621 | 0.0058  |
| BCl  | 1.3715  | -0.2663 | 0.0647  | 0.2092  | 0.0882  | -0.3696 | 0.0185  |
| BF   | 1.8768  | -0.2189 | 0.1031  | 0.3796  | 0.1319  | -0.4052 | 0.0574  |
| BrBr | -0.1476 | 3.2306  | 0.3872  | 0.0780  | 0.0915  | -0.3977 | -0.0467 |
| BrF  | -0.1265 | 2.8913  | 0.3988  | 0.0895  | 0.1025  | -0.4383 | -0.0197 |
| BrGa | 0.4715  | -0.1771 | 0.0987  | 0.1765  | 0.1200  | -0.3474 | -0.0024 |
| BrLi | 0.7339  | 0.2267  | 0.2181  | 0.3083  | 0.1778  | -0.3454 | -0.0156 |
| BrNa | 0.8600  | 0.2625  | 0.1968  | 0.2899  | 0.1559  | -0.3215 | -0.0217 |
| ClAl | 0.8363  | -0.2443 | 0.0849  | 0.2063  | 0.1123  | -0.3413 | 0.0104  |
| ClBr | -0.1306 | 2.8819  | 0.3907  | 0.0875  | 0.1007  | -0.4110 | -0.0350 |
| ClCl | -0.1380 | 2.9465  | 0.4537  | 0.0991  | 0.1150  | -0.4297 | -0.0189 |
| ClF  | -0.1561 | 2.9081  | 0.5017  | 0.1083  | 0.1284  | -0.4748 | 0.0060  |
| ClGa | 0.5813  | -0.1954 | 0.1014  | 0.1992  | 0.1260  | -0.3576 | 0.0038  |
| ClLi | 0.8119  | 0.1816  | 0.2356  | 0.3613  | 0.1994  | -0.3699 | -0.0121 |
| CO   | 0.1896  | -0.1282 | 0.2014  | 0.2748  | 0.2310  | -0.5191 | 0.0927  |
| CS   | 0.3026  | -0.1751 | 0.1014  | 0.1601  | 0.1229  | -0.4197 | 0.0132  |
| CSe  | -0.0372 | -0.2025 | 0.0847  | 0.1022  | 0.1062  | -0.4028 | -0.0047 |
| FF   | -0.2820 | 4.6396  | 0.8200  | 0.1044  | 0.1454  | -0.5934 | 0.0378  |
| FGa  | 0.9844  | -0.2215 | 0.1106  | 0.2819  | 0.1421  | -0.3812 | 0.0178  |
| FLi  | 1.2424  | 0.0862  | 0.2650  | 0.5471  | 0.2440  | -0.4254 | -0.0023 |
| GeO  | 0.1251  | -0.1698 | 0.1134  | 0.1536  | 0.1366  | -0.4002 | 0.0055  |
| GeS  | -0.0741 | -0.2584 | 0.0957  | 0.1195  | 0.1290  | -0.3730 | -0.0151 |
| GeSe | -0.1309 | -0.0884 | 0.1007  | 0.0960  | 0.1104  | -0.3528 | -0.0231 |
| HAl  | 2.3711  | -0.2972 | 0.0496  | 0.2378  | 0.0705  | -0.2985 | 0.0066  |
| HB   | 5.0474  | -0.3985 | 0.0276  | 0.2772  | 0.0458  | -0.3509 | 0.0090  |
| HBr  | 0.1284  | 0.6421  | 0.3859  | 0.2652  | 0.2350  | -0.4293 | 0.0717  |
| HCl  | 0.1646  | 0.5195  | 0.4253  | 0.3259  | 0.2799  | -0.4698 | 0.0911  |
| HF   | 0.5202  | 0.8007  | 0.6891  | 0.5818  | 0.3827  | -0.5934 | 0.1190  |
| HGa  | 2.3828  | -0.2383 | 0.0572  | 0.2539  | 0.0750  | -0.3034 | 0.0048  |
| HLi  | 16.5266 | 0.1122  | 0.1396  | 2.1997  | 0.1255  | -0.2989 | -0.0021 |
| HNa  | 11.6644 | 0.1908  | 0.1223  | 1.3011  | 0.1027  | -0.2690 | -0.0066 |
| LiLi | 41.7623 | 0.0213  | 0.0514  | 2.1522  | 0.0503  | -0.1998 | -0.0071 |
| LiNa | 25.4656 | 0.1503  | 0.0558  | 1.2844  | 0.0485  | -0.1981 | -0.0105 |
| NaCl | -30.2245| -25.0006| 0.1731  | 0.2107  | -0.0072 | 0.0263  | -0.0072 |
| NaF  | 1.5710  | 0.1350  | 0.2248  | 0.5093  | 0.1981  | -0.3732 | -0.0110 |
| NaNa | 25.2559 | 0.2563  | 0.0593  | 1.2392  | 0.0472  | -0.1968 | -0.0121 |
| NN   | -0.0668 | -0.0866 | 0.2768  | 0.2828  | 0.3030  | -0.5919 | 0.1116  |
| NP   | -0.1279 | -0.1135 | 0.1354  | 0.1332  | 0.1527  | -0.4499 | 0.0134  |
| OSi  | 0.1362  | -0.1971 | 0.1257  | 0.1779  | 0.1566  | -0.4235 | 0.0194  |
| PP   | -0.1580 | -0.2030 | 0.1026  | 0.1083  | 0.1287  | -0.3858 | -0.0083 |
| SeSi | -0.0530 | -0.1898 | 0.0932  | 0.1089  | 0.1150  | -0.3718 | -0.0133 |
| SSi  | 0.0256  | -0.1919 | 0.1009  | 0.1281  | 0.1249  | -0.3846 | -0.0046 |

## Table S5.C Data for diatomic molecules at M06-2X/aug-cc-pVDZ.

|      | $\delta^-$ | $\delta^+$ | $\Delta E_{\text{ST}}^{[v]}$ | $\Delta E_{\text{DQ}}^{[v],-}$ | $\Delta E_{\text{DQ}}^{[v],+}$ | $I$ | $A$ |
|------|-----------|-----------|-----------|-----------|-----------|-----------|-----------|
| AlBr | 0.5625 | -0.2123 | 0.0860 | 0.1707 | 0.1092 | -0.3393 | -0.0064 |
| AlF  | 1.4304 | -0.2111 | 0.1006 | 0.3100 | 0.1275 | -0.3589 | 0.0069 |
| AsAs | -0.0269 | -0.1077 | 0.0658 | 0.0717 | 0.0737 | -0.3519 | -0.0315 |
| AsN  | -0.1650 | -0.2513 | 0.1006 | 0.1122 | 0.1344 | -0.4170 | -0.0140 |
| AsP  | -0.1915 | -0.2537 | 0.0824 | 0.0893 | 0.1104 | -0.3684 | -0.0250 |
| BBr  | 0.9721 | -0.2357 | 0.0639 | 0.1649 | 0.0836 | -0.3660 | -0.0020 |
| BCl  | 1.3009 | -0.2299 | 0.0695 | 0.2076 | 0.0902 | -0.3734 | 0.0080 |
| BF   | 1.7506 | -0.1614 | 0.1087 | 0.3564 | 0.1296 | -0.4088 | 0.0267 |
| BrBr | -0.1536 | 1.5436 | 0.2196 | 0.0731 | 0.0863 | -0.4034 | -0.0606 |
| BrF  | -0.1224 | 1.2695 | 0.2238 | 0.0865 | 0.0986 | -0.4438 | -0.0385 |
| BrGa | 0.4177 | -0.1694 | 0.1000 | 0.1706 | 0.1203 | -0.3509 | -0.0104 |
| BrLi | * | * | * | * | * | * | * |
| BrNa | 0.8506 | 0.2105 | 0.1903 | 0.2909 | 0.1572 | -0.3237 | -0.0251 |
| ClAl | 0.8363 | -0.2146 | 0.0889 | 0.2079 | 0.1132 | -0.3446 | -0.0023 |
| ClBr | -0.1443 | 1.3637 | 0.2227 | 0.0806 | 0.0942 | -0.4158 | -0.0510 |
| ClCl | -0.1586 | 1.1299 | 0.2260 | 0.0893 | 0.1061 | -0.4328 | -0.0374 |
| ClF  | -0.1668 | 0.9784 | 0.2396 | 0.1009 | 0.1211 | -0.4786 | -0.0180 |
| ClGa | 0.5717 | -0.1801 | 0.1029 | 0.1972 | 0.1255 | -0.3592 | -0.0061 |
| ClLi | 0.8190 | 0.1333 | 0.2229 | 0.3578 | 0.1967 | -0.3687 | -0.0190 |
| CO   | 0.1835 | -0.0985 | 0.2067 | 0.2714 | 0.2293 | -0.5203 | 0.0569 |
| CS   | 0.2695 | -0.1644 | 0.1050 | 0.1595 | 0.1257 | -0.4235 | 0.0030 |
| CSe  | -0.0679 | -0.1892 | 0.0889 | 0.1022 | 0.1097 | -0.4078 | -0.0125 |
| FF   | -0.2917 | 1.0598 | 0.2836 | 0.0975 | 0.1377 | -0.5930 | 0.0083 |
| FGa  | 0.9237 | -0.1966 | 0.1137 | 0.2722 | 0.1415 | -0.3820 | 0.0031 |
| FLi  | 2001.0285 | -1.0000 | 0.0000 | -0.2503 | -0.0001 | -0.2391 | -0.0001 |
| GeO  | 0.1819 | -0.1708 | 0.1122 | 0.1599 | 0.1353 | -0.4025 | -0.0079 |
| GeS  | -0.1561 | -0.1038 | 0.1142 | 0.1075 | 0.1274 | -0.3722 | -0.0223 |
| GeSe | 0.0848 | -0.1011 | 0.0987 | 0.1191 | 0.1098 | -0.3536 | -0.0290 |
| HAl  | 2.3802 | -0.2633 | 0.0517 | 0.2370 | 0.0701 | -0.2981 | -0.0047 |
| HB   | 4.9021 | -0.3133 | 0.0321 | 0.2756 | 0.0467 | -0.3516 | -0.0039 |
| HBr  | 0.1463 | 0.1347 | 0.2649 | 0.2676 | 0.2335 | -0.4325 | 0.0244 |
| HCl  | 0.1958 | 0.1795 | 0.3229 | 0.3274 | 0.2738 | -0.4707 | 0.0276 |
| HF   | 0.5555 | 0.1455 | 0.4264 | 0.5790 | 0.3722 | -0.5961 | 0.0312 |
| HGa  | 2.4723 | -0.2236 | 0.0568 | 0.2541 | 0.0732 | -0.3025 | -0.0047 |
| HLi  | 17.2053 | 0.1028 | 0.1363 | 2.2502 | 0.1236 | -0.2981 | -0.0096 |
| HNa  | 12.1469 | 0.1922 | 0.1216 | 1.3408 | 0.1020 | -0.2687 | -0.0109 |
| LiLi | 43.6256 | 0.0602 | 0.0521 | 2.1930 | 0.0491 | -0.1995 | -0.0101 |
| LiNa | 26.6126 | 0.1579 | 0.0559 | 1.3337 | 0.0483 | -0.1984 | -0.0120 |
| NaCl | 0.9615 | 0.1839 | 0.2040 | 0.3381 | 0.1723 | -0.3407 | -0.0228 |
| NaF  | 1.5817 | 0.1295 | 0.2220 | 0.5074 | 0.1965 | -0.3705 | -0.0158 |
| NaNa | 25.7183 | 0.2531 | 0.0591 | 1.2609 | 0.0472 | -0.1973 | -0.0135 |
| NN   | 0.0273 | -0.0862 | 0.2720 | 0.3058 | 0.2977 | -0.5920 | 0.0831 |
| NP   | -0.0836 | -0.0865 | 0.1345 | 0.1349 | 0.1472 | -0.4502 | -0.0012 |
| OSi  | 0.0373 | -0.1831 | 0.1267 | 0.1609 | 0.1551 | -0.4256 | -0.0002 |
| PP   | -0.1592 | -0.2140 | 0.1012 | 0.1082 | 0.1287 | -0.3867 | -0.0174 |
| SeSi | -0.0712 | -0.1930 | 0.0942 | 0.1084 | 0.1167 | -0.3752 | -0.0208 |
| SSi  | 0.0126 | -0.1915 | 0.1015 | 0.1272 | 0.1256 | -0.3863 | -0.0140 |

*SCF convergence problem for ionic excited state(s)

# Table S5.D Data for diatomic molecules at M06-2X/aug-cc-pVTZ.

| | $\delta^-$ | $\delta^+$ | $\Delta E_{\text{ST}}^{[v]}$ | $\Delta E_{\text{DQ}}^{[v],-}$ | $\Delta E_{\text{DQ}}^{[v],+}$ | $I$ | $A$ |
|---|---|---|---|---|---|---|---|
| AlBr | 0.5945 | -0.2091 | 0.0849 | 0.1711 | 0.1073 | -0.3354 | -0.0054 |
| AlF | 1.4993 | -0.2055 | 0.1012 | 0.3182 | 0.1273 | -0.3579 | 0.0065 |
| AsAs | -0.2305 | -0.2865 | 0.0680 | 0.0733 | 0.0953 | -0.3517 | -0.0314 |
| AsN | -0.1569 | -0.2429 | 0.1011 | 0.1126 | 0.1335 | -0.4172 | -0.0155 |
| AsP | -0.0241 | -0.0906 | 0.0836 | 0.0898 | 0.0920 | -0.3680 | -0.0253 |
| BBr | 1.0116 | -0.2398 | 0.0618 | 0.1636 | 0.0813 | -0.3619 | -0.0022 |
| BCl | 1.3821 | -0.2300 | 0.0676 | 0.2092 | 0.0878 | -0.3695 | 0.0075 |
| BF | 1.9041 | -0.1473 | 0.1112 | 0.3788 | 0.1304 | -0.4057 | 0.0257 |
| BrBr | -0.1463 | 1.1893 | 0.1994 | 0.0777 | 0.0911 | -0.3976 | -0.0520 |
| BrF | -0.1130 | 0.8638 | 0.1876 | 0.0893 | 0.1007 | -0.4394 | -0.0336 |
| BrGa | 0.4270 | -0.1642 | 0.1000 | 0.1707 | 0.1197 | -0.3474 | -0.0087 |
| BrLi | 0.7343 | 0.1673 | 0.2076 | 0.3084 | 0.1778 | -0.3454 | -0.0202 |
| BrNa | 0.8567 | 0.2162 | 0.1898 | 0.2898 | 0.1561 | -0.3219 | -0.0248 |
| ClAl | 0.8948 | -0.2120 | 0.0880 | 0.2117 | 0.1117 | -0.3414 | -0.0012 |
| ClBr | -0.1274 | 0.9743 | 0.1973 | 0.0872 | 0.0999 | -0.4109 | -0.0425 |
| ClCl | -0.1352 | 0.8441 | 0.2107 | 0.0988 | 0.1142 | -0.4297 | -0.0288 |
| ClF | -0.1456 | 0.7831 | 0.2255 | 0.1081 | 0.1265 | -0.4759 | -0.0127 |
| ClGa | 0.6234 | -0.1779 | 0.1032 | 0.2039 | 0.1256 | -0.3578 | -0.0045 |
| ClLi | * | * | * | * | * | * | * |
| CO | 0.1992 | -0.0852 | 0.2096 | 0.2748 | 0.2291 | -0.5194 | 0.0526 |
| CS | 0.3015 | -0.1586 | 0.1037 | 0.1603 | 0.1232 | -0.4203 | 0.0016 |
| CSe | -0.0379 | -0.1871 | 0.0865 | 0.1024 | 0.1064 | -0.4032 | -0.0134 |
| FF | -0.2754 | 0.7661 | 0.2555 | 0.1048 | 0.1447 | -0.5939 | 0.0109 |
| FGa | 0.9814 | -0.1941 | 0.1144 | 0.2812 | 0.1419 | -0.3824 | 0.0040 |
| FLi | 1.2281 | 0.0849 | 0.2664 | 0.5470 | 0.2455 | -0.4262 | -0.0100 |
| GeO | 0.2065 | -0.1568 | 0.1143 | 0.1636 | 0.1356 | -0.4015 | -0.0086 |
| GeS | 0.0155 | -0.1836 | 0.0957 | 0.1191 | 0.1173 | -0.3732 | -0.0217 |
| GeSe | -0.0265 | 0.0118 | 0.0998 | 0.0961 | 0.0987 | -0.3527 | -0.0280 |
| HAl | 2.3756 | -0.2587 | 0.0522 | 0.2375 | 0.0704 | -0.2985 | -0.0046 |
| HB | 5.0472 | -0.3189 | 0.0312 | 0.2773 | 0.0459 | -0.3511 | -0.0045 |
| HBr | 0.1432 | 0.0841 | 0.2507 | 0.2643 | 0.2312 | -0.4295 | 0.0212 |
| HCl | 0.1822 | 0.0748 | 0.2940 | 0.3234 | 0.2735 | -0.4703 | 0.0235 |
| HF | 0.5513 | 0.0807 | 0.4039 | 0.5798 | 0.3737 | -0.5972 | 0.0255 |
| HGa | 2.3805 | -0.2134 | 0.0590 | 0.2535 | 0.0750 | -0.3035 | -0.0040 |
| HLi | 16.4680 | 0.1382 | 0.1427 | 2.1902 | 0.1254 | -0.2990 | -0.0093 |
| HNa | 11.5968 | 0.1714 | 0.1201 | 1.2917 | 0.1025 | -0.2692 | -0.0112 |
| LiLi | 41.7166 | 0.0272 | 0.0517 | 2.1502 | 0.0503 | -0.1998 | -0.0092 |
| LiNa | 25.3393 | 0.1454 | 0.0557 | 1.2800 | 0.0486 | -0.1982 | -0.0119 |
| NaCl | 0.9513 | 0.1872 | 0.2063 | 0.3391 | 0.1738 | -0.3416 | -0.0226 |
| NaF | 1.5416 | 0.1300 | 0.2263 | 0.5091 | 0.2003 | -0.3747 | -0.0155 |
| NaNa | * | * | * | * | * | * | * |
| NN | -0.0657 | -0.0080 | 0.3003 | 0.2829 | 0.3027 | -0.5930 | 0.0724 |
| NP | -0.1244 | -0.1143 | 0.1348 | 0.1332 | 0.1522 | -0.4508 | -0.0034 |
| OSi | 0.1411 | -0.1803 | 0.1271 | 0.1769 | 0.1551 | -0.4248 | -0.0007 |
| PP | -0.1532 | -0.2049 | 0.1017 | 0.1083 | 0.1279 | -0.3863 | -0.0179 |
| SeSi | -0.0525 | -0.1851 | 0.0936 | 0.1088 | 0.1148 | -0.3719 | -0.0205 |
| SSi | 0.0264 | -0.1856 | 0.1014 | 0.1278 | 0.1245 | -0.3848 | -0.0140 |

*SCF convergence problem for ionic excited state(s)

## Table S6.A Data for diatomic molecules at Cam-B3LYP/ cc-pVDZ.

|      | $\delta^-$ | $\delta^+$ | $\Delta E_{ST}^{[v]}$ | $\Delta E_{DQ}^{[v],-}$ | $\Delta E_{DQ}^{[v],+}$ | $I$ | $A$ |
|------|-----------|-----------|-----------|-----------|-----------|-----------|-----------|
| AlBr | 0.6227 | -0.1984 | 0.0871 | 0.1763 | 0.1087 | -0.3389 | 0.0040 |
| AlF | 1.3151 | -0.2299 | 0.0974 | 0.2928 | 0.1265 | -0.3615 | 0.0247 |
| AsAs | 0.0240 | -0.0788 | 0.0763 | 0.0849 | 0.0829 | -0.3564 | -0.0118 |
| AsN | -0.2407 | -0.2465 | 0.1061 | 0.1069 | 0.1408 | -0.4168 | 0.0145 |
| AsP | -0.1701 | -0.2506 | 0.0850 | 0.0941 | 0.1134 | -0.3688 | -0.0073 |
| BBr | 1.0795 | -0.2031 | 0.0629 | 0.1641 | 0.0789 | -0.3635 | 0.0128 |
| BCl | 1.3408 | -0.2129 | 0.0680 | 0.2021 | 0.0864 | -0.3732 | 0.0246 |
| BF | 1.7352 | -0.1794 | 0.1056 | 0.3521 | 0.1287 | -0.4088 | 0.0662 |
| BrBr | -0.1505 | 6.1825 | 0.5765 | 0.0682 | 0.0803 | -0.3956 | -0.0426 |
| BrF | -0.1731 | 5.5892 | 0.6138 | 0.0770 | 0.0931 | -0.4387 | -0.0105 |
| BrGa | 0.4122 | -0.1448 | 0.1101 | 0.1818 | 0.1288 | -0.3561 | 0.0037 |
| BrLi | 0.8173 | 0.2678 | 0.2185 | 0.3133 | 0.1724 | -0.3473 | -0.0201 |
| BrNa | 0.9457 | 0.3206 | 0.2000 | 0.2947 | 0.1515 | -0.3236 | -0.0266 |
| ClAl | 0.7836 | -0.2120 | 0.0892 | 0.2019 | 0.1132 | -0.3470 | 0.0087 |
| ClBr | -0.1478 | 5.8862 | 0.6067 | 0.0751 | 0.0881 | -0.4110 | -0.0347 |
| ClCl | -0.1485 | 5.5174 | 0.6532 | 0.0854 | 0.1002 | -0.4298 | -0.0222 |
| ClF | -0.1795 | 5.2129 | 0.6779 | 0.0895 | 0.1091 | -0.4700 | 0.0075 |
| ClGa | 0.4540 | -0.1655 | 0.1131 | 0.1970 | 0.1355 | -0.3681 | 0.0081 |
| ClLi | 0.8868 | 0.2302 | 0.2322 | 0.3561 | 0.1887 | -0.3677 | -0.0187 |
| CO | 0.2904 | -0.1034 | 0.2050 | 0.2950 | 0.2286 | -0.5195 | 0.1076 |
| CS | 0.2751 | -0.1427 | 0.1059 | 0.1574 | 0.1235 | -0.4235 | 0.0218 |
| CSe | 0.2225 | -0.1566 | 0.0896 | 0.1299 | 0.1063 | -0.4031 | 0.0061 |
| FF | -0.3150 | 9.4697 | 1.2396 | 0.0811 | 0.1184 | -0.5801 | 0.0456 |
| FGa | 0.8491 | -0.2032 | 0.1199 | 0.2783 | 0.1505 | -0.3887 | 0.0253 |
| FLi | 1.3324 | 0.1113 | 0.2596 | 0.5448 | 0.2336 | -0.4233 | -0.0084 |
| GeO | 0.1059 | -0.1458 | 0.1233 | 0.1596 | 0.1444 | -0.4056 | 0.0192 |
| GeS | -0.0842 | -0.1055 | 0.1165 | 0.1193 | 0.1303 | -0.3730 | -0.0046 |
| GeSe | -0.0763 | -0.0991 | 0.1060 | 0.1087 | 0.1177 | -0.3580 | -0.0106 |
| HAl | 2.3292 | -0.2445 | 0.0521 | 0.2297 | 0.0690 | -0.3015 | 0.0052 |
| HB | 5.4016 | -0.2679 | 0.0298 | 0.2605 | 0.0407 | -0.3552 | 0.0132 |
| HBr | 0.1580 | 1.2514 | 0.5323 | 0.2738 | 0.2364 | -0.4315 | 0.0894 |
| HCl | 0.2005 | 1.0970 | 0.5734 | 0.3283 | 0.2734 | -0.4660 | 0.1069 |
| HF | 0.5467 | 1.3659 | 0.8725 | 0.5704 | 0.3688 | -0.5793 | 0.1536 |
| HGa | 2.2516 | -0.1940 | 0.0640 | 0.2584 | 0.0795 | -0.3110 | 0.0083 |
| HLi | 17.7030 | 0.1489 | 0.1403 | 2.2834 | 0.1221 | -0.3035 | -0.0069 |
| HNa | 12.3902 | 0.2333 | 0.1248 | 1.3554 | 0.1012 | -0.2757 | -0.0100 |
| LiLi | 48.3788 | 0.0502 | 0.0456 | 2.1450 | 0.0434 | -0.1941 | -0.0100 |
| LiNa | 30.9339 | 0.1716 | 0.0491 | 1.3383 | 0.0419 | -0.1920 | -0.0112 |
| NaCl | 1.0351 | 0.2813 | 0.2109 | 0.3350 | 0.1646 | -0.3396 | -0.0253 |
| NaF | 1.7026 | 0.1744 | 0.2205 | 0.5075 | 0.1878 | -0.3700 | -0.0164 |
| NaNa | 29.1752 | 0.2849 | 0.0525 | 1.2336 | 0.0409 | -0.1899 | -0.0118 |
| NN | -0.0735 | -0.0734 | 0.2649 | 0.2649 | 0.2859 | -0.5805 | 0.1244 |
| NP | -0.1520 | -0.1239 | 0.1306 | 0.1264 | 0.1490 | -0.4456 | 0.0269 |
| OSi | 0.1620 | -0.1547 | 0.1306 | 0.1796 | 0.1545 | -0.4200 | 0.0256 |
| PP | -0.1512 | -0.2310 | 0.0958 | 0.1058 | 0.1246 | -0.3832 | -0.0016 |
| SeSi | -0.0880 | 0.0229 | 0.1155 | 0.1030 | 0.1129 | -0.3683 | -0.0082 |
| SSi | 0.0170 | -0.1503 | 0.1061 | 0.1270 | 0.1249 | -0.3860 | -0.0011 |

## Table S6.B Data for diatomic molecules at Cam-B3LYP/ cc-pVTZ.

|      | $\delta^-$ | $\delta^+$ | $\Delta E_{ST}^{[v]}$ | $\Delta E_{DQ}^{[v],-}$ | $\Delta E_{DQ}^{[v],+}$ | $I$ | $A$ |
|------|---------|---------|---------|---------|---------|---------|---------|
| AlBr | 0.6514  | -0.1959 | 0.0868  | 0.1783  | 0.1079  | -0.3383 | -0.0004 |
| AlF  | 1.4992  | -0.2315 | 0.0958  | 0.3116  | 0.1247  | -0.3599 | 0.0198  |
| AsAs | 0.0355  | -0.0687 | 0.0773  | 0.0860  | 0.0830  | -0.3574 | -0.0245 |
| AsN  | -0.0568 | -0.0779 | 0.1054  | 0.1079  | 0.1143  | -0.4199 | -0.0039 |
| AsP  | -0.1479 | -0.2330 | 0.0862  | 0.0957  | 0.1123  | -0.3696 | -0.0189 |
| BBr  | 1.1190  | -0.1809 | 0.0638  | 0.1651  | 0.0779  | -0.3620 | 0.0019  |
| BCl  | 1.3967  | -0.1825 | 0.0700  | 0.2051  | 0.0856  | -0.3709 | 0.0142  |
| BF   | 1.9014  | -0.1714 | 0.1073  | 0.3758  | 0.1295  | -0.4082 | 0.0523  |
| BrBr | -0.1180 | 3.7536  | 0.4038  | 0.0749  | 0.0849  | -0.3941 | -0.0503 |
| BrF  | -0.1276 | 3.2505  | 0.4147  | 0.0851  | 0.0976  | -0.4382 | -0.0235 |
| BrGa | 0.3728  | -0.1389 | 0.1102  | 0.1757  | 0.1280  | -0.3564 | -0.0023 |
| BrLi | 0.7817  | 0.2419  | 0.2186  | 0.3136  | 0.1760  | -0.3492 | -0.0209 |
| BrNa | 0.9097  | 0.2762  | 0.1959  | 0.2931  | 0.1535  | -0.3249 | -0.0281 |
| ClAl | 0.7771  | -0.2117 | 0.0885  | 0.1995  | 0.1123  | -0.3452 | 0.0059  |
| ClBr | -0.1174 | 3.2367  | 0.3965  | 0.0826  | 0.0936  | -0.4084 | -0.0396 |
| ClCl | -0.1217 | 3.2902  | 0.4541  | 0.0930  | 0.1059  | -0.4266 | -0.0251 |
| ClF  | -0.1401 | 3.3875  | 0.5049  | 0.0989  | 0.1151  | -0.4691 | -0.0019 |
| ClGa | 0.4658  | -0.1624 | 0.1128  | 0.1974  | 0.1346  | -0.3679 | 0.0036  |
| ClLi | 0.8507  | 0.2000  | 0.2325  | 0.3586  | 0.1938  | -0.3701 | -0.0179 |
| CO   | 0.3052  | -0.1017 | 0.2065  | 0.3000  | 0.2299  | -0.5232 | 0.0866  |
| CS   | 0.2994  | -0.1285 | 0.1069  | 0.1594  | 0.1227  | -0.4230 | 0.0082  |
| CSe  | 0.2344  | -0.1448 | 0.0903  | 0.1304  | 0.1056  | -0.4034 | -0.0084 |
| FF   | -0.2760 | 5.5090  | 0.8147  | 0.0906  | 0.1252  | -0.5837 | 0.0224  |
| FGa  | 0.8488  | -0.1970 | 0.1206  | 0.2777  | 0.1502  | -0.3914 | 0.0174  |
| FLi  | 1.2404  | 0.1086  | 0.2698  | 0.5453  | 0.2434  | -0.4316 | -0.0094 |
| GeO  | 0.2039  | -0.1355 | 0.1255  | 0.1748  | 0.1452  | -0.4089 | 0.0054  |
| GeS  | -0.0321 | -0.0425 | 0.1180  | 0.1193  | 0.1233  | -0.3739 | -0.0141 |
| GeSe | 0.0336  | 0.0227  | 0.1073  | 0.1085  | 0.1049  | -0.3587 | -0.0209 |
| HAl  | 2.3032  | -0.2342 | 0.0534  | 0.2303  | 0.0697  | -0.3022 | 0.0020  |
| HB   | 5.4538  | -0.2213 | 0.0319  | 0.2643  | 0.0409  | -0.3555 | 0.0015  |
| HBr  | 0.1545  | 0.7015  | 0.3978  | 0.2699  | 0.2338  | -0.4337 | 0.0684  |
| HCl  | 0.1851  | 0.5819  | 0.4291  | 0.3214  | 0.2712  | -0.4688 | 0.0863  |
| HF   | 0.5341  | 0.8493  | 0.6900  | 0.5724  | 0.3731  | -0.5938 | 0.1144  |
| HGa  | 2.2349  | -0.1867 | 0.0648  | 0.2577  | 0.0797  | -0.3122 | 0.0039  |
| HLi  | 16.7150 | 0.1317  | 0.1398  | 2.1883  | 0.1235  | -0.3044 | -0.0077 |
| HNa  | 11.8046 | 0.2073  | 0.1224  | 1.2981  | 0.1014  | -0.2758 | -0.0104 |
| LiLi | 47.2512 | 0.0348  | 0.0450  | 2.0990  | 0.0435  | -0.1938 | -0.0103 |
| LiNa | 30.1401 | 0.1698  | 0.0482  | 1.2831  | 0.0412  | -0.1916 | -0.0118 |
| NaCl | 1.0016  | 0.2378  | 0.2068  | 0.3344  | 0.1671  | -0.3410 | -0.0260 |
| NaF  | 1.5648  | 0.1672  | 0.2298  | 0.5049  | 0.1968  | -0.3796 | -0.0189 |
| NaNa | 29.1268 | 0.2780  | 0.0512  | 1.2071  | 0.0401  | -0.1898 | -0.0122 |
| NN   | 0.0670  | -0.0721 | 0.2725  | 0.3134  | 0.2937  | -0.5875 | 0.1056  |
| NP   | -0.1293 | -0.1120 | 0.1308  | 0.1282  | 0.1473  | -0.4491 | 0.0091  |
| OSi  | 0.1876  | -0.1547 | 0.1306  | 0.1835  | 0.1545  | -0.4234 | 0.0142  |
| PP   | -0.1274 | -0.2148 | 0.0973  | 0.1081  | 0.1239  | -0.3837 | -0.0117 |
| SeSi | -0.1060 | -0.2318 | 0.0976  | 0.1136  | 0.1271  | -0.3729 | -0.0166 |
| SSi  | 0.0376  | -0.1475 | 0.1057  | 0.1287  | 0.1240  | -0.3859 | -0.0087 |

# Table S6.C Data for diatomic molecules at Cam-B3LYP/aug-cc-pVDZ.

|      | $\delta^-$ | $\delta^+$ | $\Delta E_{ST}^{[v]}$ | $\Delta E_{DQ}^{[v],-}$ | $\Delta E_{DQ}^{[v],+}$ | $I$ | $A$ |
|------|------------|------------|-----------------------|-------------------------|-------------------------|-----|-----|
| AlBr | 0.6074 | -0.1786 | 0.0891 | 0.1744 | 0.1085 | -0.3415 | -0.0098 |
| AlF  | 1.4357 | -0.1931 | 0.1008 | 0.3042 | 0.1249 | -0.3636 | 0.0026 |
| AsAs | -0.1816 | -0.2584 | 0.0764 | 0.0843 | 0.1030 | -0.3580 | -0.0304 |
| AsN  | -0.2097 | -0.2445 | 0.1037 | 0.1084 | 0.1372 | -0.4212 | -0.0168 |
| AsP  | -0.1642 | -0.2433 | 0.0846 | 0.0934 | 0.1118 | -0.3703 | -0.0269 |
| BBr  | 1.0804 | -0.1610 | 0.0662 | 0.1641 | 0.0789 | -0.3650 | -0.0070 |
| BCl  | 1.3458 | -0.1666 | 0.0719 | 0.2024 | 0.0863 | -0.3745 | 0.0026 |
| BF   | 1.7918 | -0.1354 | 0.1086 | 0.3507 | 0.1256 | -0.4126 | 0.0219 |
| BrBr | -0.1460 | 1.7701 | 0.2183 | 0.0673 | 0.0788 | -0.3975 | -0.0639 |
| BrF  | -0.1396 | 1.5775 | 0.2366 | 0.0790 | 0.0918 | -0.4444 | -0.0469 |
| BrGa | 0.3501 | -0.1249 | 0.1132 | 0.1746 | 0.1293 | -0.3589 | -0.0106 |
| BrLi | 0.7820 | 0.1942 | 0.2090 | 0.3119 | 0.1750 | -0.3493 | -0.0262 |
| BrNa | 0.9034 | 0.2466 | 0.1925 | 0.2940 | 0.1544 | -0.3258 | -0.0315 |
| ClAl | 0.7734 | -0.1891 | 0.0914 | 0.1998 | 0.1127 | -0.3493 | -0.0061 |
| ClBr | -0.1411 | 1.5543 | 0.2225 | 0.0748 | 0.0871 | -0.4121 | -0.0560 |
| ClCl | -0.1427 | 1.3088 | 0.2287 | 0.0849 | 0.0990 | -0.4304 | -0.0448 |
| ClF  | -0.1528 | 1.1771 | 0.2337 | 0.0909 | 0.1073 | -0.4756 | -0.0329 |
| ClGa | 0.4769 | -0.1423 | 0.1166 | 0.2008 | 0.1359 | -0.3708 | -0.0072 |
| ClLi | 0.8507 | 0.1641 | 0.2238 | 0.3559 | 0.1923 | -0.3696 | -0.0241 |
| CO   | 0.3009 | -0.0956 | 0.2046 | 0.2942 | 0.2262 | -0.5253 | 0.0497 |
| CS   | 0.2642 | -0.1260 | 0.1090 | 0.1577 | 0.1247 | -0.4271 | -0.0048 |
| CSe  | 0.2095 | -0.1405 | 0.0924 | 0.1301 | 0.1075 | -0.4067 | -0.0179 |
| FF   | -0.2869 | 1.4479 | 0.2914 | 0.0849 | 0.1190 | -0.5888 | -0.0186 |
| FGa  | 0.8040 | -0.1642 | 0.1267 | 0.2734 | 0.1515 | -0.3955 | 0.0010 |
| FLi  | 1.2541 | 0.0903 | 0.2628 | 0.5434 | 0.2411 | -0.4294 | -0.0167 |
| GeO  | 0.0959 | -0.1315 | 0.1250 | 0.1577 | 0.1439 | -0.4124 | -0.0103 |
| GeS  | 0.0268 | 0.0047 | 0.1169 | 0.1195 | 0.1164 | -0.3742 | -0.0221 |
| GeSe | 0.0475 | 0.0189 | 0.1060 | 0.1089 | 0.1040 | -0.3591 | -0.0266 |
| HAl  | 2.3150 | -0.2109 | 0.0545 | 0.2291 | 0.0691 | -0.3021 | -0.0080 |
| HB   | 5.3520 | -0.1794 | 0.0338 | 0.2617 | 0.0412 | -0.3562 | -0.0113 |
| HBr  | 0.1744 | 0.1684 | 0.2696 | 0.2710 | 0.2307 | -0.4354 | 0.0191 |
| HCl  | 0.2185 | 0.2017 | 0.3188 | 0.3233 | 0.2653 | -0.4705 | 0.0217 |
| HF   | 0.5594 | 0.1805 | 0.4313 | 0.5698 | 0.3654 | -0.6006 | 0.0239 |
| HGa  | 2.2409 | -0.1697 | 0.0660 | 0.2577 | 0.0795 | -0.3116 | -0.0061 |
| HLi  | 17.3052 | 0.1278 | 0.1379 | 2.2389 | 0.1223 | -0.3039 | -0.0138 |
| HNa  | 12.2282 | 0.1869 | 0.1203 | 1.3406 | 0.1013 | -0.2752 | -0.0142 |
| LiLi | 48.3895 | 0.0429 | 0.0452 | 2.1422 | 0.0434 | -0.1940 | -0.0119 |
| LiNa | 30.8592 | 0.1532 | 0.0483 | 1.3344 | 0.0419 | -0.1920 | -0.0127 |
| NaCl | 0.9908 | 0.2147 | 0.2044 | 0.3351 | 0.1683 | -0.3419 | -0.0297 |
| NaF  | 1.5440 | 0.1485 | 0.2283 | 0.5056 | 0.1988 | -0.3809 | -0.0235 |
| NaNa | 29.0987 | 0.2467 | 0.0510 | 1.2324 | 0.0409 | -0.1900 | -0.0131 |
| NN   | 0.0568 | -0.0801 | 0.2637 | 0.3030 | 0.2867 | -0.5882 | 0.0710 |
| NP   | -0.1415 | -0.1396 | 0.1282 | 0.1279 | 0.1490 | -0.4503 | -0.0081 |
| OSi  | 0.1570 | -0.1492 | 0.1293 | 0.1759 | 0.1520 | -0.4263 | -0.0062 |
| PP   | 0.0400 | -0.0593 | 0.0950 | 0.1051 | 0.1010 | -0.3846 | -0.0225 |
| SeSi | -0.1792 | -0.0877 | 0.1147 | 0.1032 | 0.1258 | -0.3693 | -0.0245 |
| SSi  | 0.0152 | -0.1501 | 0.1054 | 0.1259 | 0.1240 | -0.3883 | -0.0192 |

## Table S6.D Data for diatomic molecules at Cam-B3LYP/aug-cc-pVTZ.

|      | $\delta^-$ | $\delta^+$ | $\Delta E_{ST}^{[v]}$ | $\Delta E_{DQ}^{[v],-}$ | $\Delta E_{DQ}^{[v],+}$ | $I$ | $A$ |
|------|--------|--------|--------|--------|--------|--------|--------|
| AlBr | 0.6506 | -0.1795 | 0.0884 | 0.1778 | 0.1077 | -0.3383 | -0.0082 |
| AlF  | 1.5076 | -0.1912 | 0.1005 | 0.3116 | 0.1243 | -0.3609 | 0.0030 |
| AsAs | 0.0380 | -0.0684 | 0.0772 | 0.0861 | 0.0829 | -0.3575 | -0.0305 |
| AsN  | -0.0497 | -0.0798 | 0.1044 | 0.1079 | 0.1135 | -0.4210 | -0.0174 |
| AsP  | 0.0463 | -0.0621 | 0.0858 | 0.0958 | 0.0915 | -0.3699 | -0.0268 |
| BBr  | 1.1213 | -0.1582 | 0.0656 | 0.1653 | 0.0779 | -0.3619 | -0.0064 |
| BCl  | 1.4050 | -0.1618 | 0.0714 | 0.2050 | 0.0852 | -0.3710 | 0.0032 |
| BF   | 1.9272 | -0.1222 | 0.1125 | 0.3750 | 0.1281 | -0.4089 | 0.0221 |
| BrBr | -0.1165 | 1.3366 | 0.1976 | 0.0747 | 0.0846 | -0.3939 | -0.0564 |
| BrF  | -0.1147 | 0.9893 | 0.1906 | 0.0848 | 0.0958 | -0.4404 | -0.0404 |
| BrGa | 0.3702 | -0.1295 | 0.1113 | 0.1753 | 0.1279 | -0.3565 | -0.0094 |
| BrLi | 0.7792 | 0.1864 | 0.2091 | 0.3136 | 0.1762 | -0.3493 | -0.0249 |
| BrNa | 0.9038 | 0.2557 | 0.1934 | 0.2932 | 0.1540 | -0.3253 | -0.0308 |
| ClAl | 0.8360 | -0.1893 | 0.0906 | 0.2053 | 0.1118 | -0.3456 | -0.0044 |
| ClBr | -0.1142 | 1.1193 | 0.1970 | 0.0823 | 0.0929 | -0.4085 | -0.0483 |
| ClCl | -0.1187 | 0.9987 | 0.2103 | 0.0928 | 0.1052 | -0.4268 | -0.0367 |
| ClF  | -0.1286 | 0.9097 | 0.2160 | 0.0985 | 0.1131 | -0.4713 | -0.0245 |
| ClGa | 0.4620 | -0.1482 | 0.1146 | 0.1966 | 0.1345 | -0.3684 | -0.0057 |
| ClLi | 0.8434 | 0.1559 | 0.2247 | 0.3584 | 0.1944 | -0.3706 | -0.0227 |
| CO   | 0.3147 | -0.0830 | 0.2091 | 0.2997 | 0.2280 | -0.5241 | 0.0472 |
| CS   | 0.2966 | -0.1177 | 0.1087 | 0.1597 | 0.1232 | -0.4239 | -0.0046 |
| CSe  | 0.2317 | -0.1346 | 0.0918 | 0.1306 | 0.1060 | -0.4041 | -0.0179 |
| FF   | -0.2677 | 1.0787 | 0.2581 | 0.0909 | 0.1242 | -0.5862 | -0.0128 |
| FGa  | 0.8406 | -0.1709 | 0.1247 | 0.2769 | 0.1504 | -0.3936 | 0.0025 |
| FLi  | 1.2248 | 0.0899 | 0.2669 | 0.5448 | 0.2449 | -0.4327 | -0.0156 |
| GeO  | 0.2059 | -0.1281 | 0.1258 | 0.1739 | 0.1442 | -0.4106 | -0.0099 |
| GeS  | 0.0611 | -0.0858 | 0.1074 | 0.1247 | 0.1175 | -0.3815 | -0.0216 |
| GeSe | -0.0727 | -0.0879 | 0.1067 | 0.1085 | 0.1170 | -0.3586 | -0.0263 |
| HAl  | 2.3030 | -0.2111 | 0.0550 | 0.2301 | 0.0697 | -0.3022 | -0.0078 |
| HB   | 5.4371 | -0.1786 | 0.0337 | 0.2644 | 0.0411 | -0.3556 | -0.0113 |
| HBr  | 0.1712 | 0.1736 | 0.2699 | 0.2693 | 0.2299 | -0.4339 | 0.0155 |
| HCl  | 0.2046 | 0.1704 | 0.3103 | 0.3193 | 0.2651 | -0.4696 | 0.0186 |
| HF   | 0.5619 | 0.1110 | 0.4058 | 0.5705 | 0.3653 | -0.5997 | 0.0201 |
| HGa  | 2.2315 | -0.1766 | 0.0656 | 0.2575 | 0.0797 | -0.3122 | -0.0055 |
| HLi  | 16.6486 | 0.1247 | 0.1389 | 2.1798 | 0.1235 | -0.3044 | -0.0132 |
| HNa  | 11.7121 | 0.1896 | 0.1207 | 1.2893 | 0.1014 | -0.2760 | -0.0142 |
| LiLi | 47.2032 | 0.0351 | 0.0450 | 2.0974 | 0.0435 | -0.1937 | -0.0115 |
| LiNa | 29.9998 | 0.1644 | 0.0481 | 1.2794 | 0.0413 | -0.1916 | -0.0127 |
| NaCl | 0.9902 | 0.2224 | 0.2054 | 0.3345 | 0.1681 | -0.3418 | -0.0292 |
| NaF  | 1.5299 | 0.1519 | 0.2297 | 0.5045 | 0.1994 | -0.3817 | -0.0230 |
| NaNa | 29.0501 | 0.3687 | 0.0549 | 1.2053 | 0.0401 | -0.1898 | -0.0129 |
| NN   | 0.0681 | -0.0754 | 0.2713 | 0.3134 | 0.2934 | -0.5894 | 0.0729 |
| NP   | -0.1248 | -0.1148 | 0.1298 | 0.1283 | 0.1466 | -0.4504 | -0.0088 |
| OSi  | 0.1918 | -0.1472 | 0.1306 | 0.1826 | 0.1532 | -0.4252 | -0.0053 |
| PP   | -0.1227 | -0.2144 | 0.0968 | 0.1081 | 0.1232 | -0.3842 | -0.0219 |
| SeSi | -0.0024 | -0.1427 | 0.0976 | 0.1136 | 0.1138 | -0.3730 | -0.0237 |
| SSi  | 0.0380 | -0.1466 | 0.1056 | 0.1284 | 0.1237 | -0.3862 | -0.0182 |

## Table S7.A Data for diatomic molecules at LC-wPBE/cc-pVDZ.

|      | $\delta^-$ | $\delta^+$ | $\Delta E_{ST}^{[v]}$ | $\Delta E_{DQ}^{[v],-}$ | $\Delta E_{DQ}^{[v],+}$ | $I$ | $A$ |
|------|---------|---------|---------|---------|---------|---------|---------|
| AlBr | 0.7478  | -0.2547 | 0.0712  | 0.1670  | 0.0955  | -0.3363 | -0.0045 |
| AlF  | 1.5393  | -0.2710 | 0.0821  | 0.2860  | 0.1126  | -0.3553 | 0.0176  |
| AsAs | -0.2248 | -0.3008 | 0.0707  | 0.0784  | 0.1011  | -0.3591 | -0.0189 |
| AsN  | -0.1807 | -0.2143 | 0.1016  | 0.1059  | 0.1293  | -0.4184 | 0.0104  |
| AsP  | 0.0150  | -0.0810 | 0.0797  | 0.0880  | 0.0867  | -0.3724 | -0.0150 |
| BBr  | 1.2905  | -0.2999 | 0.0480  | 0.1571  | 0.0686  | -0.3618 | 0.0057  |
| BCl  | 1.5803  | -0.2962 | 0.0536  | 0.1967  | 0.0762  | -0.3711 | 0.0179  |
| BF   | 2.0252  | -0.2198 | 0.0928  | 0.3598  | 0.1189  | -0.4053 | 0.0606  |
| BrBr | -0.1310 | 5.2950  | 0.5685  | 0.0785  | 0.0903  | -0.4025 | -0.0402 |
| BrF  | -0.1610 | 5.3707  | 0.6130  | 0.0807  | 0.0962  | -0.4396 | -0.0104 |
| BrGa | 0.4599  | -0.2022 | 0.0932  | 0.1705  | 0.1168  | -0.3538 | -0.0021 |
| BrLi | 0.8351  | 0.2246  | 0.2093  | 0.3136  | 0.1709  | -0.3461 | -0.0233 |
| BrNa | 0.9543  | 0.2573  | 0.1929  | 0.2998  | 0.1534  | -0.3213 | -0.0280 |
| ClAl | 0.9455  | -0.2644 | 0.0734  | 0.1941  | 0.0997  | -0.3435 | 0.0004  |
| ClBr | -0.1244 | 5.0555  | 0.5959  | 0.0862  | 0.0984  | -0.4169 | -0.0311 |
| ClCl | -0.1285 | 4.7245  | 0.6429  | 0.0979  | 0.1123  | -0.4359 | -0.0166 |
| ClF  | -0.1634 | 4.9570  | 0.6835  | 0.0960  | 0.1147  | -0.4712 | 0.0102  |
| ClGa | 0.5228  | -0.2196 | 0.0957  | 0.1867  | 0.1226  | -0.3648 | 0.0027  |
| ClLi | 0.9065  | 0.1925  | 0.2241  | 0.3583  | 0.1879  | -0.3668 | -0.0217 |
| CO   | 0.3173  | -0.1152 | 0.1972  | 0.2936  | 0.2229  | -0.5165 | 0.1038  |
| CS   | 0.3206  | -0.1669 | 0.0978  | 0.1550  | 0.1174  | -0.4225 | 0.0151  |
| CSe  | 0.2797  | -0.1849 | 0.0817  | 0.1282  | 0.1002  | -0.4010 | -0.0003 |
| FF   | -0.2798 | 8.8691  | 1.2453  | 0.0909  | 0.1262  | -0.5766 | 0.0531  |
| FGa  | 0.9907  | -0.2453 | 0.1029  | 0.2713  | 0.1363  | -0.3813 | 0.0211  |
| FLi  | 1.3524  | 0.0996  | 0.2569  | 0.5496  | 0.2336  | -0.4216 | -0.0111 |
| GeO  | 0.1471  | -0.1672 | 0.1194  | 0.1644  | 0.1434  | -0.4087 | 0.0154  |
| GeS  | -0.0277 | -0.1635 | 0.1007  | 0.1171  | 0.1204  | -0.3851 | -0.0115 |
| GeSe | -0.1123 | -0.1162 | 0.1006  | 0.1011  | 0.1139  | -0.3601 | -0.0178 |
| HAl  | 2.6595  | -0.3371 | 0.0386  | 0.2133  | 0.0583  | -0.2991 | -0.0032 |
| HB   | 6.9097  | -0.4775 | 0.0163  | 0.2471  | 0.0312  | -0.3524 | 0.0061  |
| HBr  | 0.1503  | 1.2072  | 0.5264  | 0.2744  | 0.2385  | -0.4321 | 0.0891  |
| HCl  | 0.1930  | 1.0578  | 0.5698  | 0.3303  | 0.2769  | -0.4678 | 0.1071  |
| HF   | 0.5350  | 1.3480  | 0.8765  | 0.5730  | 0.3733  | -0.5817 | 0.1559  |
| HGa  | 2.4384  | -0.2805 | 0.0504  | 0.2407  | 0.0700  | -0.3077 | 0.0034  |
| HLi  | 18.9390 | 0.1333  | 0.1289  | 2.2671  | 0.1137  | -0.2961 | -0.0099 |
| HNa  | 13.5400 | 0.2107  | 0.1130  | 1.3573  | 0.0933  | -0.2646 | -0.0130 |
| LiLi | 54.3502 | 0.0538  | 0.0407  | 2.1388  | 0.0386  | -0.1878 | -0.0170 |
| LiNa | 36.8382 | 0.1627  | 0.0410  | 1.3341  | 0.0353  | -0.1822 | -0.0173 |
| NaCl | 1.0423  | 0.2425  | 0.2081  | 0.3421  | 0.1675  | -0.3379 | -0.0266 |
| NaF  | 1.6890  | 0.1509  | 0.2212  | 0.5169  | 0.1922  | -0.3682 | -0.0174 |
| NaNa | 37.2322 | 0.2912  | 0.0418  | 1.2376  | 0.0324  | -0.1766 | -0.0177 |
| NN   | 0.0638  | -0.0742 | 0.2679  | 0.3079  | 0.2894  | -0.5833 | 0.1215  |
| NP   | -0.1571 | -0.1516 | 0.1258  | 0.1250  | 0.1483  | -0.4484 | 0.0210  |
| OSi  | 0.1242  | -0.1813 | 0.1231  | 0.1691  | 0.1504  | -0.4236 | 0.0195  |
| PP   | -0.1830 | -0.2568 | 0.0907  | 0.0997  | 0.1220  | -0.3877 | -0.0102 |
| SeSi | -0.0469 | -0.1850 | 0.0891  | 0.1042  | 0.1093  | -0.3768 | -0.0168 |
| SSi  | -0.0056 | -0.1876 | 0.0968  | 0.1185  | 0.1191  | -0.3900 | -0.0098 |

## Table S7.B Data for diatomic molecules at LC-wPBE/ cc-pVTZ.

|      | $\delta^-$ | $\delta^+$ | $\Delta E_{ST}^{[v]}$ | $\Delta E_{DQ}^{[v],-}$ | $\Delta E_{DQ}^{[v],+}$ | $I$ | $A$ |
|---|---|---|---|---|---|---|---|
| AlBr | 0.7931 | -0.2529 | 0.0704 | 0.1689 | 0.0942 | -0.3341 | -0.0074 |
| AlF | 1.7156 | -0.2739 | 0.0800 | 0.2991 | 0.1101 | -0.3528 | 0.0127 |
| AsAs | -0.1993 | -0.2816 | 0.0714 | 0.0796 | 0.0994 | -0.3583 | -0.0290 |
| AsN | -0.1644 | -0.2089 | 0.1005 | 0.1062 | 0.1270 | -0.4201 | -0.0060 |
| AsP | -0.1780 | -0.2619 | 0.0806 | 0.0897 | 0.1091 | -0.3714 | -0.0240 |
| BBr | 1.3430 | -0.2733 | 0.0488 | 0.1572 | 0.0671 | -0.3590 | -0.0029 |
| BCl | 1.6547 | -0.2610 | 0.0553 | 0.1986 | 0.0748 | -0.3678 | 0.0095 |
| BF | 2.0054 | -0.2115 | 0.0935 | 0.3564 | 0.1186 | -0.4036 | 0.0486 |
| BrBr | -0.1049 | 3.1503 | 0.3932 | 0.0848 | 0.0947 | -0.3980 | -0.0435 |
| BrF | -0.1176 | 3.1753 | 0.4168 | 0.0881 | 0.0998 | -0.4372 | -0.0200 |
| BrGa | 0.4752 | -0.1987 | 0.0924 | 0.1700 | 0.1153 | -0.3526 | -0.0062 |
| BrLi | 0.8140 | 0.2284 | 0.2138 | 0.3158 | 0.1741 | -0.3465 | -0.0234 |
| BrNa | 0.9334 | 0.2601 | 0.1952 | 0.2994 | 0.1549 | -0.3213 | -0.0289 |
| ClAl | 1.0216 | -0.2643 | 0.0722 | 0.1985 | 0.0982 | -0.3402 | -0.0013 |
| ClBr | -0.1001 | 2.8589 | 0.3992 | 0.0931 | 0.1034 | -0.4116 | -0.0321 |
| ClCl | -0.1071 | 2.8535 | 0.4525 | 0.1048 | 0.1174 | -0.4300 | -0.0161 |
| ClF | -0.1271 | 3.2810 | 0.5116 | 0.1043 | 0.1195 | -0.4685 | 0.0030 |
| ClGa | 0.6005 | -0.2179 | 0.0946 | 0.1937 | 0.1210 | -0.3631 | -0.0002 |
| ClLi | 0.8808 | 0.1663 | 0.2249 | 0.3628 | 0.1929 | -0.3682 | -0.0204 |
| CO | 0.3344 | -0.1149 | 0.1972 | 0.2974 | 0.2228 | -0.5190 | 0.0845 |
| CS | 0.1007 | -0.1516 | 0.0985 | 0.1278 | 0.1161 | -0.4206 | 0.0040 |
| CSe | 0.0190 | -0.1729 | 0.0820 | 0.1010 | 0.0991 | -0.3998 | -0.0121 |
| FF | -0.2438 | 5.1539 | 0.8231 | 0.1011 | 0.1338 | -0.5789 | 0.0325 |
| FGa | 1.0537 | -0.2410 | 0.1024 | 0.2770 | 0.1349 | -0.3828 | 0.0138 |
| FLi | 1.2643 | 0.0858 | 0.2641 | 0.5508 | 0.2432 | -0.4295 | -0.0119 |
| GeO | 0.1629 | -0.1620 | 0.1200 | 0.1666 | 0.1432 | -0.4110 | 0.0027 |
| GeS | -0.0091 | -0.1575 | 0.1001 | 0.1178 | 0.1189 | -0.3831 | -0.0189 |
| GeSe | 0.0105 | 0.0181 | 0.1017 | 0.1009 | 0.0999 | -0.3591 | -0.0257 |
| HAl | 2.6493 | -0.3201 | 0.0398 | 0.2135 | 0.0585 | -0.2994 | -0.0059 |
| HB | 7.0607 | -0.4043 | 0.0184 | 0.2495 | 0.0310 | -0.3521 | -0.0036 |
| HBr | 0.1476 | 0.6666 | 0.3916 | 0.2697 | 0.2350 | -0.4326 | 0.0708 |
| HCl | 0.1778 | 0.5537 | 0.4254 | 0.3225 | 0.2738 | -0.4691 | 0.0884 |
| HF | 0.5258 | 0.8309 | 0.6907 | 0.5756 | 0.3772 | -0.5950 | 0.1170 |
| HGa | 2.4340 | -0.2680 | 0.0511 | 0.2397 | 0.0698 | -0.3086 | -0.0004 |
| HLi | 17.7647 | 0.1191 | 0.1295 | 2.1719 | 0.1157 | -0.2972 | -0.0106 |
| HNa | 12.8779 | 0.1930 | 0.1121 | 1.3042 | 0.0940 | -0.2648 | -0.0134 |
| LiLi | 52.0767 | 0.0320 | 0.0407 | 2.0936 | 0.0394 | -0.1880 | -0.0171 |
| LiNa | 35.4055 | 0.1584 | 0.0408 | 1.2829 | 0.0352 | -0.1821 | -0.0177 |
| NaCl | 1.0205 | 0.2205 | 0.2070 | 0.3427 | 0.1696 | -0.3382 | -0.0267 |
| NaF | 1.5580 | 0.1301 | 0.2273 | 0.5144 | 0.2011 | -0.3775 | -0.0196 |
| NaNa | 36.6513 | 0.2835 | 0.0414 | 1.2157 | 0.0323 | -0.1768 | -0.0180 |
| NN | 0.0703 | -0.0728 | 0.2746 | 0.3170 | 0.2962 | -0.5889 | 0.1043 |
| NP | -0.1340 | -0.1363 | 0.1257 | 0.1260 | 0.1455 | -0.4508 | 0.0049 |
| OSi | 0.1560 | -0.1896 | 0.1204 | 0.1718 | 0.1486 | -0.4262 | 0.0086 |
| PP | -0.1553 | -0.2403 | 0.0919 | 0.1021 | 0.1209 | -0.3866 | -0.0179 |
| SeSi | -0.0259 | -0.1854 | 0.0877 | 0.1049 | 0.1077 | -0.3751 | -0.0232 |
| SSi | 0.0168 | -0.1892 | 0.0954 | 0.1196 | 0.1177 | -0.3886 | -0.0155 |

# Table S7.C Data for diatomic molecules at LC-wPBE/aug-cc-pVDZ.

|      | $\delta^-$ | $\delta^+$ | $\Delta E_{ST}^{[v]}$ | $\Delta E_{DQ}^{[v],-}$ | $\Delta E_{DQ}^{[v],+}$ | $I$ | $A$ |
|------|--------|--------|--------|--------|--------|--------|--------|
| AlBr | 0.6683 | -0.2420 | 0.0720 | 0.1584 | 0.0949 | -0.3376 | -0.0138 |
| AlF  | 1.6990 | -0.2465 | 0.0832 | 0.2979 | 0.1104 | -0.3565 | 0.0004 |
| AsAs | -0.2210 | -0.2936 | 0.0705 | 0.0777 | 0.0998 | -0.3592 | -0.0326 |
| AsN  | -0.1695 | -0.2314 | 0.0993 | 0.1073 | 0.1291 | -0.4215 | -0.0162 |
| AsP  | 0.0201 | -0.0755 | 0.0791 | 0.0873 | 0.0856 | -0.3723 | -0.0295 |
| BBr  | 1.2954 | -0.2556 | 0.0508 | 0.1567 | 0.0683 | -0.3620 | -0.0091 |
| BCl  | 1.5932 | -0.2504 | 0.0568 | 0.1964 | 0.0757 | -0.3713 | 0.0013 |
| BF   | 2.1089 | -0.1803 | 0.0942 | 0.3574 | 0.1150 | -0.4082 | 0.0239 |
| BrBr | -0.1290 | 1.5524 | 0.2253 | 0.0769 | 0.0883 | -0.4021 | -0.0568 |
| BrF  | -0.1277 | 1.5179 | 0.2372 | 0.0822 | 0.0942 | -0.4435 | -0.0423 |
| BrGa | 0.4443 | -0.1862 | 0.0952 | 0.1689 | 0.1169 | -0.3554 | -0.0118 |
| BrLi | 0.8098 | 0.1983 | 0.2075 | 0.3134 | 0.1732 | -0.3470 | -0.0276 |
| BrNa | 0.9229 | 0.2273 | 0.1914 | 0.2999 | 0.1560 | -0.3225 | -0.0314 |
| ClAl | 0.8761 | -0.2506 | 0.0740 | 0.1853 | 0.0988 | -0.3445 | -0.0097 |
| ClBr | -0.1204 | 1.3847 | 0.2306 | 0.0851 | 0.0967 | -0.4158 | -0.0479 |
| ClCl | -0.1253 | 1.1335 | 0.2350 | 0.0963 | 0.1101 | -0.4343 | -0.0349 |
| ClF  | -0.1374 | 1.1080 | 0.2361 | 0.0966 | 0.1120 | -0.4751 | -0.0262 |
| ClGa | 0.5574 | -0.2020 | 0.0978 | 0.1908 | 0.1225 | -0.3663 | -0.0078 |
| ClLi | 0.8788 | 0.1660 | 0.2230 | 0.3594 | 0.1913 | -0.3679 | -0.0254 |
| CO   | 0.3320 | -0.1088 | 0.1957 | 0.2925 | 0.2196 | -0.5207 | 0.0534 |
| CS   | 0.3109 | -0.1509 | 0.1005 | 0.1552 | 0.1184 | -0.4245 | -0.0061 |
| CSe  | -0.0067 | -0.1690 | 0.0842 | 0.1006 | 0.1013 | -0.4031 | -0.0192 |
| FF   | -0.2542 | 1.3379 | 0.2977 | 0.0950 | 0.1273 | -0.5840 | -0.0081 |
| FGa  | 0.9553 | -0.2138 | 0.1072 | 0.2665 | 0.1363 | -0.3870 | 0.0015 |
| FLi  | 1.2739 | 0.0899 | 0.2631 | 0.5489 | 0.2414 | -0.4277 | -0.0181 |
| GeO  | 0.0485 | -0.1617 | 0.1190 | 0.1489 | 0.1420 | -0.4144 | -0.0101 |
| GeS  | -0.0299 | -0.1653 | 0.0998 | 0.1160 | 0.1196 | -0.3861 | -0.0245 |
| GeSe | 0.0224 | 0.0161 | 0.1008 | 0.1014 | 0.0992 | -0.3599 | -0.0295 |
| HAl  | 2.6566 | 0.6751 | 0.0974 | 0.2126 | 0.0582 | -0.2992 | -0.0117 |
| HB   | 6.8734 | -0.3447 | 0.0206 | 0.2476 | 0.0315 | -0.3526 | -0.0124 |
| HBr  | 0.1622 | 0.1963 | 0.2791 | 0.2712 | 0.2333 | -0.4344 | 0.0227 |
| HCl  | 0.2062 | 0.1822 | 0.3183 | 0.3248 | 0.2692 | -0.4707 | 0.0241 |
| HF   | 0.5473 | 0.1699 | 0.4327 | 0.5723 | 0.3699 | -0.6017 | 0.0253 |
| HGa  | 2.4328 | 0.4828 | 0.1037 | 0.2400 | 0.0699 | -0.3079 | -0.0064 |
| HLi  | 18.4640 | 0.1325 | 0.1294 | 2.2240 | 0.1143 | -0.2967 | -0.0156 |
| HNa  | 13.3465 | 0.1888 | 0.1114 | 1.3442 | 0.0937 | -0.2642 | -0.0163 |
| LiLi | 54.2306 | 0.0424 | 0.0403 | 2.1363 | 0.0387 | -0.1879 | -0.0177 |
| LiNa | 36.7076 | 0.1929 | 0.0421 | 1.3310 | 0.0353 | -0.1822 | -0.0178 |
| NaCl | 1.0077 | 0.2054 | 0.2060 | 0.3431 | 0.1709 | -0.3393 | -0.0296 |
| NaF  | 1.5326 | 0.1364 | 0.2312 | 0.5153 | 0.2035 | -0.3792 | -0.0235 |
| NaNa | 37.0700 | 0.8088 | 0.0588 | 1.2369 | 0.0325 | -0.1768 | -0.0181 |
| NN   | -0.0841 | -0.0817 | 0.2664 | 0.2657 | 0.2901 | -0.5894 | 0.0748 |
| NP   | -0.1464 | -0.1643 | 0.1235 | 0.1262 | 0.1478 | -0.4516 | -0.0087 |
| OSi  | 0.1277 | -0.1880 | 0.1189 | 0.1651 | 0.1464 | -0.4287 | -0.0078 |
| PP   | -0.1771 | -0.2537 | 0.0898 | 0.0990 | 0.1203 | -0.3874 | -0.0257 |
| SeSi | -0.0472 | -0.1908 | 0.0877 | 0.1033 | 0.1084 | -0.3774 | -0.0286 |
| SSi  | -0.0045 | -0.1938 | 0.0949 | 0.1172 | 0.1178 | -0.3907 | -0.0232 |

# Table S7.D Data for diatomic molecules at LC-wPBE/aug-cc-pVTZ.

|      | $\delta^-$ | $\delta^+$ | $\Delta E_{ST}^{[v]}$ | $\Delta E_{DQ}^{[v],-}$ | $\Delta E_{DQ}^{[v],+}$ | $I$ | $A$ |
|------|------|------|------|------|------|------|------|
| AlBr | 0.7935 | -0.2425 | 0.0711 | 0.1684 | 0.0939 | -0.3340 | -0.0119 |
| AlF  | 1.7867 | -0.2437 | 0.0829 | 0.3056 | 0.1097 | -0.3535 | 0.0010 |
| AsAs | 0.0290 | -0.0797 | 0.0712 | 0.0797 | 0.0774 | -0.3582 | -0.0329 |
| AsN  | -0.1597 | -0.2107 | 0.0997 | 0.1062 | 0.1263 | -0.4210 | -0.0171 |
| AsP  | 0.0398 | -0.0705 | 0.0802 | 0.0898 | 0.0863 | -0.3714 | -0.0295 |
| BBr  | 1.3461 | -0.2516 | 0.0502 | 0.1573 | 0.0670 | -0.3589 | -0.0085 |
| BCl  | 1.6649 | -0.2448 | 0.0562 | 0.1985 | 0.0745 | -0.3678 | 0.0020 |
| BF   | 2.1504 | -0.1638 | 0.0980 | 0.3690 | 0.1171 | -0.4042 | 0.0246 |
| BrBr | -0.1037 | 1.1554 | 0.2030 | 0.0844 | 0.0942 | -0.3977 | -0.0483 |
| BrF  | -0.1047 | 0.9454 | 0.1904 | 0.0876 | 0.0979 | -0.4391 | -0.0352 |
| BrGa | 0.4729 | 0.2399 | 0.1428 | 0.1696 | 0.1152 | -0.3526 | -0.0106 |
| BrLi | 0.8120 | 0.1810 | 0.2059 | 0.3159 | 0.1743 | -0.3465 | -0.0262 |
| BrNa | 0.9291 | 0.2236 | 0.1901 | 0.2996 | 0.1553 | -0.3216 | -0.0308 |
| ClAl | 1.0251 | -0.2503 | 0.0732 | 0.1978 | 0.0977 | -0.3404 | -0.0076 |
| ClBr | -0.0971 | 0.9682 | 0.2019 | 0.0926 | 0.1026 | -0.4115 | -0.0393 |
| ClCl | -0.1045 | 0.8470 | 0.2153 | 0.1044 | 0.1166 | -0.4300 | -0.0259 |
| ClF  | * | * | * | * | * | * | * |
| ClGa | 0.5980 | 0.1997 | 0.1448 | 0.1929 | 0.1207 | -0.3634 | -0.0065 |
| ClLi | 0.8754 | 0.1497 | 0.2224 | 0.3627 | 0.1934 | -0.3685 | -0.0240 |
| CO   | 0.3444 | -0.0953 | 0.1999 | 0.2971 | 0.2210 | -0.5198 | 0.0513 |
| CS   | 0.0997 | -0.1416 | 0.1000 | 0.1281 | 0.1165 | -0.4213 | -0.0059 |
| CSe  | 0.0176 | -0.1626 | 0.0832 | 0.1011 | 0.0993 | -0.4003 | -0.0193 |
| FF   | -0.2366 | 0.9799 | 0.2628 | 0.1013 | 0.1327 | -0.5813 | -0.0018 |
| FGa  | 1.0067 | 0.1191 | 0.1510 | 0.2708 | 0.1349 | -0.3846 | 0.0024 |
| FLi  | 1.2484 | 0.0834 | 0.2653 | 0.5506 | 0.2449 | -0.4306 | -0.0169 |
| GeO  | 0.0787 | -0.1579 | 0.1197 | 0.1533 | 0.1421 | -0.4125 | -0.0099 |
| GeS  | -0.0083 | -0.1596 | 0.0995 | 0.1175 | 0.1184 | -0.3833 | -0.0240 |
| GeSe | -0.1051 | 0.3853 | 0.1563 | 0.1010 | 0.1128 | -0.3589 | -0.0292 |
| HAl  | 2.6486 | -0.2979 | 0.0410 | 0.2133 | 0.0585 | -0.2994 | -0.0115 |
| HB   | 7.0387 | -0.3434 | 0.0204 | 0.2495 | 0.0310 | -0.3521 | -0.0125 |
| HBr  | 0.1601 | 0.1555 | 0.2680 | 0.2691 | 0.2320 | -0.4327 | 0.0180 |
| HCl  | 0.1935 | 0.0969 | 0.2947 | 0.3206 | 0.2686 | -0.4697 | 0.0202 |
| HF   | 0.5521 | 0.0995 | 0.4066 | 0.5739 | 0.3698 | -0.6006 | 0.0212 |
| HGa  | 2.4281 | 0.3166 | 0.0920 | 0.2395 | 0.0699 | -0.3086 | -0.0063 |
| HLi  | 17.6841 | 0.1210 | 0.1298 | 2.1633 | 0.1158 | -0.2973 | -0.0151 |
| HNa  | 12.7716 | 0.1882 | 0.1118 | 1.2960 | 0.0941 | -0.2649 | -0.0166 |
| LiLi | 51.9233 | 0.0241 | 0.0405 | 2.0920 | 0.0395 | -0.1880 | -0.0174 |
| LiNa | 35.1708 | 0.1458 | 0.0405 | 1.2794 | 0.0354 | -0.1822 | -0.0179 |
| NaCl | 1.0116 | 0.1917 | 0.2032 | 0.3429 | 0.1705 | -0.3389 | -0.0291 |
| NaF  | 1.5240 | 0.1269 | 0.2296 | 0.5142 | 0.2037 | -0.3796 | -0.0231 |
| NaNa | 36.4257 | 0.9044 | 0.0618 | 1.2143 | 0.0324 | -0.1769 | -0.0181 |
| NN   | -0.0692 | -0.0775 | 0.2730 | 0.2754 | 0.2959 | -0.5905 | 0.0770 |
| NP   | -0.1297 | -0.1376 | 0.1250 | 0.1261 | 0.1449 | -0.4518 | -0.0096 |
| OSi  | 0.1612 | -0.1863 | 0.1198 | 0.1709 | 0.1472 | -0.4277 | -0.0070 |
| PP   | -0.1515 | -0.2404 | 0.0915 | 0.1022 | 0.1204 | -0.3868 | -0.0251 |
| SeSi | -0.0256 | -0.1859 | 0.0876 | 0.1048 | 0.1076 | -0.3750 | -0.0278 |
| SSi  | 0.0178 | -0.1904 | 0.0950 | 0.1194 | 0.1173 | -0.3887 | -0.0220 |

*SCF convergence problem for ionic excited state(s)